\documentclass[12pt]{article}
\usepackage{german}
\usepackage[english]{babel}
\usepackage{amsmath}
\usepackage{graphicx}
\usepackage{multirow}
\usepackage{epsf}
\usepackage{babel}
\usepackage{amssymb}
\usepackage[utf8]{inputenc}
\usepackage{float}
\usepackage{epsfig}
\usepackage{rotating}
\usepackage{subfig}
\usepackage{float}
\usepackage{dlfltxbcodetips}
\usepackage[a4paper,hmargin=1.7cm,vmargin=3cm]{geometry}
\usepackage{url}
\usepackage{versions}
\usepackage[authoryear]{natbib}
\usepackage{color}
\usepackage{pstricks-add,pst-tree,pstricks}

\graphicspath{{graphics/}}
\makeatother

\newcommand{\coefaa}{$d=2, D_2=4, D_3=6$}
\newcommand{\coefaS}{$d=2, D_2=4$}
\newcommand{\coefb}{$d=3, D_2=6, D_3=3$}
\newcommand{\coefba}{$d=3, D_2=6, D_3=6$}
\newcommand{\coefbS}{$d=3, D_2=6$}

\newcommand{\condset}{s_{ij}} 

\newcommand{\SA}{simplifying assumption}

\renewcommand{\P}{\mathbb{P}}
\newcommand{\ps}{;\kern 0.08em}
\newcommand{\unicondset}{u_{ij}} 
\newcommand{\citeN}{\cite}
\newcommand{\shortciteN}{\cite}
\newcommand{\pvc}[1]{#1^{\text{\tiny PVC}}} 
\newcommand{\hl}{e} 
\newcommand{\hltilde}{\tilde{e}}  
\newcommand{\ubasis}[1]{{B}_{#1}}

\usepackage{pifont}

\begin{document}
\title{Estimating Non-Simplified Vine Copulas Using Penalized Splines}
\author{Christian Schellhase\thanks{Centre for Statistics, Bielefeld University, Department of Business Administration and Economics, Germany, cschellhase@wiwi.uni-bielefeld.de}  \and Fabian Spanhel\thanks{Department of Statistics, Ludwig-Maximilians-Universit\"at M\"unchen, Germany, spanhel@stat.uni-muenchen.de}}
\date{17.11.2016}

\maketitle      
\begin{abstract} 
Vine copulas (or pair-copula constructions) have become an important tool for high-dimensional dependence modeling. Typically, so called simplified vine copula models are estimated where bivariate conditional copulas are approximated by bivariate unconditional copulas. We present the first non-parametric estimator of a non-simplified vine copula that allows for varying conditional copulas using penalized hierarchical B-splines. Throughout the vine copula, we test for the simplifying assumption in each edge, establishing a data-driven non-simplified vine copula estimator. To overcome the curse of dimensionality, we approximate conditional copulas with more than one conditioning argument by a conditional copula with the first principal component as conditioning argument. An extensive simulation study is conducted, showing a substantial improvement in the out-of-sample Kullback-Leibler divergence if the null hypothesis of a simplified vine copula can be rejected. We apply our method to the famous uranium data and present a classification of an eye state data set, demonstrating the potential benefit that can be achieved when conditional copulas are modeled.\par  
\end{abstract}
\section{Introduction}
Simplified vine copulas, or pair-copula constructions, have become a very active field of research over the last decade (\citet{Bedford2002,Aas-etal:09,Kurowicka2011,ScheKau:12b,Brechmann2012}, \citeauthor{Spanhel2015} \citeyearpar{Spanhel2015}). Their popularity stems from their simplicity and flexibility. Simplified vine copula models decompose the complex modeling of a high-dimensional copula into the hierarchical modeling of several bivariate unconditional copulas. These bivariate unconditional copulas are often called pair-copulas and can be chosen arbitrarily. Due to this fact, simplified vine copula models give rise to very flexible multivariate copula models which are often found to be superior to other multivariate copula models \citep{Aas2009,Fischer2009}.  Moreover, if the data generating process (dgp) satisfies the simplifying assumption \citep{HobkHaff2010}, the dgp can actually be represented by a simplified vine copula.\par
However, in general, a simplified vine copula model is only an approximation of a multivariate copula. That is because bivariate conditional copulas are typically required as building blocks  to express an arbitrary copula by a vine copula. A simplified vine copula model then approximates the conditional copulas by bivariate unconditional copulas, see \citet{Spanhel2015} for a detailed investigation of such approximations. Depending on the dgp, the goodness of such an approximation can be adequate or insufficient. For theses reasons, the modeling of conditional copulas might improve the modeling of high-dimensional copulas. \par
The modeling of a conditional copula, without reference to the vine copula framework, has been investigated by several authors. \citet{Acar2011,Abegaz2012} and \citet{Vatter2015}, investigate the semi-parametric estimation of a conditional copula. That is, a parametric copula family is specified and its dependence parameter is treated as a non-parametric function of the conditioning argument(s). \citet{Gijbels2011,Veraverbeke2011,Gijbels2012} and \citet{Fermanian2012}, apply kernel-methods to extract a conditional copula from a conditional distribution function. The literature on estimating conditional copulas within the vine copula framework is less developed. In the three-dimensional case, \citet{Acar2012} propose a local likelihood estimator for a parametric conditional copula with one conditioning argument. \citet{Lopez2013} utilize a Gaussian process to model the parametric conditional copulas of a higher-dimensional vine copula. The approaches of \citet{Acar2012} and \citet{Lopez2013} are both semi-parametric and require the choice of a parametric copula family. The selection of the copula family is computationally expensive because one has to fit the conditional copula for each copula family in order to choose the best  family. Moreover, it is also possible that the conditional copula can not be  well approximated  by a parametric copula family with a varying dependence parameter. \par
The objective of this paper is to construct a non-parametric non-simplified vine copula estimator to improve the approximation of a multivariate distribution. For this purpose, we apply penalized B-splines to model the conditional copulas of a non-simplified vine copula. We show how penalized B-splines can be used to directly estimate a conditional copula so that there is no need to extract a conditional copula from a previously fitted conditional distribution as it is the case in the kernel-based approaches of \citet{Gijbels2011,Veraverbeke2011,Gijbels2012} and \citet{Fermanian2012}. A direct model of the conditional copula is beneficial when it comes to simulations, because the extraction of a conditional copula is computationally expensive if one wants to simulate from the fitted vine copula model.

In order to tackle the curse of dimensions, we approximate a bivariate conditional copula with a vector of conditioning arguments by a bivariate conditional copula with the first principal component of the conditioning variables as the conditioning argument. Simulation studies show that our approach can outperform parametric and non-parametric simplified vine copula estimators if the null hypothesis of an unconditional copula can be rejected. Moreover, an application to two data sets demonstrates that a substantial improvement can be achieved if conditional copulas are present and modeled. To the best of our knowledge, this paper presents the first effort to model a non-simplified vine copula using non-parametric methods. The rest of the paper is structured as follows. Section \ref{Sec.A} discusses (simplified) vine copulas and conditional copulas. In Section \ref{Sec.B} we construct an estimator for conditional copula densities using penalized B-splines. We describe the estimation of non-simplified regular vine copulas using conditional copulas in Section \ref{Sec.C}. Simulation results and two applications are presented in Section \ref{Sec.D}. A discussion in Section \ref{Sec.E} closes the article.

\section{Background on (simplified) vine copulas and conditional copulas}\label{Sec.A}
Consider a continuous $p$-dimensional random vector $X=(X_1,\dots,X_p)$. Following Sklar's \citeyear{Sklar:59} theorem, we can write the distribution of $X$ as
\begin{equation} \label{eq:copula1}
 F_{1:p}(x_1,\dots,x_p)=C_{1:p}\big(F_1(x_1),\dots,F_p(x_p)\big),\notag
\end{equation}
where $C_{1:p}$ is the copula of $X$ and $F_i$ is the marginal distribution of $X_i$, $i=1,\ldots,p$. Assuming that $X$ has a density $f_{1:p}$, we can express the density $f_{1:p}$ using the copula density $c_{1:p}$ and the marginal densities $f_i$, $i=1,\ldots,p$, by 
\begin{equation} \label{eq:copula2}
f_{1:p}(x_1,\dots,x_p) = c_{1:p}\big(F_1(x_1),\dots,F_p(x_p)\big) \prod_{j=1}^p f_j(x_j).\notag
\end{equation}
Regular vine copulas \citep{Bedford2002} provide a functional decomposition of a copula into a sequence of bivariate conditional copulas. For ease of presentation, we demonstrate this decomposition by considering the so called D-vine copulas which are a subset of regular vine copulas. Let $\condset = i+1,\ldots,i+j-1$, and denote $u_{k|\condset} := F_{k|\condset}(x_k|x_{\condset})$ for $k=i,i+j$, with $u_{k|\condset}:=u_k$ for $j=1$. The copula density $c_{1:p}$ of $X$ can be represented by
\begin{align*} 
 c_{1:p}(u_1,\dots,u_p)  & =  \prod_{j=1}^{p-1}
 \underbrace{
 \prod_{i=1}^{p-j}c_{i,i+j|\condset }\big(u_{i|\condset},u_{i+j|\condset}
 |u_{\condset}\big)}_{\text{conditional copulas of the j-th tree}} 
\end{align*}
where $c_{i,i+j|\condset}$ is the density of the conditional copula of $(X_i,X_{i+j})$ given $X_{\condset}$ \citep{Patton2006b} with $c_{i,i+j|\condset}=c_{i,i+1}$ for $j=1$. For instance, in three dimensions the copula density $c_{1:3}$ can be decomposed into
\begin{align}
\label{3dex}
c_{1:3}(u_1,u_2,u_3) & = \underbrace{c_{12}(u_1,u_2)\cdot c_{23}(u_2,u_3)}
_{\text{copulas of the first tree}} \cdot \underbrace{c_{13|2}\big(u_{1|2},u_{3|2}|u_2\big),}_{
\text{conditional copula of the second tree}}
\end{align}
where $u_{1|2} = F_{1|2}(x_1|x_2)$ and $u_{3|2}=F_{3|2}(x_3|x_2)$. Vine copulas can be considered as an ordered sequence of trees, where $j$ refers to the number of the tree and a bivariate conditional copula $C_{i,i+j|\condset}$  is assigned to each of the $K-j$ edges of tree $j$. The conditional copula $C_{i,i+j|\condset}$ in terms of \citet{Patton2006b} is defined by 
\begin{align*}
C_{i,i+j|\condset}(a,b|u_{\condset}) &:= \P(U_{i|\condset}\leq a, U_{i+j|\condset}\leq b|U_{\condset}=u_{\condset}),
\end{align*}
where $(a,b,u_{\condset})\in [0,1]^{j+1}$. Note that each bivariate conditional copula $C_{i,i+j|\condset}$ is in general a function of $j+1$ variables. Due to their general form, D-vine copulas do not provide a feasible model framework without further assumptions. For this reason, one commonly assumes that the dgp satisfies the \SA{}. That is, one assumes that $C_{i,i+j|\condset}(\cdot,\cdot|x_{\condset})$  does not depend on $x_{\condset}$ for all $j=1,\ldots,p-1,i=1,\ldots,d-j$, i.e., one postulates that $C_{i,i+j|\condset}$ depends not on j+1 but only on two variables. The validity of the \SA{} is true for the multivariate Gaussian, Student-t and Clayton copula. However, it is not true in general \citep{HobkHaff2010,Stober2013b}.

If the \SA{} does not hold for the dgp, a pair-copula construction that is based on bivariate unconditional copulas only approximates the underlying copula. The question which simplified vine copula optimally approximates the non-simplified vine copula is investigated in \citet{Spanhel2015}. In practice, pair-copula constructions are commonly estimated in a stepwise fashion, i.e., after the first tree has been estimated one continues with the estimation of the second tree and so on. The resulting best simplified vine copula approximation that results from this sequential procedure and minimizes the Kullback-Leibler divergence from the true copula is the partial vine copula (PVC). The partial vine copula consists of two-dimensional $\textbf{(j-1)}$-th order partial copulas $\pvc{C}_{i,i+j \ps \condset}$ and its density is given by
\begin{align}
\label{eq:SDvine}
 \pvc{c}_{1:p}(u_1,\dots,u_p)  & =  \prod_{j=1}^{p-1}\prod_{i=1}^{p-j}\pvc{c}_{i,i+j\ps \condset }\big(  \pvc{u}_{i|\condset},\pvc{u}_{i+j|\condset}\big), 
\end{align}
where $\pvc{u}_{k|\condset} = \pvc{F}_{k|\condset}(x_k|x_{\condset}),  k=i,i+j,$ is given in Definition 4.1 in \citet{Spanhel2015} and $\pvc{c}_{i,i+j|\condset}=c_{i,i+1}$ for $j=1$. For instance, in three dimensions the partial vine copula is given by
\begin{align*}
\pvc{c}_{1:3}(u_1,u_2,u_3) & = \pvc{c}_{12}\big(u_1,u_2\big) \cdot \pvc{c}_{23}\big(u_2,u_3\big)
\cdot \pvc{c}_{13;2}\big(\pvc{u}_{1|2},\pvc{u}_{3|2}\big)\\
& = c_{12}\big(u_1,u_2\big)\cdot c_{23}\big(u_2,u_3\big) \cdot \pvc{c}_{13;2}\big(u_{1|2},u_{3|2}\big),
\end{align*}
and it holds that
\begin{align*} 
 \pvc{c}_{13;2}(u_{1|2},u_{3|2})= \int_0^1 c_{13|2}(u_{1|2},u_{3|2}|u_2)du_2.
\end{align*}
Note that $c_{1:p}=\pvc{c}_{1:p}$ if the simplifying assumption holds, i.e., each conditional copula $C_{i,i+j|\condset}$ is equal to the corresponding $\textbf{(j-1)}$-th order partial copula $\pvc{C}_{i,i+j \ps \condset}$ and the density of $C_{1:p}$ is given by a product of $p(p-1)/2$ unconditional bivariate copula densities.
\subsection{Illustrating the Partial Vine Copula and the simplifying assumption}
We illustrate the relation between a non-simplified vine copula and its corresponding partial vine copula with the following three-dimensional non-simplified vine copula $C_{1:3}$ which is also used for the simulation study in Section \ref{Sec.D}. Let $C_{12}$ and $C_{23}$ be Frank copulas \citep{Nelsen:06} with Kendall's $\tau=0.25$ and assume that $C_{13|2}(\cdot,\cdot|u_2)$ is given by a Frank copula such that Kendall's $\tau$ is given by $\tau(u_2) = 0.4-0.8u_2$. The upper left panel of Figure \ref{frankex} shows Kendall's $\tau$ of the conditional copula $C_{13|2}$ as a function of $u_2$. The variation of the conditional copula $C_{13|2}$ is further illustrated in the upper right and the lower left panel of Figure \ref{frankex}.
The upper right panel shows 1000 observations from the random vector $(U_{1|2},U_{3|2})$ if we condition on the event $U_2<0.5$ whereas the lower left panel shows 1000 realizations from $(U_{1|2},U_{3|2})$ if we condition on the opposite event $U_2>0.5$. In other words, the upper right panel shows realizations from the copula 
\begin{equation*}
C^{\star}(u_{1|2},u_{3|2}) = \int_0^{0.5} C_{13|2}(u_{1|2},u_{3|2}|u_2)du_2
\end{equation*}
whereas the lower left panel shows realizations from the copula
\begin{equation*}
C^{\dagger}(u_{1|2},u_{3|2}) = \int_{0.5}^1 C_{13|2}(u_{1|2},u_{3|2}|u_2)du_2
\end{equation*}
Since Kendall's $\tau$ of the conditional copula is positive for $0\leq u_2< 0.5$ and negative for $0.5<u_2\leq 1$, we observe a slightly positive dependence ($\hat{\tau} \approx 0.2)$ in the upper right panel and a slightly negative dependence ($\hat{\tau} \approx -0.2)$ in the lower left panel. If one uses the partial vine copula to approximate this dgp by a simplified vine copula, one tries to model the data given in the lower right panel of Figure \ref{frankex} which shows 2000 observations from the random vector $(U_{1|2},U_{3|2})$. Expressed differently, these realizations are drawn from the first-order partial copula
\begin{equation*}
\pvc{C}_{13;2}(u_{1|2},u_{3|2}) = \int_0^1 C_{13|2}(u_{1|2},u_{3|2}|u_2)du_2.
\end{equation*}
We see that the positive and negative dependence is averaged out and that the sample resembles a realization from the independence copula with no clear direction of dependence ($\hat{\tau} \approx 0$). Although the partial copula $\pvc{C}_{13;2}$ is not the independence copula, it is very close to the independence copula in terms of Kullback-Leibler divergence, so that the partial vine copula approximation of the data generating process \eqref{3dex} may be written as
\begin{align*}
\pvc{c}_{1:3}(u_1,u_2,u_3) & \approx c_{12}(u_1,u_2)\cdot c_{23}(u_2,u_3).
\end{align*}
That is, the dependence between $(U_1,U_3)$ conditional on $U_2$ is pretty much ignored and only the dependence in the first tree is modeled.

Up to now, it is still unclear how appropriate the simplifying assumption is in practical applications. Although it is implausible that the simplifying assumption is strictly true in applications, it may be adequate because the distance between the partial vine copula and the data generating non-simplified vine copula is negligible. For instance, if the null hypothesis of a simplified vine copula can not be rejected, the additional effort of modeling bivariate conditional copulas may not pay off. However, if the null hypothesis of a simplified vine copula can be rejected, the approximation may not be sufficient and one could improve the modeling of the dgp by incorporating conditional copulas. Moreover, the interpretation of a partial vine copula approximation is not so clear in this case because spurious (un)conditional (in)dependencies may emerge, meaning that two random variables seem to be (un)conditionally independent when actually they are not \citep{Spanhel2015}. Therefore, we introduce a framework that tests for the \SA{} and accounts for the conditional copulas of a vine copula if the simplifying assumption can be rejected.
\subsection{{Overcoming the Curse of Dimensions}}
It is not useful to model all conditional copulas without imposing any restrictions. Without any constraints, a vine copula density decomposes a p-dimensional copula density $c_{1:p}$ into $p-j$ functions of $j+1$ variables, where $j=1,\ldots,p-1$. Thus, the conditional copula density in the last tree has the same dimensionality as $c_{1:p}$ and it would be more sensible to directly estimate the $p$-dimensional copula density $c_{1:p}$ in this case. Moreover, if one tries to consistently estimate each conditional copula $C_{i,i+j|\condset}$, then one ultimately runs into the curse of dimensions. Consequently, we have to impose constraints that are weaker than the \SA{} but still yield a modeling framework which overcomes the curse of dimensions.\par
We propose to approximate a conditional copula $C_{i,i+j|\condset}(\cdot,\cdot|x_{\condset})$ with $j-1$ conditioning arguments $x_{\condset}$ by a conditional copula $\tilde{C}_{i,i+j|\unicondset}(\cdot,\cdot|v_{ij})$ with one conditioning argument $v_{ij}$, where $v_{ij}$ is a function of $x_{\condset}$. We follow this approach because non-parametric estimators of a bivariate conditional copula suffer greatly from the curse of dimensions if the number of conditioning arguments is larger than one \citep{Scott2008,Nagler2015}. Once the conditional copula density has more than one conditioning argument, we use the first principal component of the $j-1$ conditioning arguments $x_{\condset}$ as the scalar conditioning argument $v_{ij}$.\par
To fix ideas, let $k\in s_{ij}$ and $l=1,\ldots,n$, where $n$ is the size of the observed sample, so that $u_{k,l} = (n+1)^{-1}\sum_{i=1}^n 1\!\!\!1_{x_{k,i}\leq x_{k,l}}$ is the standardized rank of observation $x_{k,l}$. Moreover, let $\bar{u}_k$ be the sample mean of $u_k = (u_{k,l})_{l=1,\ldots,n}$ so that $\tilde{u}_k = u_k-\bar{u}_k$ has zero mean. Applying a principal component analysis to $\tilde{u}_{s_{ij}}= (\tilde{u}_{k})_{k\in s_{ij}}$ gives the values of the first principal component $(v_{ij,l})_{l=1,\ldots,n}$. The standardized ranks of the first principal component $(v_{ij,l})_{l=1,\ldots,n}$ are used to approximate $C_{i,i+j|\condset}(\cdot,\cdot|x_{\condset,l})$ by $\tilde{C}_{i,i+j|\condset}(\cdot,\cdot|v_{ij,l})$ for $l=1,\ldots,n$. Note that if $j=2$ then $v_{ij} = u_{\condset}$. 
We use the first principal component as the scalar conditioning variable for the conditional copula, since the first principal component typically accounts for much variability of the conditioning variables. Modeling the conditional copula density with the standardized ranks of the first principal component covers the variation of the conditioning arguments successfully as presented in simulations and applications in Section \ref{Sec.D}. Moreover, the approach is computationally cheap. Alternative approaches that reduce the dimension of the conditioning vector to a scalar are deliberately left open for further research. 

\section{Penalized B-Spline Estimation of a Conditional Copula Density} \label{Sec.B}
The estimation of $p$-dimensional unconditional copula densities by means of penalized (hierarchical) B-splines has been investigated in \citeN{KauScheRup:12}. We now extend this approach to a bivariate conditional copula density with one conditioning argument. For this purpose, we consider exemplary the conditional copula $C_{12|3}$ of $F_{12|3}$, where $F_{12|3}$ is the conditional cumulative distribution function (cdf) of $(X_1,X_2)$ given $X_3$. Recall that the conditional copula density $c_{12|3}$ has to satisfy the following constraints
\begin{align} \label{eq:constraint-1}
\forall (u_1,u_2,u_3)\in(0,1)^3\colon& c_{12|3}(u_1,u_2|u_3) \geq 0, \\
\label{eq:constraint-2}
\forall (u_1,u_2,u_3)\in(0,1)^3\colon& \int_0^1 c_{12|3}(u_1,u_2|u_3)\text{ d}u_1 =  \int_0^1 c_{12|3}(u_1,u_2|u_3)\text{ d}u_2 = 1.
\end{align}
The first constraint ensures that the conditional copula density is non-negative while the second constraint guarantees that each marginal density is the density of the uniform distribution. Note that the second constraint implies that $c_{12|3}$ integrates to one, i.e.,
\begin{equation*}
\forall u_3\in(0,1)\colon \int_{[0,1]^2} c_{12|3}(u_1,u_2|u_3)\text{ d}u_1\text{ d}u_2=1.
\end{equation*}
The constraint given in \eqref{eq:constraint-2} can be imposed by extracting the conditional copula density from the underlying conditional density using
\begin{align} \label{extract}
c_{12|3}(u_1,u_2|u_3) =\frac{f_{12|3}(F_{1|3}^{-1}(u_1|u_3),F_{2|3}^{-1}(u_2|u_3)|u_3)} {f_{1|3}(F_{1|3}^{-1}(u_{1}|u_3)|u_3)f_{2|3}(F_{2|3}^{-1}(u_{2}|u_3)|u_3)}.\notag
\end{align}
For the extraction of a conditional copula we have to estimate the conditional densities \linebreak $(f_{1|3}, f_{2|3}, f_{12|3})$ and compute the quantile functions $F_{1|3}^{-1}$ and $F_{2|3}^{-1}$. Note that the quantile functions $F_{1|3}^{-1}$ and $F_{2|3}^{-1}$ must be consistent with the conditional densities $f_{1|3}$ and $f_{2|3}$, i.e., they can not be estimated separately. Therefore, the quantile functions are evaluated by inverting the conditional cdfs $F_{1|3}$ and $F_{2|3}$. In general, the inversion of these conditional cdfs can only be accomplished by numerical inversion methods.  Thus, a repeated evaluation of the conditional copula density which is extracted from a conditional density can computationally be rather expensive. For this reason, we aim at modeling the conditional copula $c_{12|3}$ directly using a mixture of B-spline basis densities.\par
\subsection{Sparse B-Spline Density Basis}
\label{bsplines}
The construction of our estimator is based on the idea to estimate (conditional) copula densities as a weighted sum of linear B-splines, constructed on a set of $K$ equidistant knots. Let $\mathbf{u}_j = (u_{j,1},\ldots,u_{j,n})'$, $\tau= (\tau_k)_{k=1,\ldots,K}$ be a tuple of equidistant knots with $\tau_1=0$ and $\tau_K=1$, and $\phi_{\tau_k}(x), x\in [0,1]$ be a regular linear univariate B-spline normalized to be a density, i.e., $\int {\phi_{\tau_k}}(x) \text{  d}x =1$. The $K$-dimensional univariate B-spline density basis for  $\mathbf{u}_j$ is given by
\begin{equation*} \ubasis{}^{(\tau)}(\mathbf{u}_j):=\begin{pmatrix} \phi_{\tau_1}(u_{j,1}) & \dots & \phi_{\tau_K}(u_{j,1})\\
\vdots & & \vdots\\
\phi_{\tau_1}(u_{j,n}) & \dots & \phi_{\tau_K}(u_{j,n})\end{pmatrix}.
\end{equation*}
The three-dimensional full tensor product for $(\mathbf{u}_1,\mathbf{u}_2,\mathbf{u}_3)$ follows as
\begin{equation*}
{\Phi}_{K} (\mathbf{u}_1,\mathbf{u}_2,\mathbf{u}_3) := \bigotimes_{j=1}^3 \ubasis{}^{(\tau)}(\mathbf{u}_j). 
\end{equation*}
and the corresponding approximation of the conditional copula density using B-splines is given by
\begin{align} \label{eq:basisrep}
c_{12|3}(\mathbf{u}_1,\mathbf{u}_2|\mathbf{u}_3; \mathbf{b}) := \Phi_{K}(\mathbf{u}_1,\mathbf{u}_2,\mathbf{u}_3)\mathbf{b} = \sum_{1 \leq k_1,k_2,k_3\leq K} \!\!b_{k_1,k_2,k_3}\ {\prod_{j=1}^3\phi_{\tau_{k_j}}(\mathbf{u}_j)},
\end{align}
where $\mathbf{b}:=(b_{k_1,k_2,k_3}\colon 1\leq k_1,k_2,k_3\leq K)$ are the spline basis coefficients. The goodness of the approximation depends on $K$ which determines the dimension of the basis ${\Phi}_{{K}}$ and therefore the number of coefficients in $\mathbf{b}$. Obviously, $K$ can not be chosen too large as the increasing amount of coefficients  boosts the computational demand dramatically.\footnote{ For instance, choosing $K=9$ in \eqref{eq:basisrep} results in 729 coefficients (see Table \ref{Table1}).} In order to handle the increasing amount of coefficients for increasing $K$, we follow \citet{KauScheRup:12} and make use of Zenger's so called sparse grids \citep{Zenger:91} to obtain a reduced basis to establish numerical feasibility. \par
To implement the approach, let the {linear univariate} B-spline density basis {$B^{(\tau(d))}$} be built upon $2^d+1$ equidistant knots {which are collected in the tuple $\tau(d)= (\tau_i(d))_{i=1,\ldots,2^d+1}=(k 2^{-d})_{k=0,\ldots,2^d}$}. Note that the basis $B^{(\tau(d))}$ has dimension $K=2^d+1$, i.e., the number of knots depend on $d$. The full tensor product of the corresponding B-spline basis is given by
\begin{equation}\label{eq:fullbase}
\boldsymbol{\Phi}^{(d)}(\mathbf{u}_1,\mathbf{u}_2,\mathbf{u}_3)=\bigotimes_{j=1}^3 B^{(\tau(d))}(\mathbf{u}_j).
\end{equation}
Following the construction principle of a hierarchical B-spline basis for copula estimation in \citet{KauScheRup:12}, we reduce the spline coefficients of the full tensor product in \eqref{eq:fullbase}. To this end, we omit some knots of the full tensor product given in \eqref{eq:fullbase} to construct the sparse B-spline basis $\mathbf{\tilde{\Phi}}^{(d,D)}(\mathbf{u}_1,\mathbf{u}_2,\mathbf{u}_3)$. 
The construction of $\mathbf{\tilde{\Phi}}^{(d,D)}(\mathbf{u}_1,\mathbf{u}_2,\mathbf{u}_3)$ and the corresponding notation are presented in detail in Appendix \ref{App:Bspline} (see \eqref{eq:sparsegrid}). The index $d$ is the degree of the univariate B-spline basis and of the univariate hierarchical B-spline basis. The index $D, d\leq D\leq 3d,$ refers to the maximum \emph{cumulated} hierarchy level and determines the sparsity. The lower $d$ the sparser the basis. Moreover, we have that $\tilde{\mathbf{\Phi}}^{(d,3d)}=\tilde{\mathbf{\Phi}}^{(d)}$, i.e., setting $D=3d$ yields the full tensor product basis, while $D<3d$ results in a sparse basis. According to the construction of $\mathbf{\tilde{\Phi}}^{(d,D)}(\mathbf{u}_1,\mathbf{u}_2,\mathbf{u}_3)$, the corresponding spline coefficient vector $\boldsymbol{\tilde{b}}^{(d,D)}$ is a reordered and reduced form of $\mathbf{b}$. The reduction of the basis decreases the number of spline coefficients and thereby the numerical effort as can be seen in Table \ref{Table1} where the dimension of $\mathbf{\tilde{\Phi}}^{(d,D)}$ is shown for various values of $d$ and $D$. The approximation of the conditional copula density using the sparse B-spline basis $\mathbf{\tilde{\Phi}}^{(d,D)}(\mathbf{u}_1,\mathbf{u}_2,\mathbf{u}_3)$ is given by
\begin{align}
\label{condcopsparse}
c_{12|3}(\mathbf{u}_1,\mathbf{u}_2|\mathbf{u}_3;\boldsymbol{\mathbf{\tilde{b}}}_{}^{(d,D)}) := \tilde{{\mathbf{\Phi}}}^{(d,D)}(\mathbf{u}_1,\mathbf{u}_2,\mathbf{u}_3)  \tilde{\mathbf{b}}^{(d,D)} = \sum_{\substack{1\leq k_1,k_2,k_3 \leq K\\  \!\!\!\! \sum_{i=1}^3 e_{k_i}\leq D}} \tilde{\mathbf{b}}_{k_1,k_2,k_3}^{(d)} \prod_{j=1}^3 \tilde{\mathbf{B}}_{k_j}^{(\tau(d))}(\mathbf{u}_j),
\end{align}
where $\tilde{\mathbf{B}}_{k_j}^{(\tau(d))}(\mathbf{u}_j)$ is the $k_j$-th column of the hierarchical B-spline basis $\tilde{\mathbf{B}}^{(\tau(d))}(\mathbf{u}_j)$, which is defined by the construction of the sparse B-spline basis (see \eqref{eq:hierBspl} in  Appendix \ref{App:Bspline}). To achieve a conditional copula density, some constraints on the coefficients $\boldsymbol{\tilde{b}}^{(d,D)}$ have to be formulated, which we discuss in the next sections.
\subsection{Estimation and Penalization}
To construct the likelihood for the spline coefficients $\mathbf{b,}$ and later on for $\boldsymbol{\tilde{b}}^{(d,D)}$, assume we observe an iid random sample ${\bf x}_i=(x_{i1},x_{i2},x_{i3})$, $i=1,\dots,n$, from which we obtain ${ u}_i=(u_{i1},u_{i2},u_{i3})$ using $u_{ij}=\hat{F}_j(x_{ij})$, where $\hat{F}_j(\cdot)$ is a (consistent) estimate of the marginal distribution function. Let $\mathbf{1}_n=(1\ldots,1)' \in \mathbb{R}^{n\times 1}$. According to \eqref{eq:basisrep}, the log likelihood for the tensor product depending on $\mathbf{b}$ is then
\begin{align} \label{eq:penlik}
 l(\mathbf{b})= \mathbf{1}^\top_n \log\{\Phi_K(\mathbf{u}_1,\mathbf{u}_2,\mathbf{u}_3)  \mathbf{b}\}  = \sum_{i=1}^n \log \left\{ \sum_{1\leq k_1,k_2,k_3\leq K}\!\!\! b_{k_1,k_2,k_3}\ \prod_{j=1}^3\phi_{\tau_{k_j}}({u}_{j,i})\right\},
\end{align}
which needs to be maximized subject to the constraints \eqref{eq:constraint-1} and \eqref{eq:constraint-2}. The accuracy of the spline approximation in \eqref{eq:basisrep} improves for large $K$, but the corresponding fit will suffer from estimation variability due to over-parameterization of the data. We briefly present the concept of penalization used in \shortciteN{KauScheRup:12}, who impose a penalty on the spline coefficients ${\mathbf b}$ to achieve a smooth fit. \citeN{EilMar:96} suggest to penalize $r$-th order differences for the B-spline coefficients and this framework can easily be extended to the multivariate setting as shown in \cite{MarxEilers:05}. Let $L \in \mathbb{R}^{(K-r)\times K}$ be a difference matrix of order $r$, e.g. for $r=1$ we get
\begin{equation*}L=
\begin{pmatrix}
1 & -1 & 0 & \cdots & 0\\
0 &  1 &-1 & \cdots & 0 \\
\vdots & \vdots &\vdots & \ddots & \vdots \\
0 & 0 &0 &1&-1
\end{pmatrix},
\end{equation*}
and let $W = \textrm{diag}(w_1,\dots,w_K)$ be the weight matrix linking a regular B-spline basis to a B-spline density basis, i.e. $w_l$ is the integral from 0 to 1 of the $l$-th regular B-spline. By means of $L$ we can now penalize differences in neighbouring spline coefficients and define the penalty matrix $P=W L^\top L W$, see also \cite{WandOrme:08} and \cite{Ruppert-etal:03}. This penalty applies only to a single dimension. To achieve smoothness of the fitted copula density for all variables, we use the Kronecker product yielding the entire penalty matrix $\mathbf{P}(\boldsymbol\lambda)=\lambda \sum_ {j=1}^3  \mathbf{P}_j$, where $\mathbf{P}_j=\left( \bigotimes_{l=1}^{j-1} I_K \right) \otimes P \otimes \left( \bigotimes_{l=j+1}^3 I_K \right)$, and $\bigotimes_{l=1}^{j-1}$ is the component-by-component tensor product (with $\bigotimes_{l=1}^0 I_K = 1 =$ $\bigotimes_{l=3+1}^3 I_K$). The coefficient $\lambda$ is the penalty parameter which needs to be selected in a data driven manner, as discussed later. Incorporating the penalty into the log-likelihood gives the penalized log-likelihood
\begin{align} \label{eq:penlik2}
 l_{pen}(\mathbf{b},\boldsymbol\lambda) = l({\mathbf{b}})-\frac12 \mathbf{b}^\top \mathbf{P}(\boldsymbol\lambda)\mathbf{b},
\end{align}
which is maximized for given $\lambda$ with respect to $\mathbf{b}$. Note that $\lambda$ determines the amount of smoothness for the fitted coefficients and setting $\lambda=0$ gives the usual ML estimate without any penalty.\par
The penalized log-likelihood \eqref{eq:penlik2} is now reformulated by replacing the B-spline basis in \eqref{eq:basisrep} with their hierarchical representation in $\mathbf{\tilde{\Phi}}^{(d,D)}(\mathbf{u}_1,\mathbf{u}_2,\mathbf{u}_3)$. Reported in the Appendix \ref{App:Bspline}, $\tilde{\mathbf{b}}^{(d)}={(\otimes_{j=1}^3\tilde{A})}^{-1}\mathbf{b}$ denotes the corresponding spline coefficient vector for the hierarchical basis \linebreak $\tilde{\mathbf\Phi}^{(d)}(\mathbf{u}_1,\mathbf{u}_2,\mathbf{u}_3)$. The penalized log-likelihood \eqref{eq:penlik2} as a function of the coefficients of the hierarchical B-spline basis $\tilde{\mathbf{b}}{^{(d)}}$ takes the form
\begin{equation*}
\tilde{l}_{pen}(\tilde{\mathbf{b}}{^{(d)}},\boldsymbol\lambda)=\tilde{l} (\tilde{\mathbf{b}}{^{(d)}})-\frac12 \tilde{\mathbf{b}}{^{(d)}}^\top \tilde{\mathbf{P}}(\boldsymbol\lambda) \tilde{\mathbf{b}}{^{(d)}},
\end{equation*}
where the log-likelihood is defined as 
\begin{equation*}
\tilde{l}(\tilde{\mathbf{b}}{^{(d)}})= {\mathbf{1}^{\top}_n\log\{\tilde{\mathbf{\Phi}}^{(d)}(\mathbf{u}_1,\mathbf{u}_2,\mathbf{u}_3)  \tilde{\mathbf{b}}{^{(d)}}\}}.
\end{equation*}
Furthermore, we define the penalty matrix $\tilde{\mathbf{P}}(\boldsymbol\lambda)=\lambda \sum_{j=1}^3 \tilde{\mathbf{P}}_j$ with
\begin{equation*}
\tilde{\mathbf{P}}_j=\left(\bigotimes_{l=1}^{j-1}\tilde{I}_{(d)}\right) \otimes \{{\tilde A}^\top P {{\tilde A}}\} \otimes \left(\bigotimes_{l=j+1}^3 \tilde{I}_{(d)}\right),
\end{equation*}
and $\tilde{I}_{(d)} = (W{\tilde A})^\top (W{\tilde A})$. Finally, the penalized log-likelihood using a sparse B-spline basis equals
\begin{align} \label{eq:penlikhier}
 \tilde{l}_{pen}^{(D)}(\tilde{\mathbf{b}}^{(d,D)},\boldsymbol{\lambda})=\tilde{l}^{(D)}(\tilde{\mathbf{b}}^{(d,D)})-\frac12 \tilde{\mathbf{b}}^{{(d,D)}^\top} \tilde{\mathbf{P}}^{(D)}(\boldsymbol{\lambda})  \tilde{\mathbf{b}}^{(d,D)}
\end{align}
with obvious definition for $\tilde{l}^{(D)}(\tilde{\mathbf{b}}^{(d,D)})$ and $\tilde{\mathbf{P}}^{(D)}(\boldsymbol{\lambda})= {\mathcal{E}({\cal O}_D)^\top} \tilde{\mathbf{P}}(\boldsymbol{\lambda})\mathcal{E}({\cal O}_D).$  $\mathcal{E}({\cal O}_D)$ is a orthogonal matrix extracting the corresponding columns of the full penalty matrix according to $\tilde{\mathbf{b}}^{(d,D)}$ (see Appendix \ref{App:Bspline}). Note that since $\tilde{\mathbf{b}}^{({d},3d)}=\tilde{\mathbf{b}}^{(d)}$ we have $\tilde{l}^{(3d)}=\tilde{l}$.
\subsection{Constraints on the Parameters}
We have to formulate constraints on the coefficient vector ${\mathbf{\tilde{b}}^{(d,D)}}$ such that $c_{12|3}(\mathbf{u}_1,\mathbf{u}_2|\mathbf{u}_3;{\mathbf{\tilde{b}}^{(d,D)}})$ in \eqref{condcopsparse} becomes a conditional copula density. The first marginal density of $c_{12|3}(\mathbf{u}_1,\mathbf{u}_2|\mathbf{u}_3)$ (see \eqref{eq:constraint-2}) follows as
\begin{align*}
\int \tilde{\mathbf{\Phi}}^{(d,D)}(\mathbf{u}_1,\mathbf{u}_2,\mathbf{u}_3) \tilde{\mathbf{b}}^{(d,D)} \text{d}\mathbf{u}_1 = \sum_{\substack{1\leq k_2,k_3\leq K}} \tilde{b}_{k_2,k_3}^{(d,D)} \prod_{j=2}^3 \tilde{\mathbf{B}}_{k_j}^{(\tau(d))}(\mathbf{u}_j)
\end{align*}
where $\tilde{b}^{(d,D)}_{k_2,k_3} = \sum_{\substack{1\leq k_1 \leq K\\ \!\!\! \sum_{i=1}^3\hl_{k_i}\leq D}} \tilde{b}^{(d)}_{k_1,k_2,k_3}.$
%
%
To guarantee that the first marginal density is one, we impose the following constraint on the spline coefficients evaluated at the knots $\tau(d)$
\begin{eqnarray} \label{eq:margd3}
\sum_{\substack{1\leq k_2,k_3\leq K}} \tilde{b}_{k_2,k_3}^{(d)} \prod_{j=2}^3 \tilde{\mathbf{B}}_{k_j}^{(\tau(d))}(\tau(d)) &=&\mathbf{1}_K. \label{eq:constraint03}
\end{eqnarray}
Note that \eqref{eq:margd3} yields linear constraints for $\tilde{\mathbf{b}}^{(d,D)}$ which are easy to implement. Identical calculations are done for the second  marginal density of $c_{12|3}(\mathbf{u}_1,\mathbf{u}_2|\mathbf{u}_3)$ which is given by \linebreak $\int \tilde{\mathbf{\Phi}}^{(d,D)}(\mathbf{u}_1,\mathbf{u}_2,\mathbf{u}_3) \tilde{\mathbf{b}}^{(d,D)} \text{d}\mathbf{u}_2$. To ensure that the density is non-negative we impose that
\begin{equation} \label{eq:constraint02}
 c_{12|3}(\mathbf{u}_1,\mathbf{u}_2|\mathbf{u}_3;\tilde{\mathbf{b}}^{(d,D)})\geq 0
\end{equation}
holds at the knots locations of the sparse B-spline density basis.
\subsection{Selection of the Penalty Parameter}
\label{penaltypar}
The penalty parameter $\lambda$ in \eqref{eq:penlikhier} needs to be selected adequately, that is the selection should be driven based on the data at hand. We borrow results from the spline smoothing literature and comprehend the penalty as normal prior, which is imposed on the spline coefficient vector (see e.g. \citet{Wahba:85}, \citet{Stein:90} or \citet{Efron:01}. The idea has been extended to penalized spline estimation presented in \citet{Ruppert-etal:03} 
and is being used here as well. To do so we adopt a Bayesian viewpoint and comprehend the penalty as parameter of {\sl 'a priori'} normal distribution on the spline coefficient such that
\begin{equation} \label{prior}
\boldsymbol{\tilde{b}^{(d,D)}} \sim N(0, \lambda^{-1} \mathbf{\tilde{P}}^{-}) 
\end{equation}
where $\mathbf{\tilde{P}}^{-}$ denotes the (generalized) inverse of the penalty matrix $\tilde{\mathbf{P}}^{(D)}$. The penalty parameter now plays the role of a (hyper) parameter in the prior distribution which can be estimated by maximizing the resulting likelihood. As sketched in \citet{ScheKau:12b} for estimations of bivariate copula density, we construct the estimating equation for a new $\hat{\lambda}$ using
\begin{equation} \label{eq:mlik3}
 \hat{\lambda}^{-1}=\frac{{\hat{\tilde{\mathbf{b}}}^{(d,D)}}^\top \tilde{\mathbf{P}}^{(D)} (\boldsymbol\lambda) \hat{\tilde{\mathbf{b}}}^{(d,D)}}{\mbox{tr}(S(\lambda))}
\end{equation}
where the smoothing matrix $S(\lambda)$ and the construction of \eqref{eq:mlik3} are explained in Appendix \ref{A.lam}.
Apparently, both sides of equation \eqref{eq:mlik3} depend on $\lambda$ but an iterative solution is possible by fixing $\lambda$ on the right hand side in \eqref{eq:mlik3}, update $\lambda$ on the left hand side and iterate this step by updating the right hand side of \eqref{eq:mlik3}. This estimation scheme yields the so called REML estimate and has been suggested in generalized linear mixed models by \citeN{Schall:91}. For penalized spline smoothing \citeN{Wood:11} shows that the selection of smoothing parameter $\lambda$ based in the mixed model approach behaves superior to AIC selected values.
\subsection{Practical Implementation}
The constraints \eqref{eq:constraint03} and \eqref{eq:constraint02} can be accommodated as side conditions in a quadratic programming optimization routine to maximize the likelihood given in \eqref{eq:penlikhier}. For this purpose, we make use of the {\ttfamily quadprog} package in {\ttfamily R}. The starting values for $\tilde{\mathbf{b}}^{(d,D)}$ are chosen such that the resulting copula density is the density of the independence copula and the initial value of $\lambda$ is set to a moderate size. In each step we estimate new coefficients $\hat{\tilde{\mathbf{b}}}^{(d,D)}$, keeping $\lambda$ fixed and then refit $\lambda$ using \eqref{eq:mlik3}. This estimation scheme is repeated until convergence. We discuss in Appendix \ref{C.Time} the computing time of our approach and present the results in Table \ref{comp.time}.
\section{Estimation of Vine Copulas}\label{Sec.C}
\citet{Dissmann:13} propose a selection algorithm that seeks to capture most of the dependence in the first couple of trees. We follow this approach and use the corrected Akaike information criterion (cAIC) as edge weight to find the corresponding minimum spanning tree. Therefore, we estimate all $\binom{p}{2}$ bivariate unconditional copula densities $c_{ij}, 1\leq i<j \leq p$. The resulting maximized log-likelihood value of a copula is denoted by  $\tilde{l}_{pen}^{(D)}(\hat{\tilde{\boldsymbol{b}}}^{(d,D)},\hat{\boldsymbol{\lambda}})$. For each copula, we calculate the corrected Akaike information criterion (cAIC) (\citet{Hurvich1989})
\begin{align} \label{eq:AIC}
 \text{AIC}_c(\boldsymbol\lambda)=-2\tilde{l}(\hat{\tilde{\mathbf{b}}}^{(d,D)},\boldsymbol\lambda) + 2 \text{df}(\boldsymbol\lambda)+ \frac{2 \text{df}(\boldsymbol\lambda) (\text{df}(\boldsymbol\lambda) +1)}{n- \text{df}(\boldsymbol\lambda)-1},
\end{align}
where df$(\boldsymbol\lambda)$ are the equivalent degrees of freedom of the model defined by
\begin{equation*}
 \text{df}(\boldsymbol\lambda)=\text{tr}\left[\left\{ 
 {\tilde{\mathbf{H}}_{pen}^{(D)}}(\hat{\tilde{\mathbf{b}}}^{(d,D)}, \boldsymbol{\lambda})\right\}^{-1} {\tilde{\mathbf{H}}_{pen}^{(D)}}\left(\hat{\tilde{\mathbf{b}}}^{(d,D)}, \boldsymbol{\lambda}=0\right) \right],
\end{equation*}
where $\tilde{\mathbf{H}}_{pen}^{(D)}(\hat{\tilde{\mathbf{b}}}^{(d,D)},
 \boldsymbol{\lambda}) 
$ denotes the second-order partial derivative of the penalized log-likelihood
 $\tilde{l}_{pen}^{(D)}(\tilde{\mathbf{b}}^{(d,D)},\boldsymbol{\lambda})$ (9) with respect to $\hat{\tilde{\mathbf{b}}}^{(d,D)}$, see also Appendix \ref{A.lam}.
The first tree is chosen such that the sum of the cAIC values is minimized and the corresponding minimum spanning tree is found. From the second tree on we repeat these steps and construct the minimum spanning tree from the set of trees that are possible in a tree of a regular vine. We consider three different non-parametric estimators for the building blocks of the vine copula. The first estimator 'SimpA' has been proposed in \citet{ScheKau:12b} for estimating $p$-dimensional copulas and uses unconditional copulas as building blocks for a simplified vine copula model, i.e., for a D-vine we obtain the density
\begin{align} \label{simpaeq}
\hat{c}^{\text{SimpA}}_{1:p}(u_1,\ldots,u_p) =  \prod_{j=1}^{p-1}\prod_{i=1}^{p-j}\hat{c}_{i,i+j\ps s_{ij} }  (u_{i|\condset},u_{i+j|\condset}),
\end{align} 
where $\hat{c}_{i,i+j\ps s_{ij} }$ is an estimator for the  density of the $j$-th order partial copula of $F_{i,i+j|s_{ij}}$ \citep[see][]{Spanhel2015}. Thus, this estimator estimates the density of the partial vine copula of $C_{1:p}$. The second estimator 'Cond', assuming that the vine structure is a D-vine, is given by
\begin{align}
\label{condeq}
\hat{c}_{1:p}^{\text{Cond}}(x_1,\ldots,x_p) =  \prod_{j=1}^{p-1}\prod_{i=1}^{p-j}\hat{c}_{i,i+j| v_{ij} } (u_{i|\condset},u_{i+j|\condset}|v_{ij}),
\end{align}
where $\hat{c}_{i,i+j| v_{ij} }$ is an estimator for the bivariate conditional copula with the standardized ranks of the first principal component of $U_{s_{ij}}$ as conditioning variable (see Section \ref{Sec.A}). The third estimator 'Test' only estimates a conditional copula if the simplifying assumption is rejected for a particular edge of the vine copula. That is, from the second tree on, we test whether the simplifying assumption is adequate for a conditional copula using a 5\% significance level. In particular, if the vine structure is a D-vine, the hypotheses for the $i$-th copula in the $j$-th tree, $j\geq 2$, are given by
\begin{align*}\label{eq:test}
&H_0 : \P\{C_{i,i+j|\condset}(u,v|X_{\condset})= C_{i,i+j\ps \condset}(u,v)\}=1, &\text{ for all } (u,v)\in(0,1)^2,
\\
\text{ vs } 
&H_1 : \P\{C_{i,i+j|\condset}(u,v|X_{\condset})= C^p_{i,i+j\ps \condset}(u,v)\}\neq 1, &\text{ for some } (u,v)\in(0,1)^2,
\end{align*}
where 
\begin{align*}
C_{i,i+j\ps \condset}^p(u,v) := \P(F_{i|\condset}(X_i|X_{\condset})\leq a, F_{i+j|\condset}(X_{i+j}|X_{\condset})\leq b)    
\end{align*}
is the partial copula of $C_{i,i+j|\condset}$ \citep{Spanhel2015c}. Note that the non-simplified vine copula estimator provides estimates for the unknown conditional cdfs $F_{i|\condset}$ and $F_{i+j|\condset}$ so that pseudo-observations from the $i$-th copula in the $j$-th tree are available. In order to test whether the conditional copula $C_{i,i+j|\condset}$ collapses to its partial copula $C_{i,i+j\ps \condset}^p$, we apply the testing procedure  proposed by \citet{Kurz2015} for which a R-Package {\tt pacotest} is in preparation. Consequently, the estimator 'Test' approximates a D-vine copula by
\begin{align*}
\hat{c}_{1:p}^{\text{Test}}(u_1,\ldots,u_p)  & =  \prod_{j=1}^{p-1}\prod_{i=1}^{p-j}\tilde{c}_{i,i+j| v_{ij} }  (u_{i|\condset},u_{i+j|\condset}|v_{ij})
\end{align*}
where $\tilde{c}_{i,i+j| v_{ij}}= \hat{c}_{i,i+j| v_{ij}}$ (see \eqref{condeq}) if the simplifying assumption is rejected and $\tilde{c}_{i,i+j| v_{ij}}= \hat{c}_{i,i+j\ps s_{ij}}$ (see \eqref{simpaeq}) if the simplifying assumption can not be rejected. The entire procedure is implemented in the {\ttfamily R} package {\ttfamily pencopulaCond} to be provided on the {\ttfamily CRAN} server ({\textit{http://cran.r-project.org/}).
\section{Simulations and Application}\label{Sec.D}
In order to investigate the performance of our non-parametric estimators for non-simplified regular vine copulas, we conduct an extensive simulation study. For that purpose, we generate $N=100$ data sets of sample size $n=500$ and $n=2000$ for three- and five-dimensional distributions, which are later explained in Section \ref{desc_ex}. Throughout the simulation study, we estimate copula densities with a varying degree $d$, which determines the amount of univariate knots, and different maximum cumulated hierarchy levels, which determine the amount of sparsity, see Table \ref{Table1} for an overview. In the following, $D_2$ and $D_3$ refer to the maximum cumulated hierarchy levels that are used for the estimation of a bivariate unconditional copula density (two-dimensional function) and a bivariate conditional copula density (three-dimensional function), respectively. For instance, \coefaa \ refers to an estimation where the degree of the univariate hierarchical B-spline basis is two $(d=2)$ and the maximum cumulated hierarchy level is four for the unconditional copula densities $(D_2=4)$ and six for the conditional copula densities $(D_3=6)$. In this case, we estimate 25 basis coefficients for a bivariate copula density and 125 basis coefficients for a conditional copula density with one conditioning argument. Changing the values of $d,D_2$, and $D_3$, varies the basis size, which is increased throughout the simulation study up to \coefba, which corresponds to 81 basis coefficients for a two-dimensional density estimation and 473 basis coefficients for a three-dimensional density estimation.\par
We use the three non-parametric vine copula estimators 'Test', 'Cond', and 'SimpA', defined in Section \ref{Sec.C}, and the  R-package {\tt VineCopula} (\citet{VineCopula}) to estimate a parametric vine copula. The functions of the R-package {\ttfamily VineCopula} are used to select the vine structure and the parametric copula families. To mitigate the possibility of finding a local maximum, we start our copula estimations with three different starting values for $\lambda$ and choose the estimation with the lowest cAIC. During the main part of the simulations, the results for the different starting values for $\lambda$ are nearly identical or comparable. \par
In order to assess the out-of-sample performance, we simulate $n$ new observations  from the underlying $p$-dimensional copula family and evaluate the out-of-sample log-likelihood and the average out-of-sample KL divergence which is given by
\begin{equation*}
KL\left(c_{1:p},\hat{c}_{1:p}\right)=\frac1{n}
 \sum_{i=n+1}^{2n}{\ln 
 \frac{c_{1:p}({u}_{1,i},\ldots,{u}_{p,i})}{\hat{c}_{1:p}({u}_{1,i},\ldots,{u}_{p,i})}},
\end{equation*}
where $c_{1:p}$ is the true copula density and $\hat{c}_{1:p}$ is the density that was estimated using the observations $(u_{1,i},\ldots,u_{p,i})_{i=1,\ldots,n}$.
\subsection{Description of the Examples} \label{desc_ex}
In three dimensions, we generate data sets from non-simplified vine copulas to analyze the performance of our estimator for non-simplified vine copulas. The first example is given by
\begin{align}
\label{fr3d}
c_{1:3}(u_1,u_2,u_3)= c_{1,2}^{\text{Fr}}(u_1,u_2;\tau_{1,2}) \cdot 
c_{2,3}^{\text{Fr}}(u_2,u_3;\tau_{2,3}) \cdot c_{1,3|2}^{\text{Fr}}(u_{1|2},u_{3|2};\tau(u_2)),
\end{align}
where $C_{1,2}^\text{Fr}(\cdot,\cdot;\tau_{1,2})$ and $C_{2,3}^\text{Fr}(\cdot,\cdot;\tau_{2,3})$ are Frank copulas with Kendall's $\tau$ equal to 0.25.
The conditional copula $C_{1,3|2}$ in the second tree is given by a bivariate Frank copula such that Kendall's $\tau$ depends on $u_2$. We use the following linear and quadratic functions to construct the variation of the conditional copula:
\begin{align*}
\text{case (a): \quad} & \tau(u_{2})=8 \beta  (u_{2} - 0.5)^2 - \beta,\\
\text{case (b): \quad} & \tau(u_{2})=\beta-2\beta u_{2},
\end{align*}
where $\beta=\{0.2,0.4,0.6\}$. We present plots of these functions for Kendell's $\tau$ in Figure \ref{fig:generator}. Thus, we consider six scenarios altogether.\par
Additionally, we investigate the performance of our estimator in three dimensions for an equally weighted mixture of two normal distributions with cdf given by
\begin{align}
F_{1:3}(x;\mu_1,\mu_2,\Sigma_1,\Sigma_2) = 0.5\Phi_{1:3}(x;\mu_1,\Sigma_1)+0.5\Phi_{1:3}(x;\mu_2,\Sigma_2), \label{norm3d}
\end{align}
where $\Phi_{1:p}(\cdot,\mu,\Sigma)$ denotes the cdf of the $p$-dimensional normal distribution with mean $\mu$ and covariance matrix $\Sigma$ and
\begin{align*}
\mu_1 =& \mathbf{1}_3,\ \  \Sigma_1 = -\frac{2}{5}\mathbf{1}_3\mathbf{1}_3'+\frac{7}{5}I_3, \\ 
\mu_2 =& -\mathbf{1}_3,\ \ \Sigma_2 = \frac{2}{5}\mathbf{1}_3\mathbf{1}_3' + \frac{3}{5}I_3.  
\end{align*}
In five dimensions, we analyze the performance for the following five-dimensional non-simplified vine copula
\begin{align}\label{fr5d}
c_{1:5}&(u_1,u_2,u_3,u_4,u_5)=c_{1,2}^\text{Fr}(u_1,u_2;\tau_{1,2}) \cdot c_{2,3}^\text{Fr}(u_2,u_3;\tau_{2,3}) \cdot c_{3,4}^\text{Fr}(u_3,u_4;\tau_{3,4}) \cdot c_{4,5}^\text{Fr}(u_4,u_5;\tau_{4,5}) \notag\\
& \quad\times \prod_{j=2}^5\prod_{i=1}^{5-j} c_{i,i+j;\condset}^{\text{Fr}}(u_{i|\condset},u_{i+j|\condset};
\tau(u_{\condset}))
\end{align}
which is a straight forward generalization of \eqref{fr3d} to the five-dimensional case. In the first tree we choose $\tau_{1,2}=\tau_{2,3}=\tau_{3,4}=\tau_{4,5}=0.25$. The variation of Kendall's $\tau$ and therefore the conditional copulas in the higher trees is specified by
\begin{align*}
\text{case (a): \quad} & \tau(u_{\condset})=8 \beta  \left(\frac1{j-1} 
\sum_{k=1}^{j-1} u_{k} - 0.5\right)^2 - \beta
\\
\text{case (b): \quad} & \tau(u_{\condset})=\beta-2\beta\left(\frac1{j-1} \sum_{k=1}^{j-1} u_{k}\right)
\end{align*}
each for $j=2,3,4$ with $\beta=\{0.2,0.4,0.6\}$. Again, there are six scenarios under consideration. Simulations from all non-simplified vine copulas are performed using algorithm 17 in \cite{Joe2015}.\par
Moreover, we investigate results for a equally weighted mixtures of two five-dimensional normal distributions with cdf given by 
%
\begin{align}
\label{norm5d}
F_{1:5}(x;\mu_1,\Sigma_1,\mu_2,\Sigma_2) = 0.5\Phi_{1:5} (x;\mu_1,\Sigma_1)+0.5\Phi_{1:5}(x;\mu_2,\Sigma_2),
\end{align}
where 
\begin{align*}
\mu_1 &= -\mathbf{1}_5, \Sigma_1 = \frac{2}{5}\mathbf{1}_5\mathbf{1}_5'+ \frac{3}{5}I_5, \mu_2 = \mathbf{1}_5,
\\
\text{vech}(\Sigma_2)& = \Big(1,- \frac{2}{5}, - \frac{2}{5}, - \frac{1}{5}, - \frac{1}{10}, 1, - \frac{2}{5}, - \frac{1}{5}, - \frac{1}{10},1, - \frac{1}{5}, -  \frac{1}{10}, 1, - \frac{1}{10},1\Big).
\end{align*}
\subsection{Results for the three-dimensional Examples}
To illustrate the results, we present in Figure \ref{figp3} box plots of the differences $KL_{non-par}-KL_{par}$, where $KL_{non-par}$ is the out-of-sample KL divergence of a non-parametric estimation ('SimpA', 'Cond' or 'Test') and $KL_{par}$ is the out-of-sample KL divergence of the parametric estimation 'VineCopula' for the three-dimensional non-simplified vine copulas. The length of the whiskers is chosen such that the whiskers cover 95\% of the differences, i.e., if the value zero is not contained within the whiskers, the difference between the out-of-sample KL divergences is significant at the 5\% level. The boxes containing results of the 'Test' estimation are plotted in blue, while the results of the 'Cond' estimation are illustrated in pink and the boxes with results of the 'Simp' estimation are presented in green. The box plots of $KL_{non-par}-KL_{par}$ thus visualize the finite-sample distribution of Vuong's closeness test if the procedure is applied to out-of-sample data.%
\footnote{
We do not use the asymptotic normal distribution of Vuong's test because both over-rejection and under-rejection can be severe when the competing models contain a rather large number of parameters, see \citet{Shi2015}. Moreover, we use an out-of-sample version of Vuong's test because the in-sample version does not account for the number of parameters.  
}
We first consider for both cases (a) and (b) the setup $\beta=0.2$ which corresponds to a minor variation of the conditional copula. Indeed, the null hypothesis that the simplifying assumption holds is not always rejected if the sample size is low ($n=500$) which can be seen from the fact that the estimators 'Cond' and 'Test' are not not always identical. We also observe that the imposed estimation of conditional copula for the estimator 'Cond' does not yield a significantly better model for both cases (a) and (b). That is because the conditional copula is very close to the partial copula and so the difference between the partial vine copula and the data generating copula is rather negligible. However, if the sample size is increased to $n=2000$ observations, the null hypothesis that the simplifying assumption is true is much more often rejected and the conditional copula can be more accurately estimated. As a result, the estimator 'Cond' and 'Test' perform significantly better than the parametric estimator VineCopula for case (b). The variation of the conditional copula seems to be harder to detect for case (a), implying that the estimators 'Cond' and 'Test' are not always identical. Only the estimator 'Cond' is significantly better than the parametric estimator but the estimator 'Test' is pretty close to being significantly better. Note that the non-parametric estimator 'SimpA' is statistically not distinguishable from the parametric simplified vine copula estimator 'VineCopula'. Thus, the reduction in the out-of-sample KL divergence that we observe for the estimators 'Test' and 'Cond' can be attributed to the modeling of the conditional copula in the second tree.\par
We now address the cases (a) and (b) when $\beta=0.4$. This setup corresponds to a moderate variation of the conditional copula which is comparable to the variation that has empirically been estimated in \citet{Gijbels2011}. For setup (a) we see that the estimator 'Cond' performs significantly better in terms of out-of-sample KL divergence than the parametric vine copula estimator, even if the sample size is low. For a small sample size ($n=500$) the null hypothesis that the simplifying assumption is true is not always rejected. Consequently, the estimator 'Test' collapses to the estimator 'SimpA' in these cases. Since the estimator 'SimpA' is comparable to the parametric estimator, the estimator 'Test' is thus not significantly better for small sample sizes, although the out-of-sample KL divergence is much lower. However, the estimator 'Test' yields a significantly lower out-of-sample KL divergence if the sample size is increased to $n=2000$ observations and the simplifying assumption is always rejected. 
For setup (b) we see that, for both sample sizes, the estimator 'Cond' and 'Test' are always identical and that the variation of the conditional copula is always detected by the testing procedure of \citet{Kurz2015}. Moreover, the estimator 'Cond' and 'Test' yield a substantially lower out-of-sample KL divergence which is also highly significant even for small sample sizes. Additionally, we observe decreasing values of $KL_{non-par}$ with a larger basis size, which is in line with the expected performance.\par
A qualitatively similar picture emerges for the case $\beta=0.6$ which corresponds to a rather strong variation of the conditional copula.%
\footnote{Note that $\beta$ could be increased up to the value of one, so that the rather strong variation for $\beta=0.6$ is by far not the strongest possible variation.}
For the setup (a) and a small sample size, the variation of the conditional copula is not detected in about 7\% of the cases. As a consequence, the estimator 'Test' does not perform significantly better than the parametric or non-parametric simplified vine copula estimators 'SimpA' and 'VineCopula'. Nonetheless, the median of the out-of-sample KL divergence is substantially lower and the estimator 'Cond', which always model a conditional copula, performs significantly better. If the sample size is increased to $n=2000$ observations, the results of the estimators 'Cond' and 'Test' coincide and are significantly better than the results of the simplified vine copula estimators. The violation of the simplifying assumption is always detected in setup (b). In addition, we also observe a remarkable decrease in the out-of-sample KL divergence if the conditional copula is estimated which is highly significant even for a small sample size. Thus, the modeling of a conditional copula leads to a substantially lower out-of-sample KL divergence if the null hypothesis of a partial copula in the second tree and therefore the simplifying assumption can be strongly rejected.%
\footnote{We also present tables with all results in the supplementary material. Table 1 in the supplementary material show the simulation results for the three-dimensional non-simplified vine copulas for the scenario case $(a)$ while Table 2 in the supplementary material presents the results for case $(b)$, where the lowest mean of the out-of-sample KL divergence measure for the non-parametric estimations 'Test' is set in bold.}

The results for the three-dimensional mixture of normal distributions are plotted in the top row of Figure \ref{fignorm}. The non-parametric estimator 'SimpA', which approximates conditional copulas by unconditional copulas, performs similarly with a little improvement as compared to the parametric estimator 'VineCopula' in terms of  out-of-sample KL divergence. Modeling the conditional copula in the second tree of the three-dimensional normal mixture ('Cond' and 'Test') greatly reduces the  out-of-sample KL divergence and results in a significantly better model even if the sample size is $n=500$. Note that the simplifying assumption is always rejected in the second tree of this example so that the estimators 'Cond' and 'Test' are identical. Increasing the sample size to $n=2000$ only slightly improves the performance of the non-parametric simplified vine copula estimator 'SimpA' as compared to 'VineCopula'. In contrast, the  out-of-sample KL divergence is much more reduced by applying the non-parametric estimators 'Cond' and 'Test'. Thus, if the null hypothesis of the validity of the simplifying assumption is strongly rejected, our proposed approach yields a substantial improvement. The estimated conditional copula density is illustrated in Figure \ref{fignormp3-ex} and an animated picture is provided in the supplementary material.%
\footnote{Table 5 in the supplementary material contains the simulation results for the three-dimensional normal distribution and five-dimensional normal distribution, where the lowest mean of the out-of-sample KL divergence measure for the non-parametric estimations 'Test' is set in bold.}
\subsection{Results for the five-dimensional Examples}
The box plots of the KL differences are plotted in Figure \ref{figp5} for the example given in \eqref{fr5d} and for the normal mixture given in \eqref{norm5d} at the bottom of Figure \ref{fignorm}. The results for the non-simplified vine copulas in five dimensions are similar to the three-dimensional case. For all sample sizes and values of $\beta$, it is still more difficult to detect the violation of the simplifying assumption for case (a), so that the estimators 'Cond' and 'Test' are often not identical for this case. Thus, although the median of the out-of-sample KL divergence difference is much smaller, the estimator 'Test' is not significantly better than the parametric estimator 'VineCopula' for the case (a) with $\beta=0.2,0.4$ and a small sample size. Increasing the sample size to $n=2000$ leads to a further decrease in the median of the out-of-sample KL divergence difference and the difference  is now highly significant for both estimators 'Cond' and 'Test'. For case (b) we see that for all sample sizes and values of $\beta$ the estimators 'Cond' and 'Test' are almost identical and perform significantly better than the simplified vine copula estimators 'SimpA' and 'VineCopula'. As expected, the improvement increases the more $\beta$ and thus the variation of the conditional copulas increases. Even for a moderate variation of the conditional copulas, i.e., $\beta=0.4$, the modeling of the three conditional copulas results in a substantial reduction of the out-of-sample KL divergence.\par
The box plots at the bottom in Figure \ref{fignorm} show that the parametric and the non-parametric estimator of the simplified vine copula model perform equally well for the five-dimensional normal mixture and both sample sizes. There is no significant difference in the  out-of-sample KL divergences for $n=500$. Increasing the sample size to $n=2000$ does also not result in a significant difference in the KL divergences of 'SimpA' and 'VineCopula'. On the contrary, the estimators 'Cond' and 'Test' both yield a substantially lower out-of-sample KL divergence which is highly significant even for small sizes ($n=500$). Since the simplifying assumption is also almost always rejected for each conditional copula, this is consistent with the fact that the variation in the conditional copulas is not negligible. Increasing the sample size to $n=2000$ observations further increases the substantial difference between the  out-of-sample KL divergence of 'Cond' or 'Test' and the parametric simplified vine copula model 'VineCopula'. Consequently, the consideration of conditional copulas may result in a remarkable performance improvement if the partial vine copula does not yield an adequate approximation.
\footnote{Table 3 in the supplementary material show the simulation results for the five-dimensional non-simplified vine copulas for the scenario case $(a)$ while Table 4 in the supplementary material presents the results for case $(b)$, where the lowest mean of the out-of-sample KL divergence measure for the non-parametric estimations 'Test' is set in bold.} 
\subsection{Application: Classifying Eye State}
Classification tasks of high dimensional data have been considered in many aspects. We present an example based on the EEG Eye State Data Set from the UCI Machine Learning Repository.%
\footnote{https://archive.ics.uci.edu/ml/datasets/EEG+Eye+State} The data set consists of 14 EEG variables with 14980 observations, which have been measured using an Emotiv EEG Neuroheadset. We denote these variables by $X$. Additionally, there is an indicator for the corresponding eye state with the states eye-closed (1) or eye-opened (0). We separate the data set into a training- and an evaluation set, selecting the first 66\% data points as training set and the rest as evaluation set. We focus on the classification of the eye-opened state (0) using a Bayes classifier. Let $\hat{f}_{(0)}$ and $\hat{f}_{(1)}$ be the densities of the 14 EEG variables if the class is (0) or (1) that have been estimated using the training data set. The posterior probability for the evaluation data set that the class is (0) is then given by
\begin{equation*}
Pr(Class=(0)|X=x)=\frac{\pi_{(0)}{\hat{f}_{(0)}(x)}}
{\pi_{(0)}{\hat{f}_{(0)}(x)}+\pi_{(1)}{\hat{f}_{(1)}(x)}},
\end{equation*}
where $\pi_{(0)}$ and $\pi_{(1)}$ are the prior probabilities of the classes. For simplicity, we assume $\pi_{(0)}=\pi_{(1)}=0.5$. For the estimation of $\hat{f}_{(0)}$ and $\hat{f}_{(1)}$ we combine the estimation of marginal densities with the estimation of regular (simplified) vine copulas.\par
To obtain the data for the vine copulas, we estimate the marginal distributions of the  14 EEG values by the standard kernel density estimator {\ttfamily kde} in the {\ttfamily R} package {\ttfamily ks} (see \citet{KS}). The  bandwidth is chosen by the plug-in bandwidth selector {\ttfamily hpi} (see \citet{Wand-Jones:95}) and then doubled to improves the performance of the estimator in areas of low density mass.  We estimate the partial vine copula using the estimator 'SimpA' from the simulation section with \coefbS \ (simp d=3) and also consider the estimator 'kdevine' (see \citet{Nagler2015}) which estimates the partial vine copula non-parametrically based on kernel techniques. Moreover, we apply the parametric vine copula estimator {\tt VineCopula} to obtain a simplified vine copula. As non-simplified vine copula estimator we use the estimator 'Test' from the simulation section with \coefba \ (nonsimp d=3).  Note that all estimators consist 91 (un)conditional copulas.\par
We evaluate the fitted densities using the remaining evaluation data set. Usually, a fitted posterior probability above $\alpha=0.5$ would indicate the class (0). Varying the value $\alpha$ allows to control the number of observations classified as (0) and (1) and influences the false positive rate (FDR) and the true positive rate (TPR). The FPR is the ratio between the number of false positives, that is, the number of (1) events classified as (0), divided by the total number of all negative (1) events, while the TRP is the ratio between number of true positive events, that is, (0) events classified as (0), divided by the total number of (0) events. Figure \ref{figclass} presents the resulting receiver operating characteristic (ROC) curves, which plot the FPR against the TPR, for $\alpha \in [0,1]$ and for the considered four estimators.\par
First of all, we see that the non-parametric partial vine copula estimators 'SimpA' and 'kdevine' perform better than the parametric estimator {\tt VineCopula}, indicating that the parametric copula families do not provide an adequate fit for the pair-copula of the partial vine copula. The estimators 'SimpA' and 'kdevine' perform very similarly if the FPR is below 10\%, but if the FPR is above 10\% the estimator 'SimpA' yields a slightly better TPR. The best performance is obtained with the non-simplified estimator  'Test', which exhibits the best TPR for any FPR. Especially if the FRP is between 5\% and 40\%, the TPR is substantially increased by up to 22\%. The major reason is that the estimator 'Test' rejects the simplifying assumption for the training data set for class (0) in 57 of 91 cases and for class (1) in 68 of 91 cases. As a result, there is a substantial improvement compared to the simplified vine copula estimators.
\subsection{Application: Uranium Data Set}
We illustrate the potential gains of a non-simplified vine copula density estimation using the seven-dimensional uranium data set included in the R package {\ttfamily copula} and originally introduced by \citet{Uranium}. The data set contains 655 observations which measure the log-concentration of 7 chemicals in water samples from the Montrose quadrangle of western Colorado. This particular data set is often discussed in articles related to copulas, see for instance \citet{Gijbels2012} or \citet{Killichies2016} who both investigate the simplifying assumption for three-dimensional subsets. Both studies detect that a non-simplified vine copula improves the fit of the data subset. For the first time, we now estimate a non-simplified vine copula model for the complete seven-dimensional uranium data set and investigate if conditional copulas are also present in the higher trees of the vine copula.\par
We use the standardized ranks to estimate the marginal distributions and obtain pseudo-observations from the copula. For the estimation of the vine copulas, we consider the estimators 'Test' with the setups \coefaa\ and \coefba, and the estimator 'SimpA' with the setups \coefaS\ and \coefb. For a comparison with a parametric vine copula, we also estimate a parametric regular vine copula using the R-package {\tt VineCopula}. The fitted log-likelihood values and the number of conditional copulas are reported in Table \ref{tab.uranium}.
We see that the estimator 'Test' rejects in 6 of 15 cases the null hypothesis that the partial copula equals the conditional copula and thus indicates a violation of the simplifying assumption for many building blocks of the vine copula.  The number of estimated conditional copula densities increases the log-likelihood compared to the parametric approach by up to 15 \% and demonstrates the improvement that can be gained if a non-parametric non-simplified vine copula estimator is applied.
\section{Discussion}\label{Sec.E}
We propose the first non-parametric estimator of a non-simplified regular vine copula that features varying conditional copulas. The use of a reduced hierarchical B-spline basis allows us to maintain numerical feasibility and to directly estimate conditional copulas. Since non-parametric estimators suffer greatly from the curse of dimensionality, we approximate the conditioning vector by a monotone function of its first principal component. This approach results in a computationally fast approximation of the conditional copula and performs well in the considered simulations and the two applications. The simulation study shows that the modeling of conditional copulas can yield a substantial decrease in the out-of-sample KL divergence if the null hypothesis of the simplifying assumption can be often rejected. Another result of the simulation study that we want to stress is that a non-simplified vine copula estimator does not, a priori, result in an improvement. If the data generating non-simplified vine copula and its partial vine copula are rather close, the simplifying assumption might not be rejected and the modeling of conditional copulas does not result in a significant improvement. This is quite similarly to the case of regression. We can not argue that an additive model is always superior to a linear model when it comes to out-of-sample prediction, it depends on to what extent non-linearities are present. The same argument holds for the estimation of vine copulas. Indeed, if the simplifying assumption can not be rejected, the estimator 'Test' collapses to an estimator of the partial vine copula. However, if the simplifying assumption can be rejected, the estimator 'Cond' and the estimator 'Test' are identical and their application can result in a significant and substantial decrease of the out-of-sample KL divergence.\par
The application of the non-parametric simplified vine copula estimator demonstrated that the simplifying assumption might be not valid in applications and that the modeling of the conditional copula is worth pursuing. The TPR of the eye state can be substantially increased for a given FPR as compared to partial vine copula estimators if conditional copulas are modeled. Moreover, the analysis of the seven-dimensional uranium data set confirmed the evidence that was previously found for three-dimensional subsets. Namely, that the simplifying assumption is not adequate for this data set and that the modeling of conditional copulas improves the out-of-sample log-likelihood. All in all, the paper presents the first step in estimating a non-simplified vine copula model non-parametrically and illustrates potential gains that can be achieved by the modeling of conditional copulas. More sophisticated approaches along the lines of \citet{hall2005} that reduce the dimension of the conditioning vector are left open for further research.\par
\vspace{0.25cm}
\textbf{Acknowledgements}\\
We thank G\"oran Kauermann (LMU Munich) for discussions at the beginning of this project.
\begin{appendix}
\section{Construction of sparse B-spline Density Basis}\label{App:Bspline}
In the following, objects signed with superscript $\tilde{\ }$ and $^{(d)}$ are associated with hierarchical B-spline basis functions, whereas objects signed with superscript $\tilde{\ }$ and $^{(d,D)}$ are linked to sparse B-spline basis functions. In order to transform $B^{(\tau(d))}(\mathbf{u}_j)$ in \eqref{eq:fullbase} into its hierarchical representation (see \citeauthor{ForBar:88} \citeyear{ForBar:88}, \citeyear{ForBar:95}), we define the hierarchical index sets $\mathcal{I}_0=\{1, 2\}$ and $\mathcal{I}_l = \{2j\mid  j\in\mathbb{N},1\leq j \leq 2^{l-1}\}$,  $l=1,\dots,d$. Let $B^{(\tau(l))}_{\mathcal{I}_l}(\mathbf{u}_j)$ denote the columns $\mathcal{I}_l$ of $B^{\tau(l)}(\mathbf{u}_j)$.
\footnote{For $l=2$, we get ${\cal I}_2 = \{2,4\}, \tau(2)=(0,0.25,0.5,0.75,1)$, and
\begin{align*}
{B}^{(\tau(2))}_{\mathcal{I}_2}(\mathbf{u}_j)&= B^{(\tau(2))}(\mathbf{u}_j)
\begin{pmatrix}0 & 0 & 0 & 1 & 0 \\ 0 & 1 & 0 & 0 & 0  \end{pmatrix}^\top = \begin{pmatrix}
\phi_{0.25}(u_{j,1}) & \phi_{0.75}(u_{j,1})
\\
\vdots & \vdots
\\
\phi_{0.25}(u_{j,n}) & \phi_{0.75}(u_{j,n})
\end{pmatrix}.
\end{align*}
}
The univariate hierarchical B-spline basis of degree $d$ is defined as
\begin{equation}\label{eq:hierBspl}
\tilde{\mathbf{B}}^{(\tau(d))}(\mathbf{u}_j)=\left( {B}^{(\tau(0))}_{\mathcal{I}_0}(\mathbf{u}_j), {{B}^{(\tau(1))}_{\mathcal{I}_1}(\mathbf{u}_j)}, \dots , \ {B}^{(\tau(d))}_{\mathcal{I}_d}(\mathbf{u}_j)\right).
\end{equation}
Figure \ref{Figure2} presents the univariate B-spline basis $B^{(\tau(d))}(\mathbf{u}_j)$ and the building parts of the corresponding hierarchical basis $\tilde{\mathbf{B}}^{(\tau(d))}(\mathbf{u}_j)$ for $d=2$. Let $\hltilde$ be a $K$-dimensional row vector such that its k-th element is given by $\tilde{e}_k:= \min\{l=0,\dots,d: k\leq |\tau(l)|\}$, e.g., $\hltilde = (0,0,1,2,2)$ if $d=2$. The vector $\hltilde$ denotes the hierarchical level of $\tilde{\mathbf{B}}^{(\tau(d))}(\mathbf{u}_j)$ and its $k$-th element identifies the hierarchical level of the $k$-th column of $\tilde{\mathbf{B}}^{(\tau(d))}(\mathbf{u}_j)$. By construction, $B^{(\tau(d))}(\mathbf{u}_j)$ and $\tilde{\mathbf{B}}^{(\tau(d))}(\mathbf{u}_j)$  have full rank, i.e. $B^{(\tau(d))}(\mathbf{u}_j)\tilde{A}=\tilde{\mathbf{B}}^{(\tau(d))}(\mathbf{u}_j)$ for some invertible $K \times K$ matrix $\tilde{A}$, so that both univariate bases $\tilde{\mathbf{B}}^{(\tau(d))}(\mathbf{u}_j)$ and $B^{(\tau(d))}(\mathbf{u}_j)$ span the same space. The three-dimensional hierarchical B-spline basis follows as
\begin{equation} \label{eq:hierachB}
\tilde{\boldsymbol{\Phi}}^{(d)}(\mathbf{u}_1,\mathbf{u}_2,\mathbf{u}_3):=\bigotimes_{j=1}^3 \tilde{\mathbf{B}}^{(\tau(d))}(\mathbf{u}_j) = \bigotimes_{j=1}^3 B^{(\tau(d))}(\mathbf{u}_j) \tilde{A}
\end{equation}
and the corresponding approximation of the conditional copula density is given by 
\begin{align*}
c_{12|3} (\mathbf{u}_1,\mathbf{u}_2|\mathbf{u}_3; \tilde{\mathbf{b}}^{(d)})=\tilde{\boldsymbol{\Phi}}^{(d)}(\mathbf{u}_1,\mathbf{u}_2,\mathbf{u}_3)\mathbf{\tilde{b}}^{(d)},
 \end{align*}
where $\mathbf{\tilde{b}}^{(d)} =  (\bigotimes_{j=1}^3\tilde{A})^{-1}\mathbf{b} $.\par
To overcome the exponential increase in the number of spline coefficients, we use a three-dimensional sparse B-spline basis which reduces the dimension by deleting the columns from the full tensor product basis whose \emph{cumulated} hierarchy level exceeds $D$, where $d\leq D\leq 3d$. The \emph{cumulated} hierarchy level of the full tensor product basis is defined as follows. For $\alpha\in\mathbb{R}^{1\times n}$ and $\beta\in\mathbb{R}^{1\times q}$,  define the $l$-th element of $(\alpha\oplus \beta)\in\mathbb{R}^{1\times nq}$ by $(\alpha\oplus \beta)_{l} = \alpha_{\left \lceil\frac{l}{q}\right \rceil}+ \beta_{l-q(\left \lceil\frac{l}{q}\right \rceil-1)}$, where $\lceil \cdot \rceil$ is the ceil function. Note that the operation $\oplus$ is associative. Recall that the $k$-th element of $\hltilde$ identifies the hierarchy level of the $k$-th column of $\tilde{\mathbf{B}}^{(\tau(d))}(\mathbf{u}_j)$, so that the  $l$-th element of $\epsilon = \hltilde \oplus \hltilde \oplus \hltilde\in \mathbb{R}^{1\times {K}^3}$ denotes the \emph{cumulated} hierarchy level of the $l$-th column of the hierarchical B-spline basis $\tilde{\boldsymbol{\mathbf{\Phi}}}^{(d)}(\mathbf{u}_1,\mathbf{u}_2,\mathbf{u}_3)$.
Define ${\cal O}_D = \{j=1,\ldots, {K_{}}^3 \colon \epsilon_j\leq D \}$, i.e., { ${\cal O}_D$ contains the position of the columns of $\tilde{\boldsymbol{\mathbf{\Phi}}}^{(d)}(\mathbf{u}_1,\mathbf{u}_2,\mathbf{u}_3)$ whose \emph{cumulated} hierarchy level does not exceed $D$.} Let ${\cal O}_D(j)$ be the $j$-th smallest element of ${\cal O}_D$ and define the orthogonal matrix $\mathcal{E}(\mathcal{O}_D)\in \mathbb{R}^{{K}^3\times|{\cal O}_D|}$ such that its $(\mathcal{O}_D(j),j)$-entry is one for $j=1,\dots,|{{\cal O}}_D|,$ and the other entries are zero. The three-dimensional sparse B-spline basis follows as
\begin{equation} \label{eq:sparsegrid}
\tilde{\boldsymbol{\mathbf{\Phi}}}^{(d,D)}(\mathbf{u}_1,\mathbf{u}_2,\mathbf{u}_3) = \left[\bigotimes_{j=1}^3 \tilde{\mathbf{B}}^{(\tau(d))} (\mathbf{u}_j)\right]
\mathcal{E}({\cal O}_D),
\end{equation}
where the lower index $d$ is the degree of the univariate hierarchical B-spline basis and the upper index $D, d\leq D\leq 3d,$ refers to the maximum \emph{cumulated} hierarchy level. Only the columns in the  hierarchical B-spline basis $\tilde{\mathbf{\Phi}}^{(d)}(\mathbf{u}_1,\mathbf{u}_2,\mathbf{u}_3)$ whose \emph{cumulated} hierarchy level does not exceed $D$ constitute the three-dimensional sparse B-spline basis $\tilde{\boldsymbol{\mathbf{\Phi}}}^{(d,D)}(\mathbf{u}_1,\mathbf{u}_2,\mathbf{u}_3)$. Figure \ref{Figure3} shows the construction principle for a bivariate sparse B-spline basis with hierarchy level $d=2$.
Figure \ref{Figure4} presents the placements of the knots for the full tensor product of B-splines and the sparse B-spline basis for $d=2$. The corresponding spline coefficients are given by $\tilde{\mathbf{b}}^{(d,D)} =  {\cal E(O_D)}^\top \tilde{\mathbf{b}}^{(d)}$.\par

\section{Marginal Likelihood} \label{A.lam}
The prior \eqref{prior} is degenerated, which needs to be corrected as follows. For simplicity,  we write $\mathbf{b}:= \mathbf{\tilde{b}^{(d,D)}}$ in this section. We decompose $\mathbf{b}$ into the two components $\mathbf{b}_{\sim}$ and $\mathbf{b}_{\bot}$, respectively, such that $\mathbf{b}_{\sim}$ is a normally distributed random vector with non degenerated variance and $\mathbf{b}_{\bot}$ contains the remaining components treated as parameters, see also \citet{WandOrme:08}. Applying a singular value decomposition we have $\mathbf{\tilde{P}^{(D)}}=\mathbf{\tilde{U}} \mathbf{\tilde{\Lambda}} \mathbf{\tilde{U}}^T$, where $\mathbf{\tilde{\Lambda}}$ is a diagonal matrix with positive eigenvalues and $\mathbf{\tilde{U}} \in \mathbb{R}^{(K+1) \times h}$ are the corresponding eigenvectors with $K+1$ being the number of elements in $\mathbf{b}$ and $h=K+1-4$ being the rank of $\mathbf{\tilde{P}^{(D)}}$. Extending $\mathbf{\tilde{U}}$ to an orthogonal basis by $\mathbf{\check{U}}$ gives 
$\mathbf{b}_{\sim}=\mathbf{\tilde{U}}^T\mathbf{b}$.
With the a priori assumption $\mathbf{b}_{\sim} \sim N(0, 
\lambda^{-1}\mathbf{\tilde{\Lambda}}^{-1})$  and $\mathbf{U}=(\mathbf{\tilde{U}}, \mathbf{\check{U}})$ as orthogonal basis we get $\mathbf{b}_{\bot}={\mathbf{\check{U}}}^T \mathbf{b}$. Conditioning on $\mathbf{b}_{\sim}$,
we get the mixed model log likelihood
\begin{equation*} 
l_m(\lambda, \mathbf{b}^{\bot})=\log \int |\lambda \mathbf{\tilde{\Lambda}}|^\frac12 \exp \left\{l^{(D)}_p(\mathbf{b}, \lambda)\right\} \mbox{d}{\mathbf{b}_{\sim}}.
\end{equation*}
The integral can be approximated by a Laplace approximation (see also \citet{Rue-etal:09})
\begin{align} \label{eq:mlik}
l_m(\lambda,\mathbf{b}^{\bot}) \approx \frac12 \log |\lambda \mathbf{\tilde{\Lambda}}|+ l_p^{(D)}(\mathbf{\hat{b}}, \lambda)-
&\frac12 \log |\mathbf{\tilde{U}}^T {\mathbf H}_{pen}^{(d,D)}(\mathbf{\hat{b}},\lambda) \mathbf{\tilde{U}}|
\end{align}
where $\mathbf{\hat{b}}$ denotes the penalized maximum likelihood estimate. We can now differentiate \eqref{eq:mlik} with respect to $\lambda$ which gives
\begin{flalign} \label{eq:mlik2}
\frac{\partial l_m(\lambda,\mathbf{\hat{b}}^{\bot})} {\partial \lambda} = - \frac12 \mathbf{\hat{b}}^T \tilde{\mathbf{P}}^{(D)}(\mathbf \lambda) \mathbf{\hat{b}}^{\bot} + \frac{1}{2 \lambda}  \cdot\mbox{tr} \underbrace{\left\{ (\mathbf{\tilde{U}}^\top \mathbf{H}_{pen}^{(d,D)}(\mathbf{\hat{b}},\lambda=0) \mathbf{\tilde{U}} + \lambda \mathbf{\tilde{\Lambda}})^{-1} \mathbf{\tilde{U}}^\top \mathbf{H}_{pen}^{(d,D)}(\mathbf{\hat{b}},\lambda=0) \mathbf{\tilde{U}} \right\}}_{:=S(\lambda)}.
\end{flalign}
and ${\mathbf H}_{pen}^{(d,D)}(\mathbf{\hat{b}},\lambda)$ denotes the second-order partial derivative of \eqref{eq:penlikhier} with respect to $\mathbf{\hat{b}}$, i.e.,
\begin{align*}\label{eq:Fisher}
 {\mathbf{H}}_{pen}^{(d,D)}(\mathbf{\hat{b}},\mathbf{\lambda})=-\sum_{i=1}^{n} \frac{\mathbf{\tilde{\Phi}}^{(d,D)}(u_{1,i},u_{2,i},u_{3,i}) {\boldsymbol{\tilde{\Phi}}^{(d,D)}}^\top(u_{1,i},u_{2,i},u_{3,i})}{c_{12|3}(u_{1,i},u_{2,i}|u_{3,i};\mathbf{\hat{b}})} - \tilde{\mathbf{P}}^{(D)}(\mathbf \lambda).
\end{align*}
\section{Computing Time} \label{C.Time}
We present some computing times, defined as elapsed time on the system, measured by {\tt R} on a machine with an Intel Core i7-2600 CPU $@ 3.40$Ghz x 4 using R.3.3.1 on Linux Mint 17.2 (64-bit). Therefore, we take the first ten data set of the three-dimensional non-simplified vine-copulas and the five-dimensional non-simplified vine-copulas form our simulation study, each in case (b) with $\beta=0.6$ for both sample sizes $n=500$ and $n=2000$. Computing time is measured for i) bivariate unconditional copula densities $c_{12}$ for the first two marginal arguments from the selected data, ii) bivariate conditional copula densities $c_{12|3}$ with one conditioning argument and iii) the estimation of the non-simplified vine copula. The computing times are realized with one starting value of $\lambda$. Choosing three starting values for each copula estimation as in the simulation study increases the computing time linearly. See Table \ref{comp.time} for the results. The five-dimensional non-simplified vine copulas are computed in approximately 3 to 7 minutes, depending on the basis size and sample size, while the computing time for a three-dimensional non-simplified vine-copula is less than one minute. It might seem counterintuitive that the computing time decreases if the sample size increases from $n=500$ to $n=2000$ observations. That is because less iteration steps in the quadratic programming are required.
\end{appendix}
\bibliographystyle{chicago}
\bibliography{Literatur}
\onecolumn

\begin{figure}[!ht]
\centering
\caption{Exemplary plots for three-dimensional non-simplified vine copula with $\tau(u_2)=0.4-0.8 u_2$. Upper left: function of Kendell's $\tau(u_2)$, upper right: random vector $(U_{1|2},U_{3|2}), U_2<0.5$, lower left: random vector $(U_{1|2},U_{3|2}), U_2>0.5$ and lower right: complete random vector $(U_{1|2},U_{3|2})$.}
\includegraphics[width=16cm]{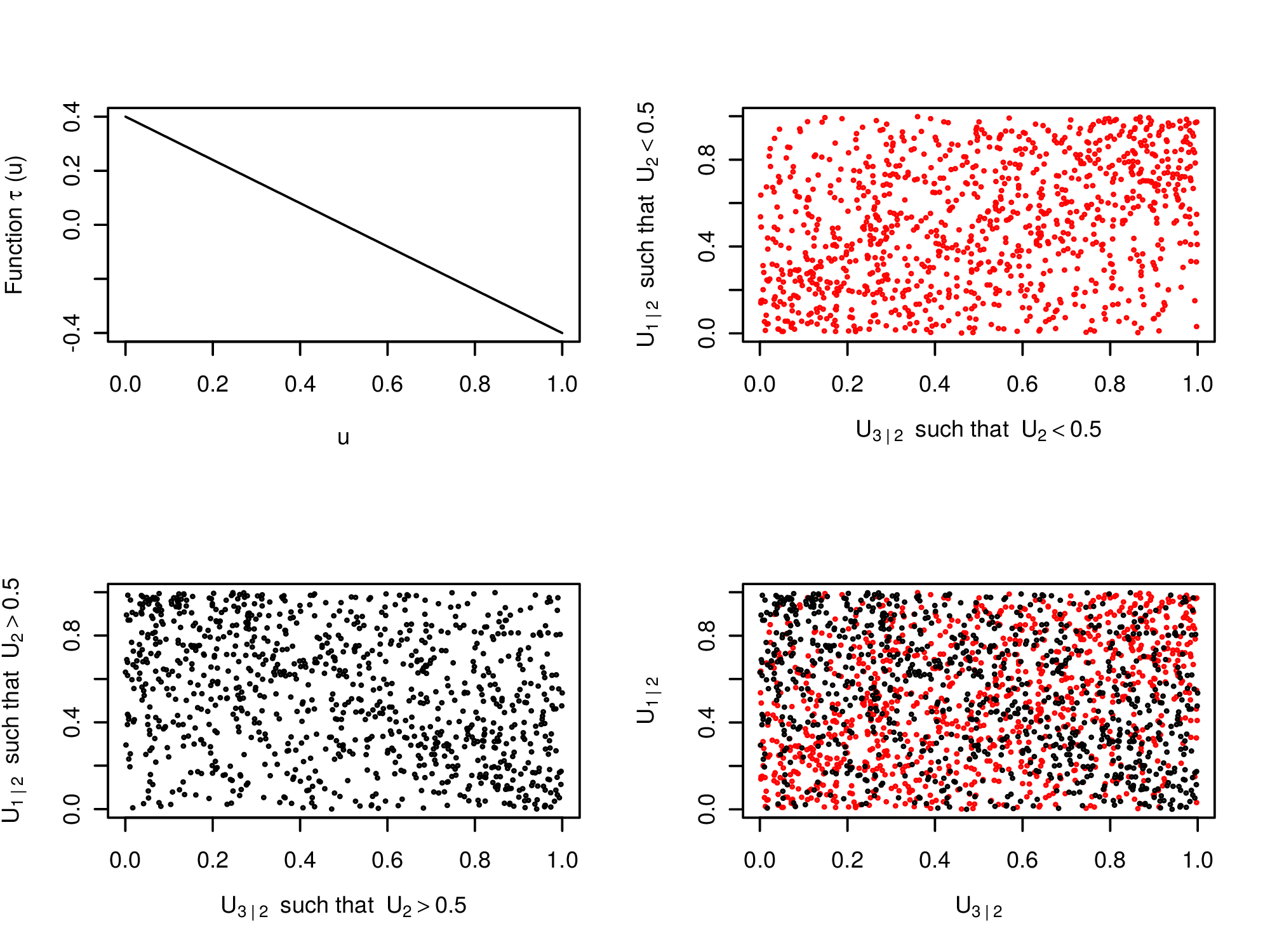}
\label{frankex}
\end{figure}

\begin{figure}[!ht]
\centering
\caption{(a) B-spline density basis $B^{(\tau(2))}(\mathbf{u}_j)$ and corresponding building blocks of the univariate hierarchical B-spline density basis $B^{(\tau(0))}_{\mathcal{I}_0}(\mathbf{u}_j), B^{(\tau(1))}_{\mathcal{I}_1}(\mathbf{u}_j)$ and $B^{(\tau(2))}_{\mathcal{I}_2}(\mathbf{u}_j)$ as graphics (b), (c) and (d).}
\subfloat{\includegraphics[width=7cm]{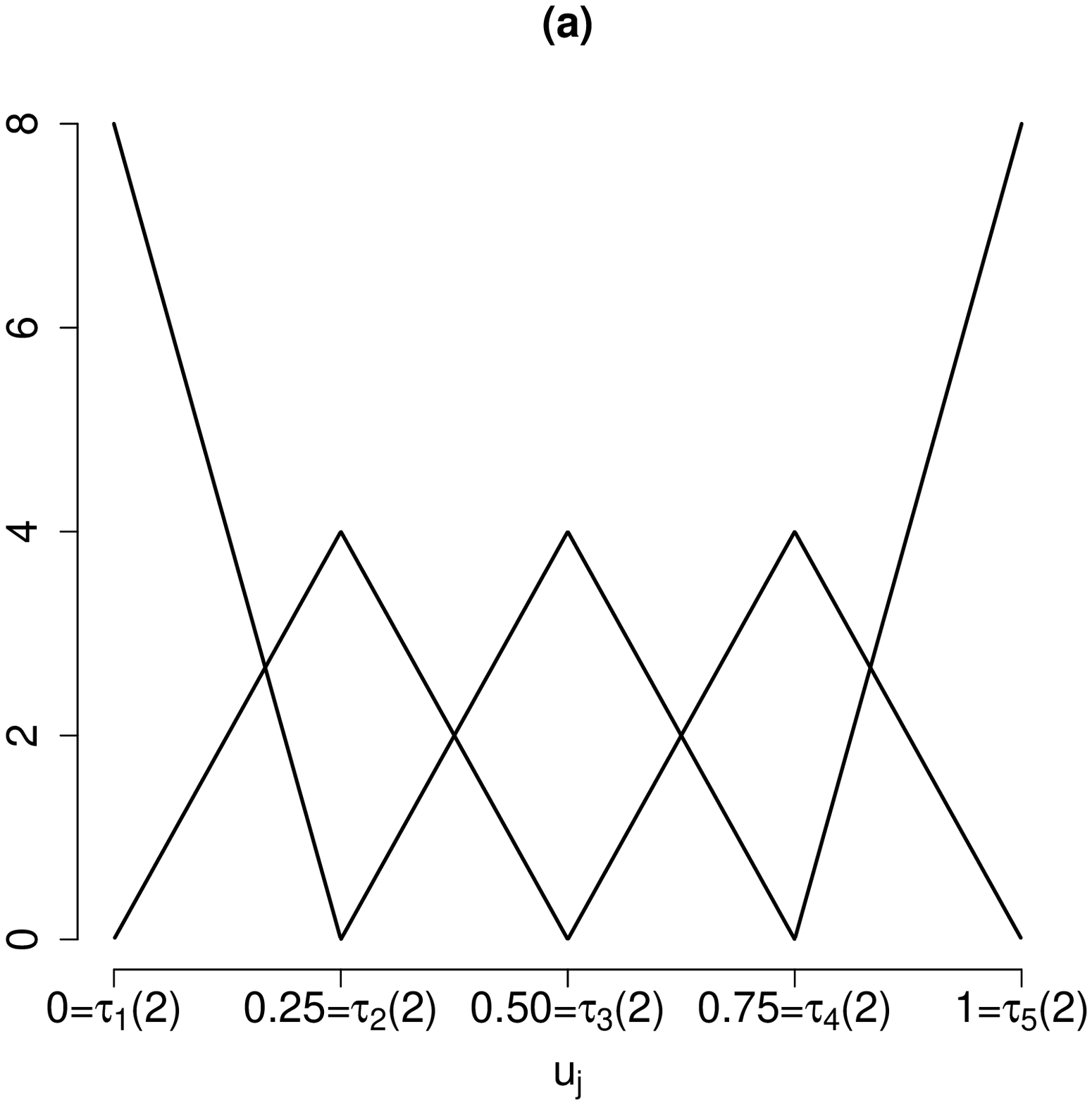}}
\subfloat{\includegraphics[width=7cm]{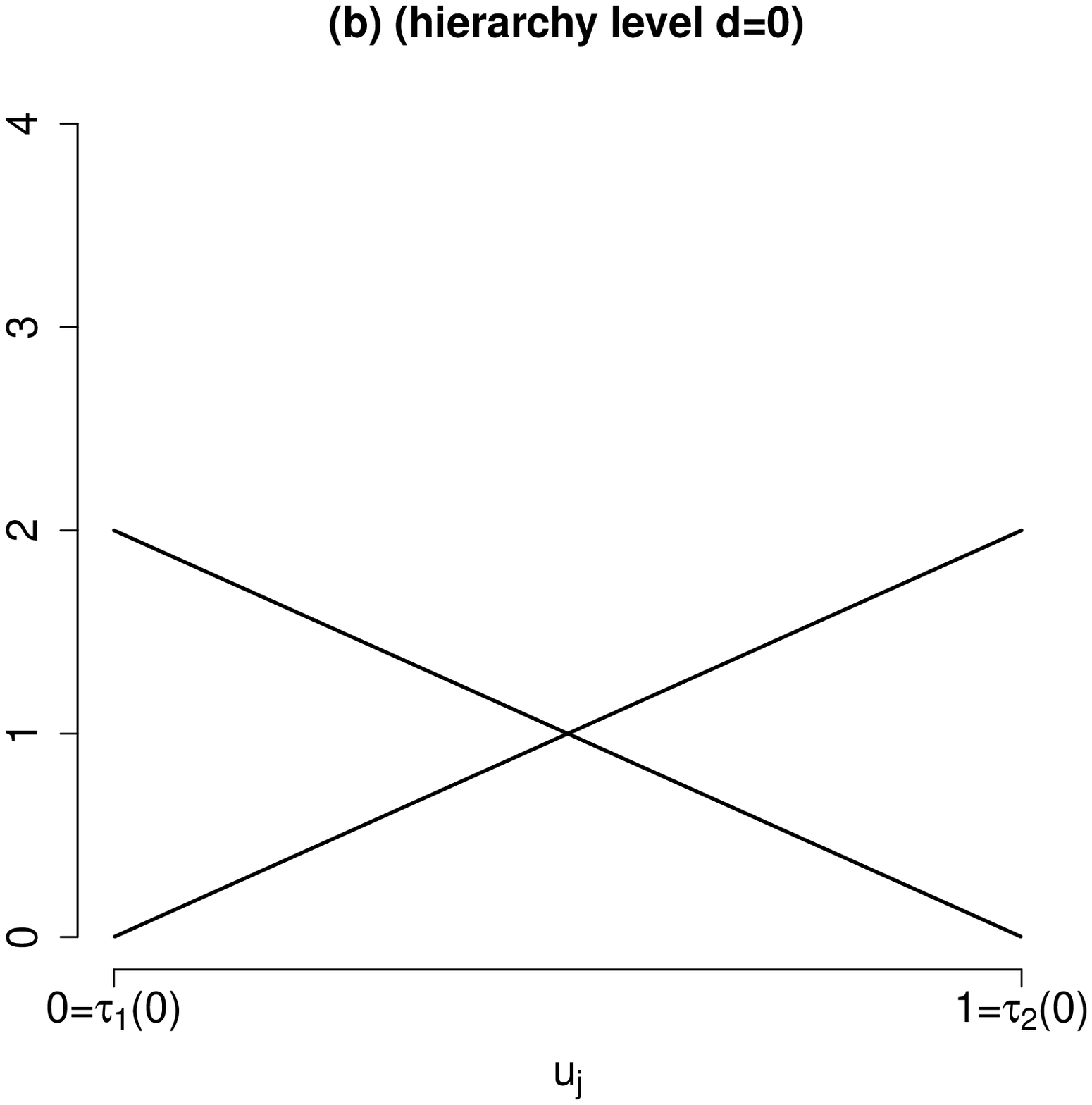}}\\
\subfloat{\includegraphics[width=7cm]{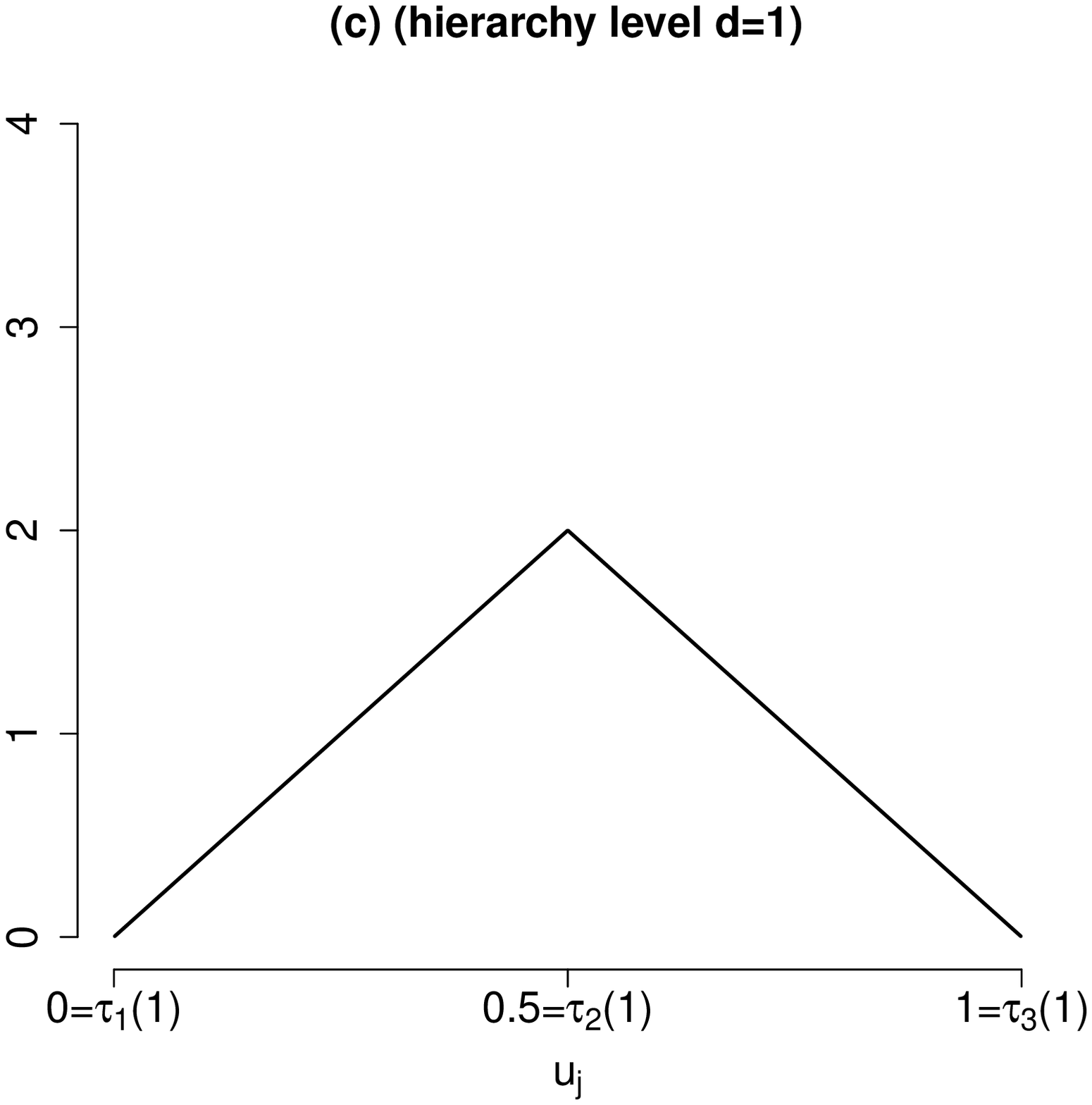}}
\subfloat{\includegraphics[width=7cm]{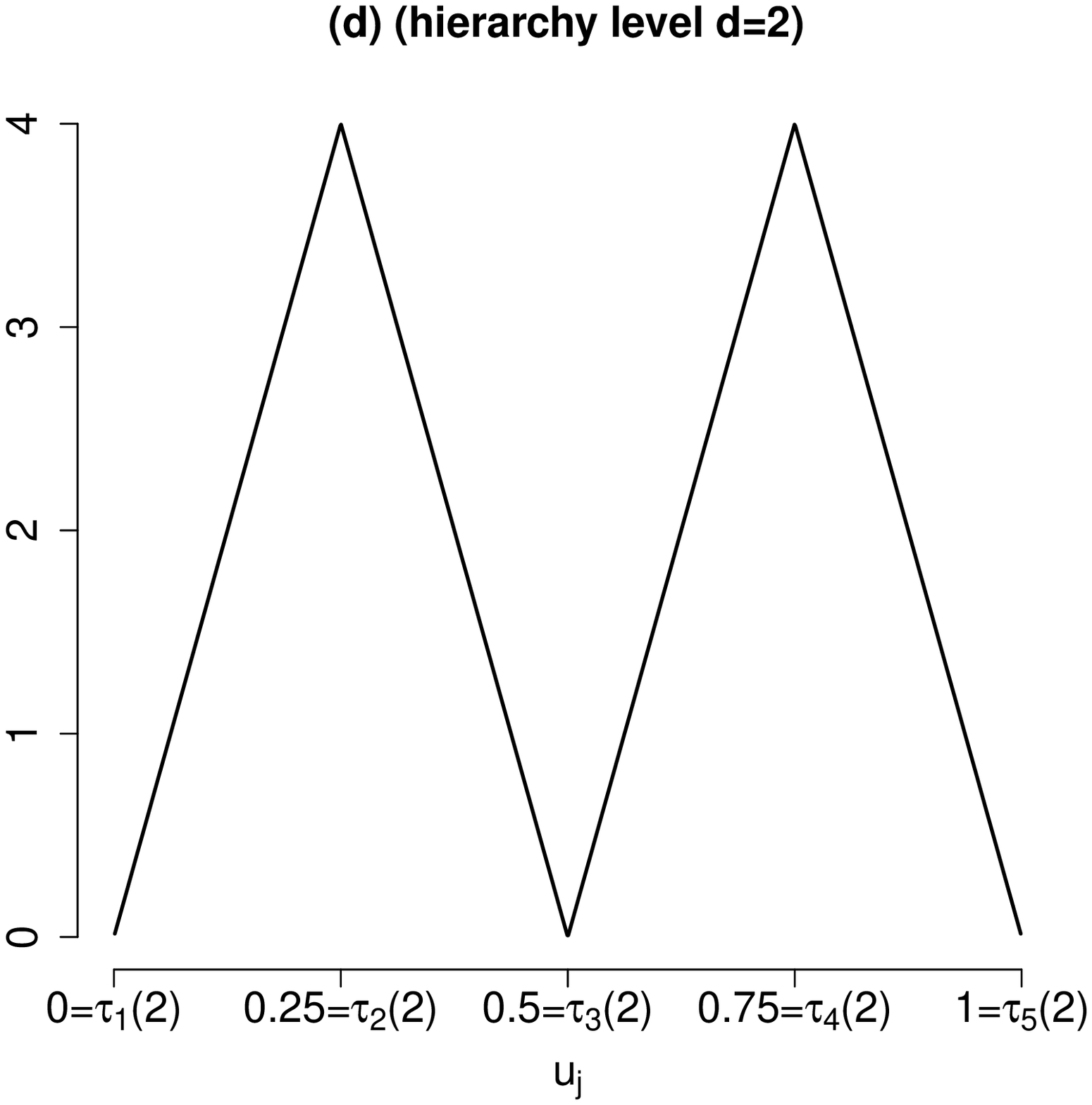}}\\
\label{Figure2}
\end{figure}

\begin{figure}[!ht]
\centering
\caption{$\tilde{\mathbf{\Phi}}^{(2,4)}(u_1,u_2)$ is the full tensor product of two univariate B-spline bases, each margin consists of ${B}^{(\tau(0))}_{\mathcal{I}_0}(\mathbf{u}_j)$, ${B}^{(\tau(1))}_{\mathcal{I}_1}(\mathbf{u}_j)$ and  ${B}^{(\tau(2))}_{\mathcal{I}_2}(\mathbf{u}_j)$ for $j=1,2$. The construction principle of the sparse B-spline basis $\tilde{\mathbf{\Phi}}^{(2,2)}(u_1,u_2)$ is to remove columns from the full tensor product, reducing the number of spline bases from 25 to 17 in this bivariate example for $d=2$.}
\subfloat{\begin{tabular}{c|c|c|c}
&${B}^{(\tau(0))}_{\mathcal{I}_0}(u_2)$ & ${B}^{(\tau(1))}_{\mathcal{I}_1}(u_2)$ & ${B}^{(\tau(2))}_{\mathcal{I}_2}(u_2)$\\
\hline
${B}^{(\tau(0))}_{\mathcal{I}_0}(u_1)$ & ${B}^{(\tau(0))}_{\mathcal{I}_0}(u_1) \otimes {B}^{(\tau(0))}_{\mathcal{I}_0}(u_2)$ &
 ${B}^{(\tau(0))}_{\mathcal{I}_0}(u_1) \otimes {B}^{(\tau(1))}_{\mathcal{I}_1}(u_2)$ &
 ${B}^{(\tau(0))}_{\mathcal{I}_0}(u_1) \otimes {B}^{(\tau(2))}_{\mathcal{I}_2}(u_2)$ \\
\hline  ${B}^{(\tau(1))}_{\mathcal{I}_1}(u_1)$ & ${B}^{(\tau(1))}_{\mathcal{I}_1}(u_1) \otimes {B}^{(\tau(0))}_{\mathcal{I}_0}(u_2)$ &
 ${B}^{(\tau(1))}_{\mathcal{I}_1}(u_1) \otimes {B}^{(\tau(1))}_{\mathcal{I}_1}(u_2)$ & {\textbf{removed}} \\
\hline  ${B}^{(\tau(2))}_{\mathcal{I}_2}(u_1)$ & ${B}^{(\tau(2))}_{\mathcal{I}_2}(u_1) \otimes {B}^{(\tau(0))}_{\mathcal{I}_0}(u_2)$ & {\textbf{removed}}& {\textbf{removed}}\\
\end{tabular}}\\
\label{Figure3}
\end{figure}

\begin{figure}[!ht]
\caption{(left) Full tensor product ${\mathbf{\Phi_5}}(u_1,u_2)$ consists of $5^2=25$ basis functions located at each dot. (right) $\tilde{\mathbf{\Phi}}^{(2,2)}(u_1,u_2)$  consists of 17 basis functions located at each dot.} 
\subfloat{\begin{pspicture}[showgrid=FALSE](8,8)
\psline{->}(1,0.5)(1,8)
\psline{->}(0.5,1)(8,1)
\pscircle(1.5,1.5){0.1}
\pscircle(3,1.5){0.1}
\pscircle(4.5,1.5){0.1}
\pscircle(6,1.5){0.1}
\pscircle(7.5,1.5){0.1}
\pscircle(1.5,3){0.1}
\pscircle(3,3){0.1}
\pscircle(4.5,3){0.1}
\pscircle(6,3){0.1}
\pscircle(7.5,3){0.1}
\pscircle(1.5,4.5){0.1}
\pscircle(3,4.5){0.1}
\pscircle(4.5,4.5){0.1}
\pscircle(6,4.5){0.1}
\pscircle(7.5,4.5){0.1}
\pscircle(1.5,6){0.1}
\pscircle(3,6){0.1}
\pscircle(4.5,6){0.1}
\pscircle(6,6){0.1}
\pscircle(7.5,6){0.1}
\pscircle(1.5,7.5){0.1}
\pscircle(3,7.5){0.1}
\pscircle(4.5,7.5){0.1}
\pscircle(6,7.5){0.1}
\pscircle(7.5,7.5){0.1}
\psline{-}(0.85,1.5)(1.15,1.5)
\psline{-}(0.85,3)(1.15,3)
\psline{-}(0.85,4.5)(1.15,4.5)
\psline{-}(0.85,6)(1.15,6)
\psline{-}(0.85,7.5)(1.15,7.5)
\psline{-}(1.5,0.85)(1.5,1.15)
\psline{-}(3,0.85)(3,1.15)
\psline{-}(4.5,0.85)(4.5,1.15)
\psline{-}(6,0.85)(6,1.15)
\psline{-}(7.5,0.85)(7.5,1.15)
\rput(0.7,1.5){$\tau_1$}
\rput(0.7,3){$\tau_2$}
\rput(0.7,4.5){$\tau_3$}
\rput(0.7,6){$\tau_4$}
\rput(0.7,7.5){$\tau_5$}
\rput(1.5,0.7){$\tau_1$}
\rput(3,0.7){$\tau_2$}
\rput(4.5,0.7){$\tau_3$}
\rput(6,0.7){$\tau_4$}
\rput(7.5,0.7){$\tau_5$}
\rput(0,4){$\mathbf{u_2}$}
\rput(4,0){$\mathbf{u_1}$}
\end{pspicture}} \hspace{1.05cm} \subfloat{\begin{pspicture}[showgrid=FALSE](8,8)
\psline{->}(1,0.5)(1,8)
\psline{->}(0.5,1)(8,1)
\pscircle(1.5,1.5){0.1}
\pscircle(4.5,1.5){0.1}
\pscircle(6,1.5){0.1}
\pscircle(3,1.5){0.1}
\pscircle(7.5,1.5){0.1}
\pscircle(1.5,3){0.1}
\pscircle(1.5,4.5){0.1}
\pscircle(4.5,4.5){0.1}
\pscircle(1.5,6){0.1}
\pscircle(7.5,4.5){0.1}
\pscircle(7.5,3){0.1}
\pscircle(7.5,6){0.1}
\pscircle(1.5,7.5){0.1}
\pscircle(3,7.5){0.1}
\pscircle(4.5,7.5){0.1}
\pscircle(6,7.5){0.1}
\pscircle(7.5,7.5){0.1}
\psline{-}(0.85,1.5)(1.15,1.5)
\psline{-}(0.85,3)(1.15,3)
\psline{-}(0.85,4.5)(1.15,4.5)
\psline{-}(0.85,6)(1.15,6)
\psline{-}(0.85,7.5)(1.15,7.5)
\psline{-}(1.5,0.85)(1.5,1.15)
\psline{-}(3,0.85)(3,1.15)
\psline{-}(4.5,0.85)(4.5,1.15)
\psline{-}(6,0.85)(6,1.15)
\psline{-}(7.5,0.85)(7.5,1.15)
\tiny
\rput(0,1.5){${B}^{(\tau(0))}_{\mathcal{I}_0}(u_2)$}
\rput(0,7.5){${B}^{(\tau(0))}_{\mathcal{I}_0}(u_2)$}
\rput(0,4.5){${B}^{(\tau(1))}_{\mathcal{I}_1}(u_2)$}
\rput(0,3){${B}^{(\tau(2))}_{\mathcal{I}_2}(u_2)$}
\rput(0,6){${B}^{(\tau(2))}_{\mathcal{I}_2}(u_2)$}
\rput(1.5,0.25){${B}^{(\tau(0))}_{\mathcal{I}_0}(u_1)$}
\rput(7.5,0.25){${B}^{(\tau(0))}_{\mathcal{I}_0}(u_1)$}
\rput(4.5,0.25){${B}^{(\tau(1))}_{\mathcal{I}_1}(u_1)$}
\rput(3,0.25){${B}^{(\tau(2))}_{\mathcal{I}_2}(u_1)$}
\rput(6,0.25){${B}^{(\tau(2))}_{\mathcal{I}_2}(u_1)$}
{\small\rput(0.75,3.75){$\mathbf{u_2}$}}
{\small\rput(3.75,0.75){$\mathbf{u_1}$}}
\psline{|-|}(1.5,0)(7.5,0)
\rput(5,-0.35){$\tilde{\mathbf{B}}^{(\tau(2))}(\mathbf{u}_1)$}
\psline{|-|}(-0.75,1.5)(-0.75,7.5)
\rput[tr]{90}(-1.1,5){$\tilde{\mathbf{B}}^{(\tau(2))}(\mathbf{u}_2)$}
\end{pspicture}}
\label{Figure4}
\end{figure}
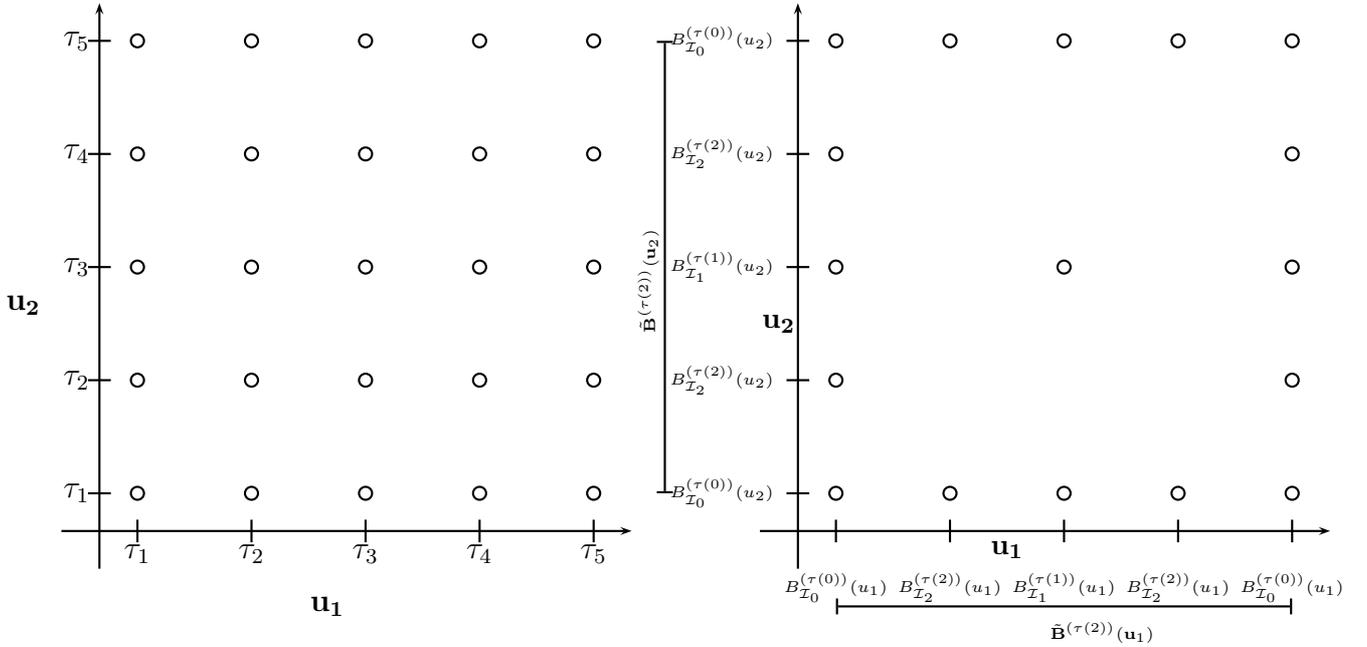

\newcommand{\fignote}{$KL_{non-par}$ is the out-of-sample KL divergence of a non-parametric estimation ('SimpA', 'Cond' or 'Test') and $KL_{par}$ is the out-of-sample KL divergence of the parametric estimation using the VineCopula package. Blue refers to the estimator 'Test', pink corresponds to the estimator 'Cond' and green refers to the estimator 'SimpA'. The whiskers cover 95\% of the data.}

\begin{figure}[!ht]
\centering
\caption{Linear and quadratic functions of kendell's $\tau$ for the construction of the conditional copula in higher trees.}
\includegraphics[width=16cm]{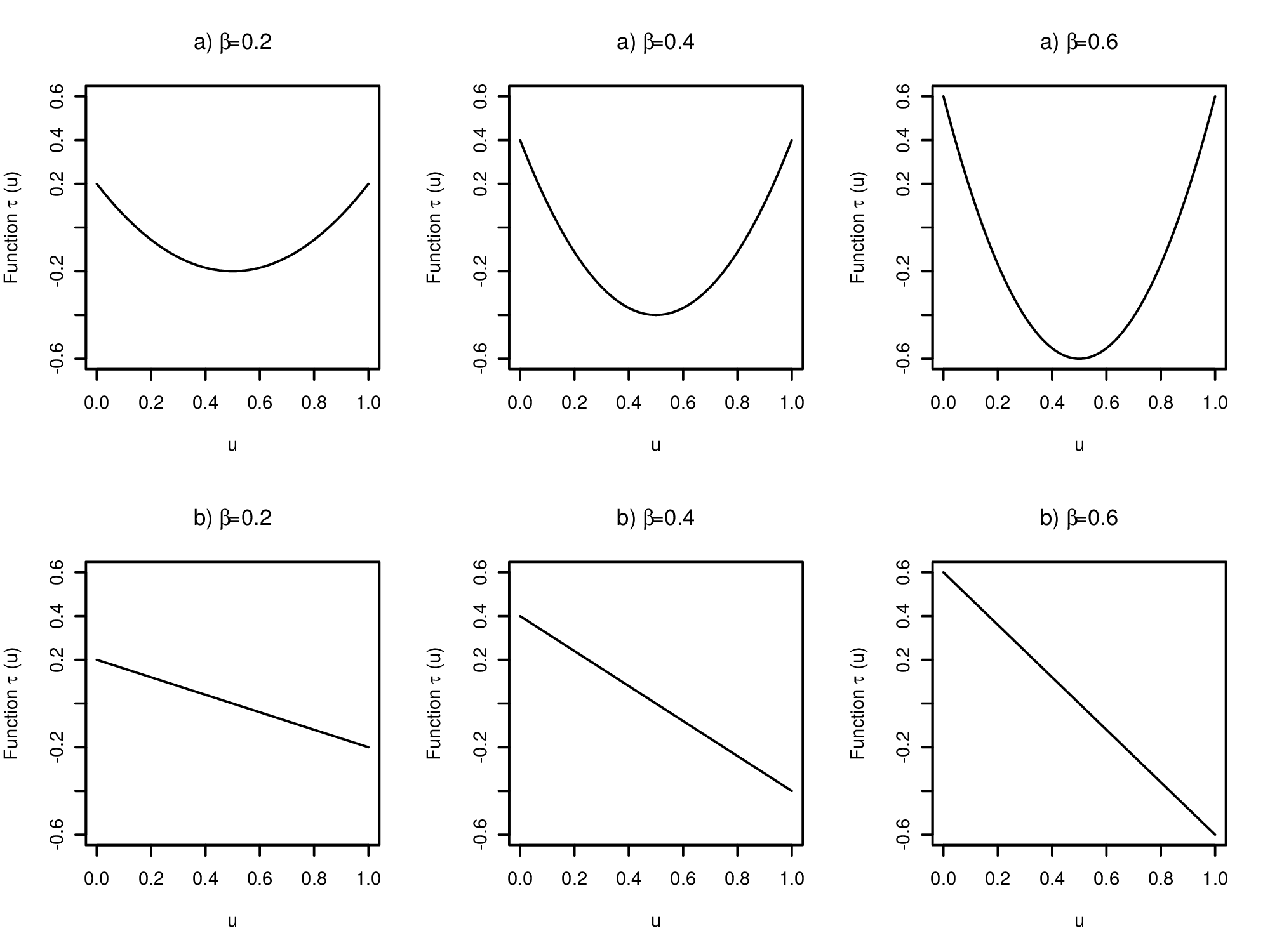}
\label{fig:generator}
\end{figure}

\begin{figure}[!ht]
\centering
\subfloat{\includegraphics[width=8.25cm]{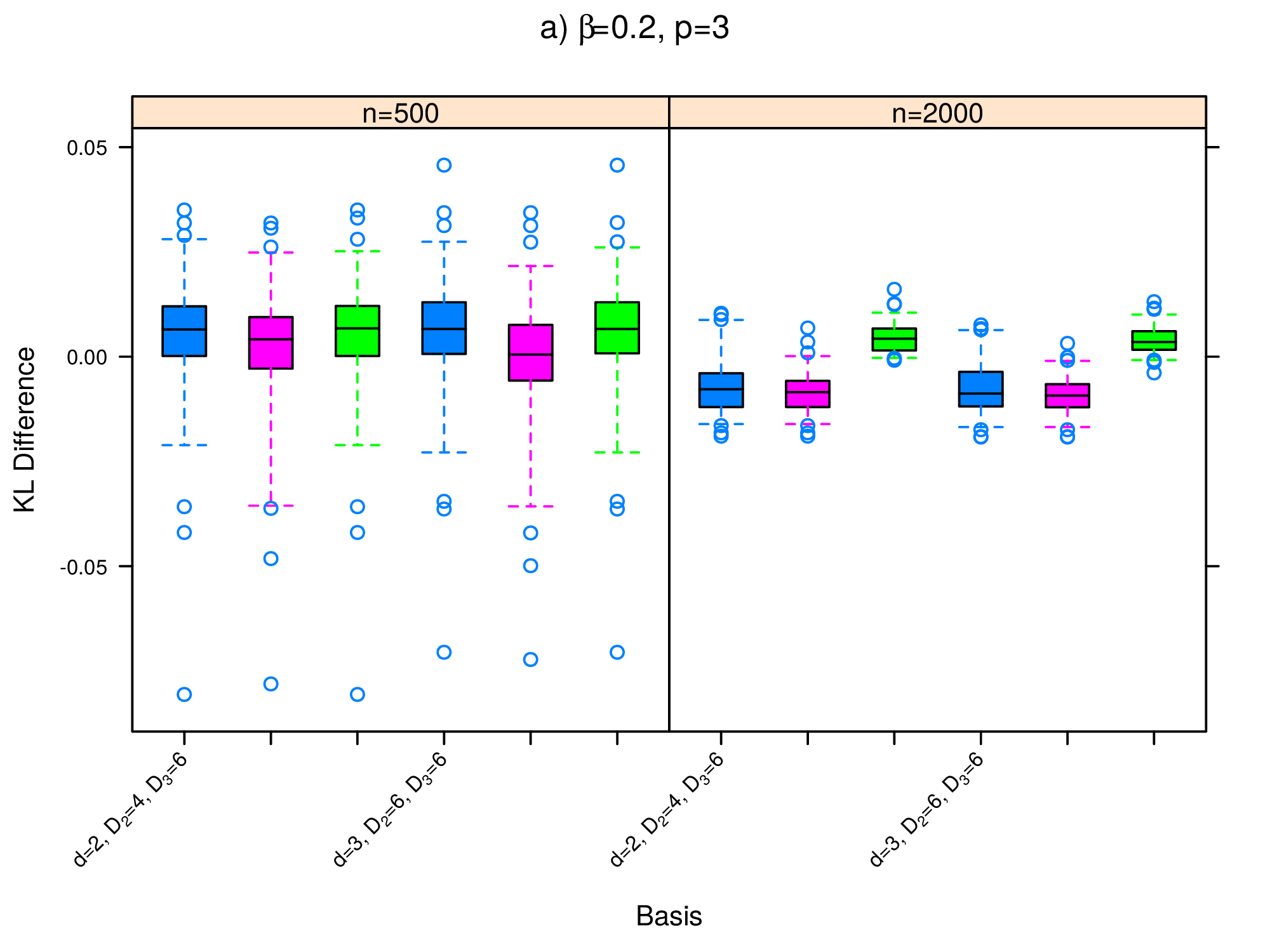}}
\subfloat{\includegraphics[width=8.25cm]{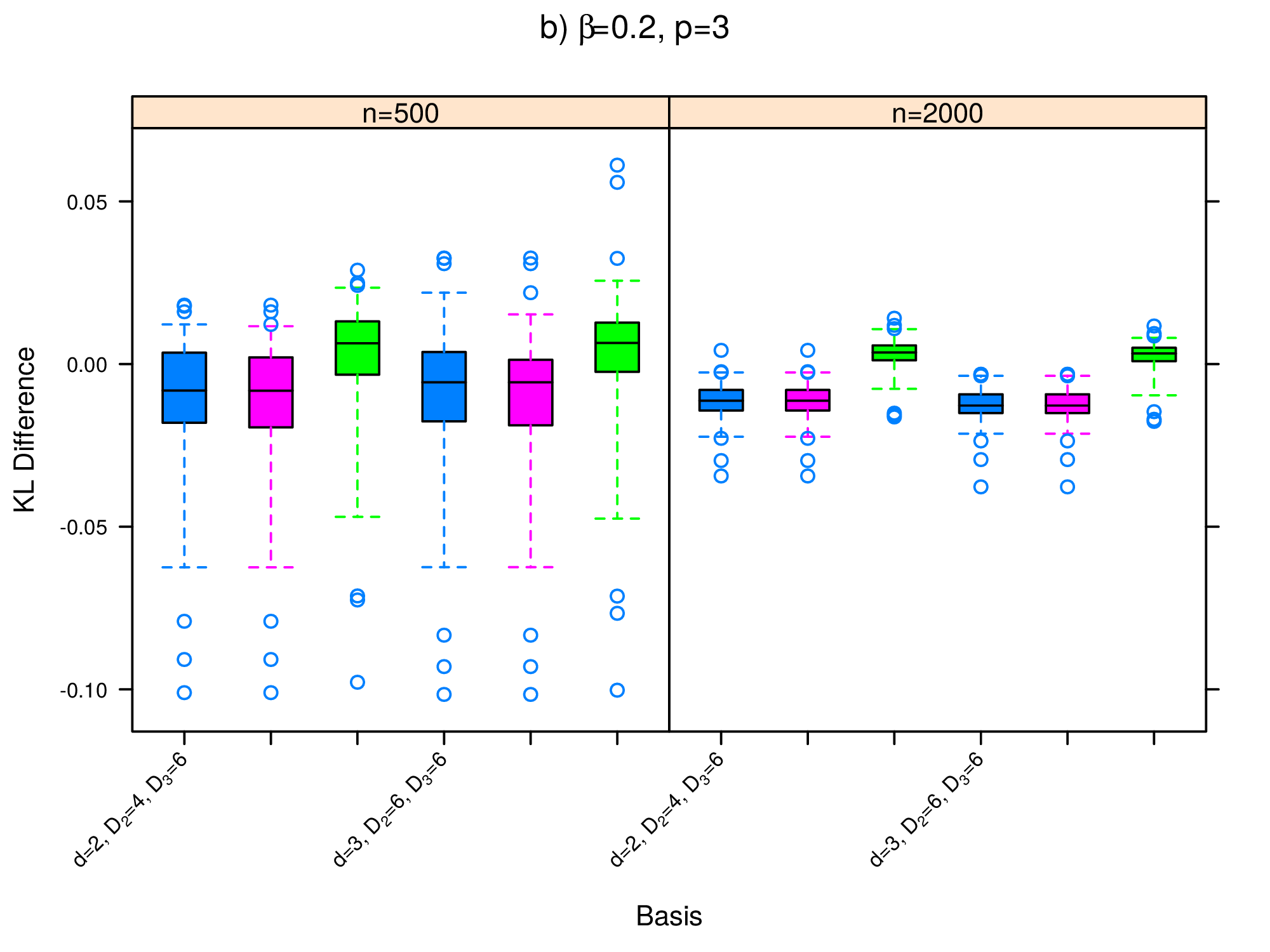}}\\
\subfloat{\includegraphics[width=8.25cm]{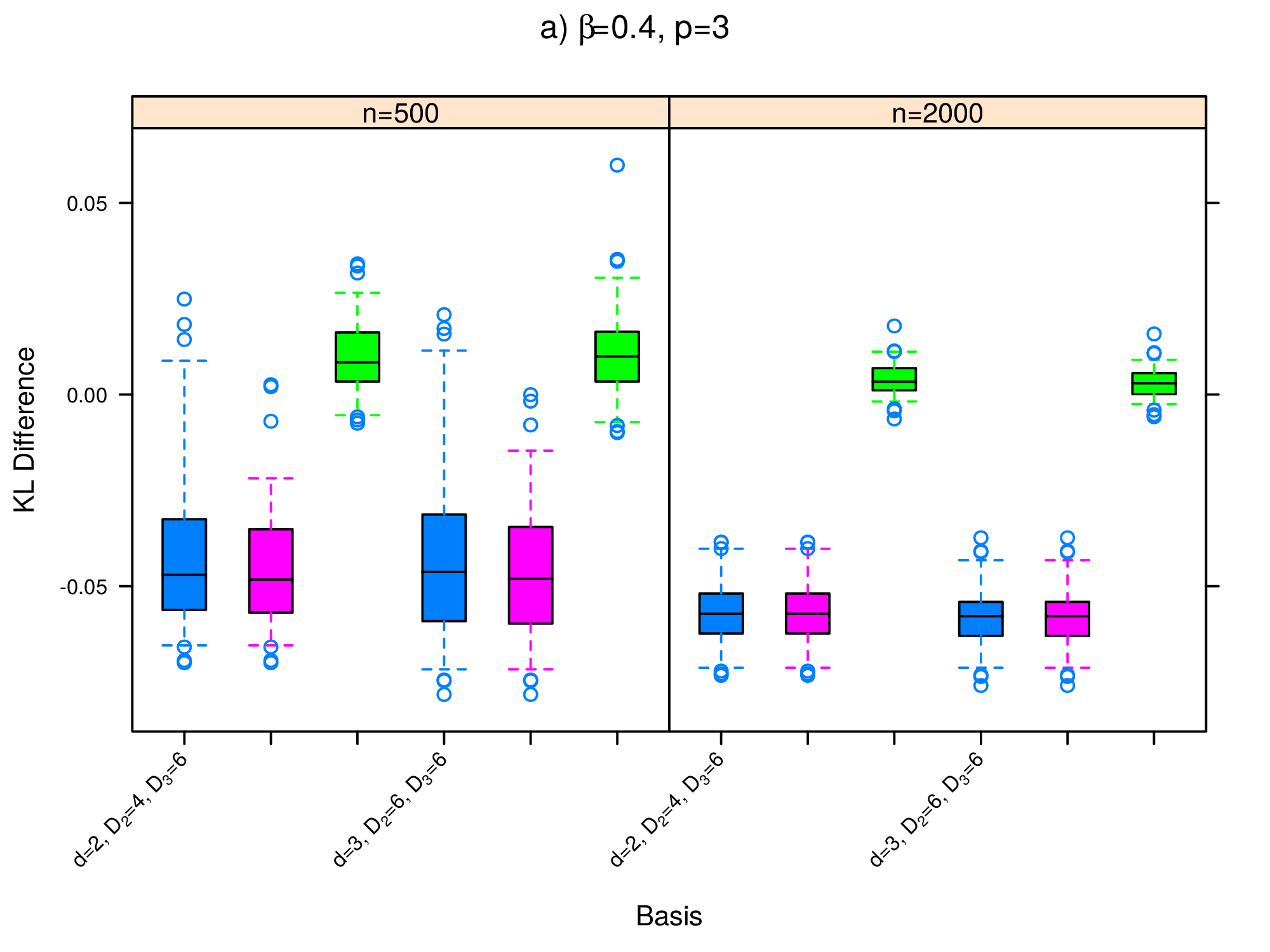}}
\subfloat{\includegraphics[width=8.25cm]{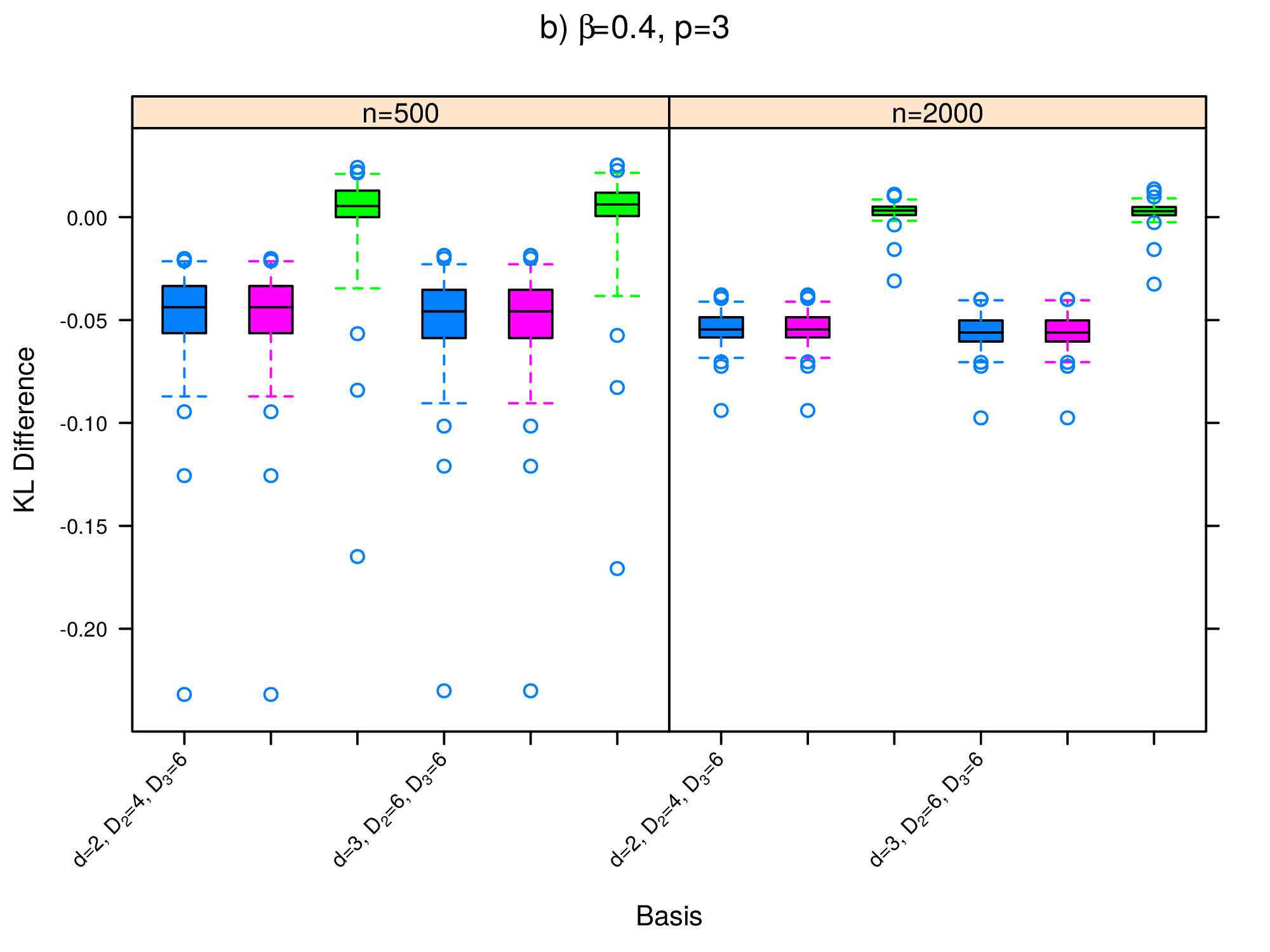}}\\
\subfloat{\includegraphics[width=8.25cm]{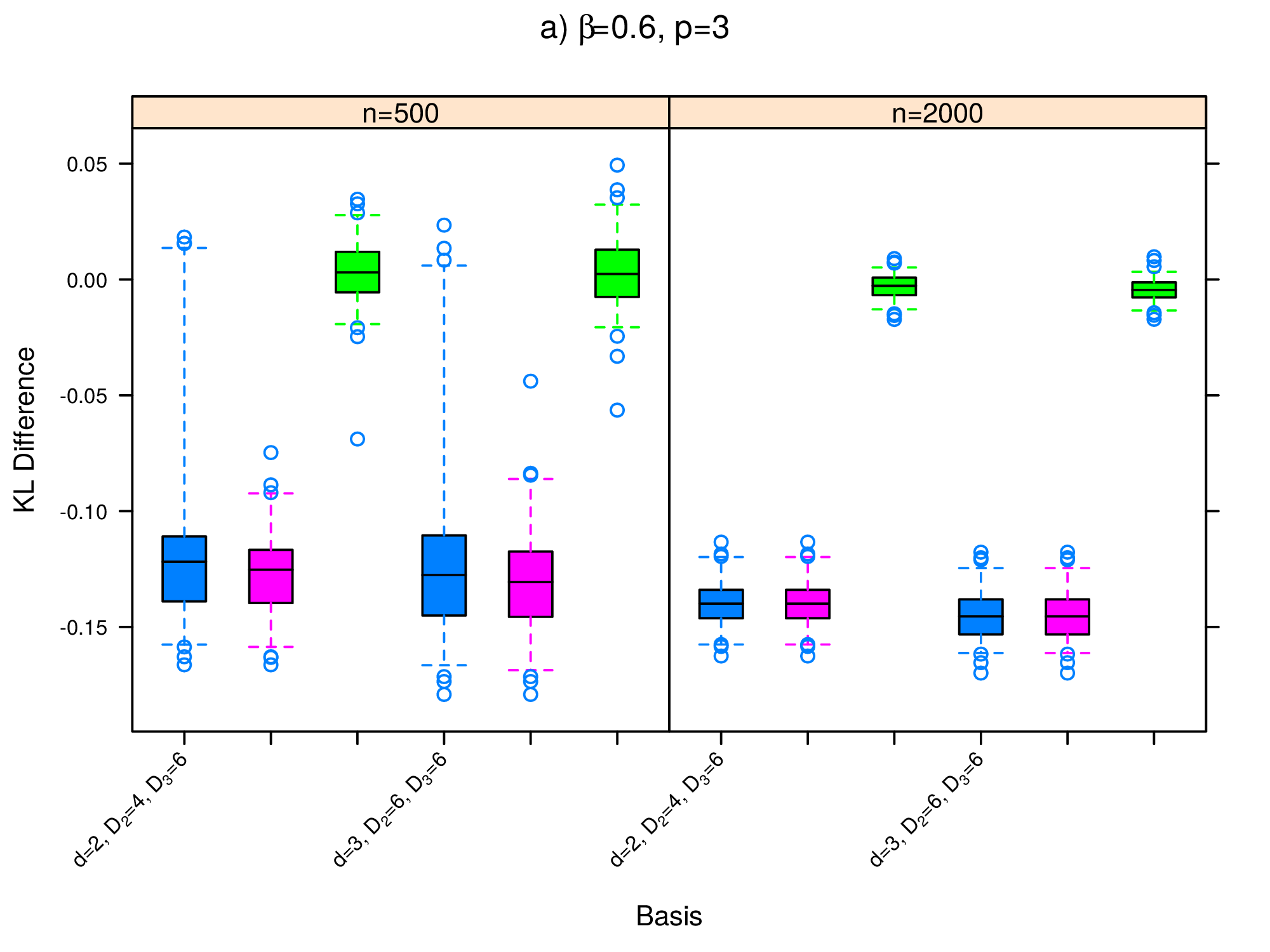}}
\subfloat{\includegraphics[width=8.25cm]{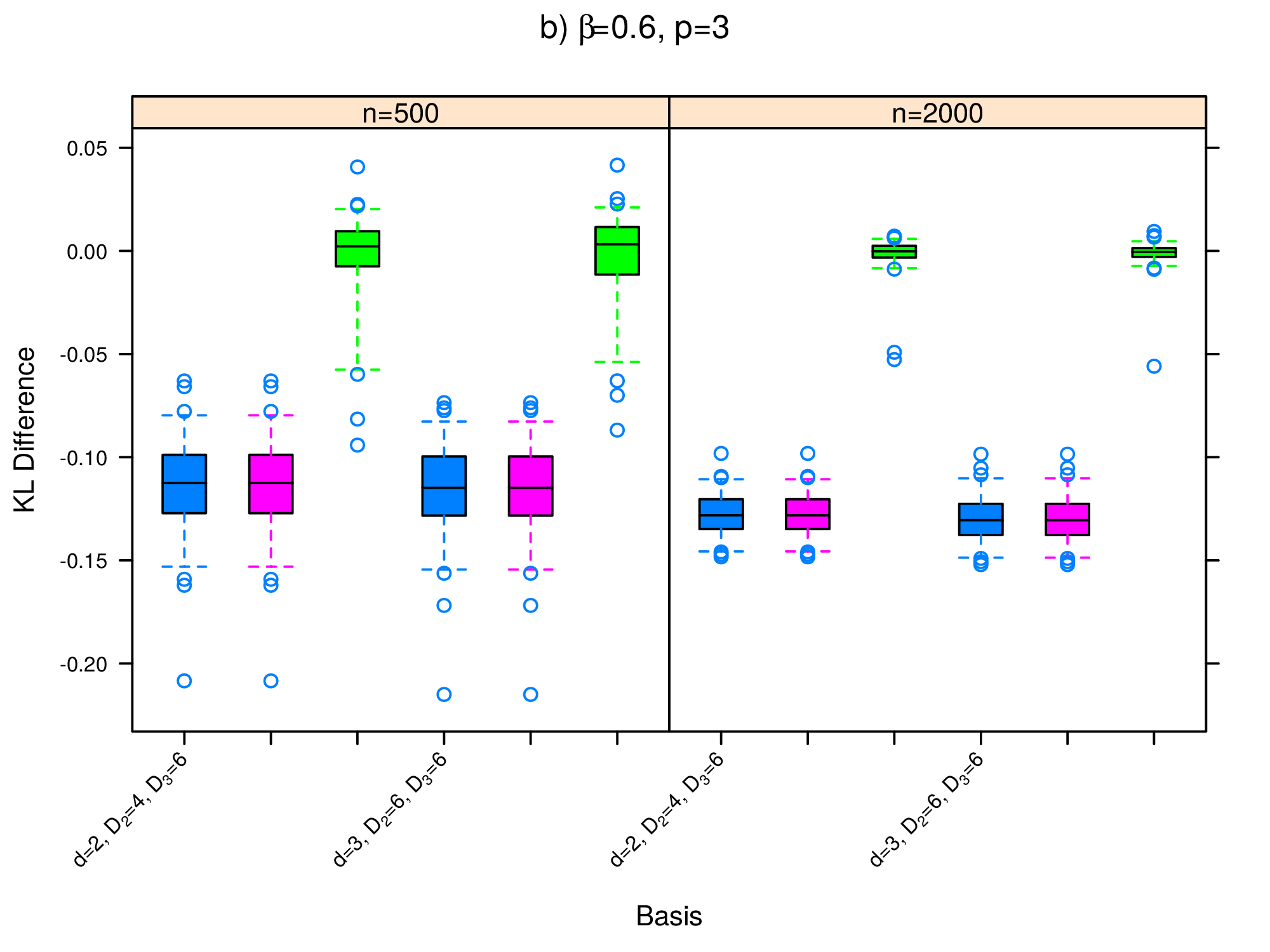}}\\
\caption{Box plots of the differences $KL_{non-par}-KL_{par}$ for three-dimensional non-simplified vine copulas. \fignote{} }\label{figp3}
\end{figure}

\begin{figure}[!ht]
\centering
\subfloat{\includegraphics[width=8.25cm]{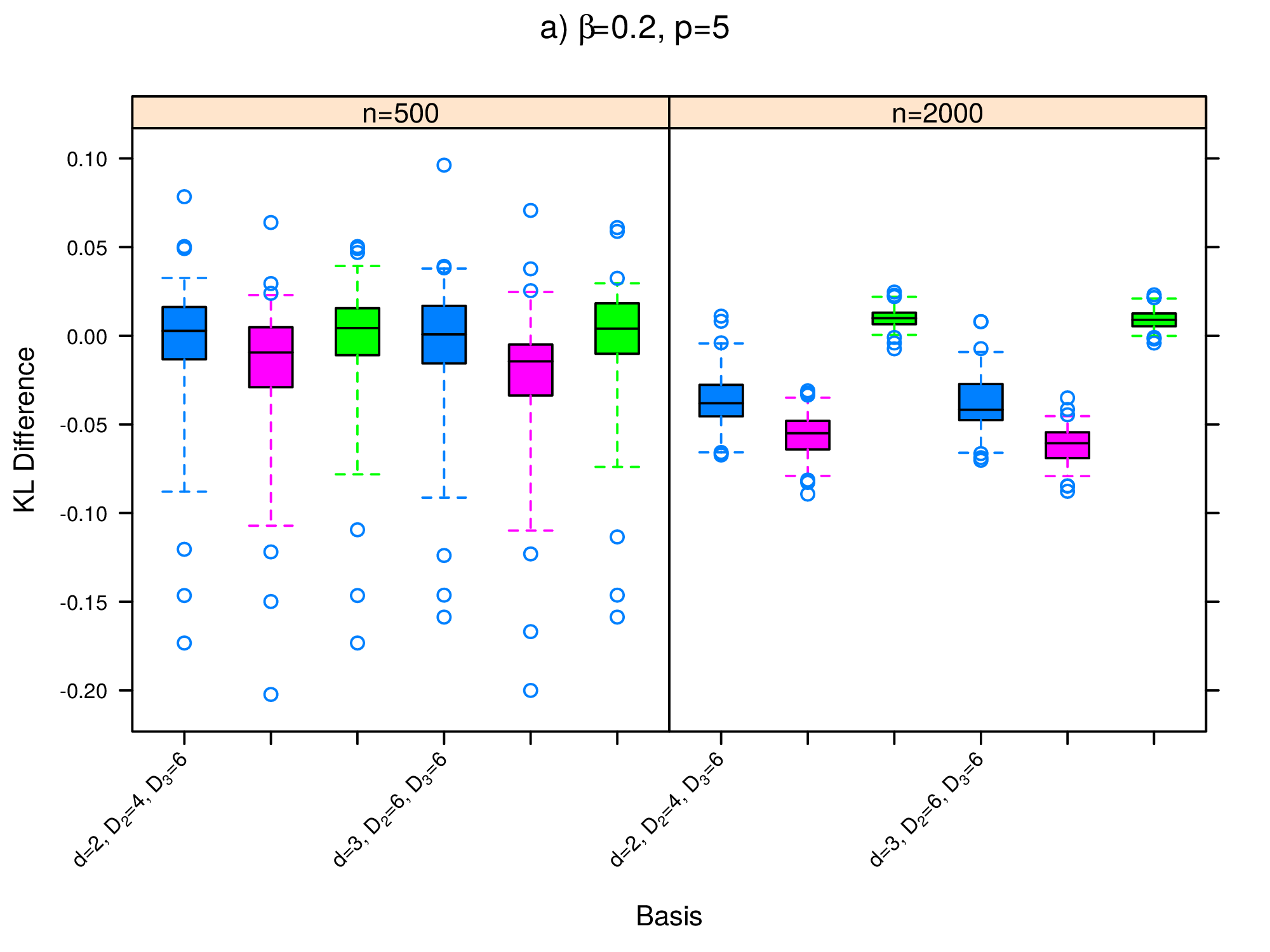}}
\subfloat{\includegraphics[width=8.25cm]{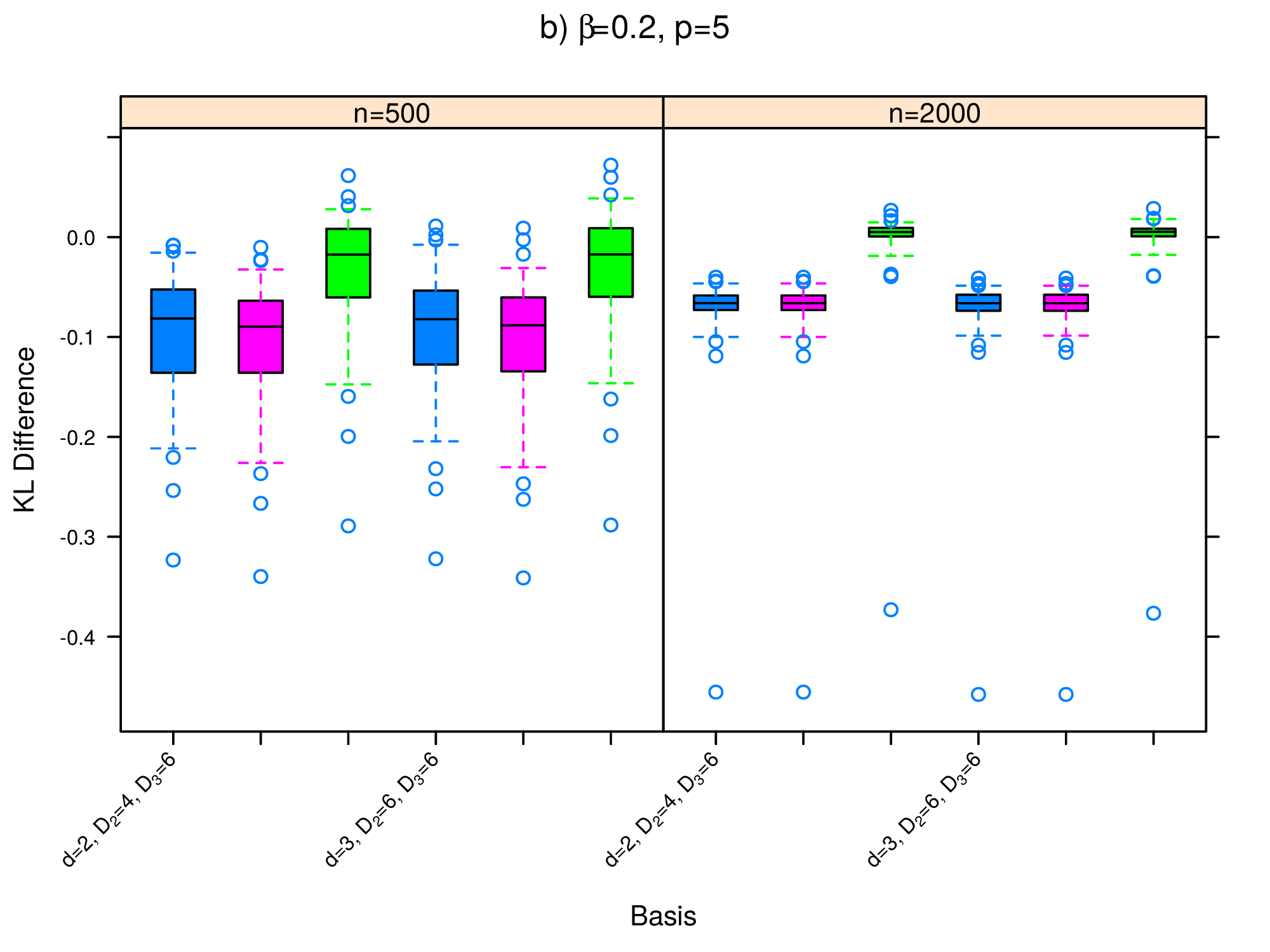}}\\
\subfloat{\includegraphics[width=8.25cm]{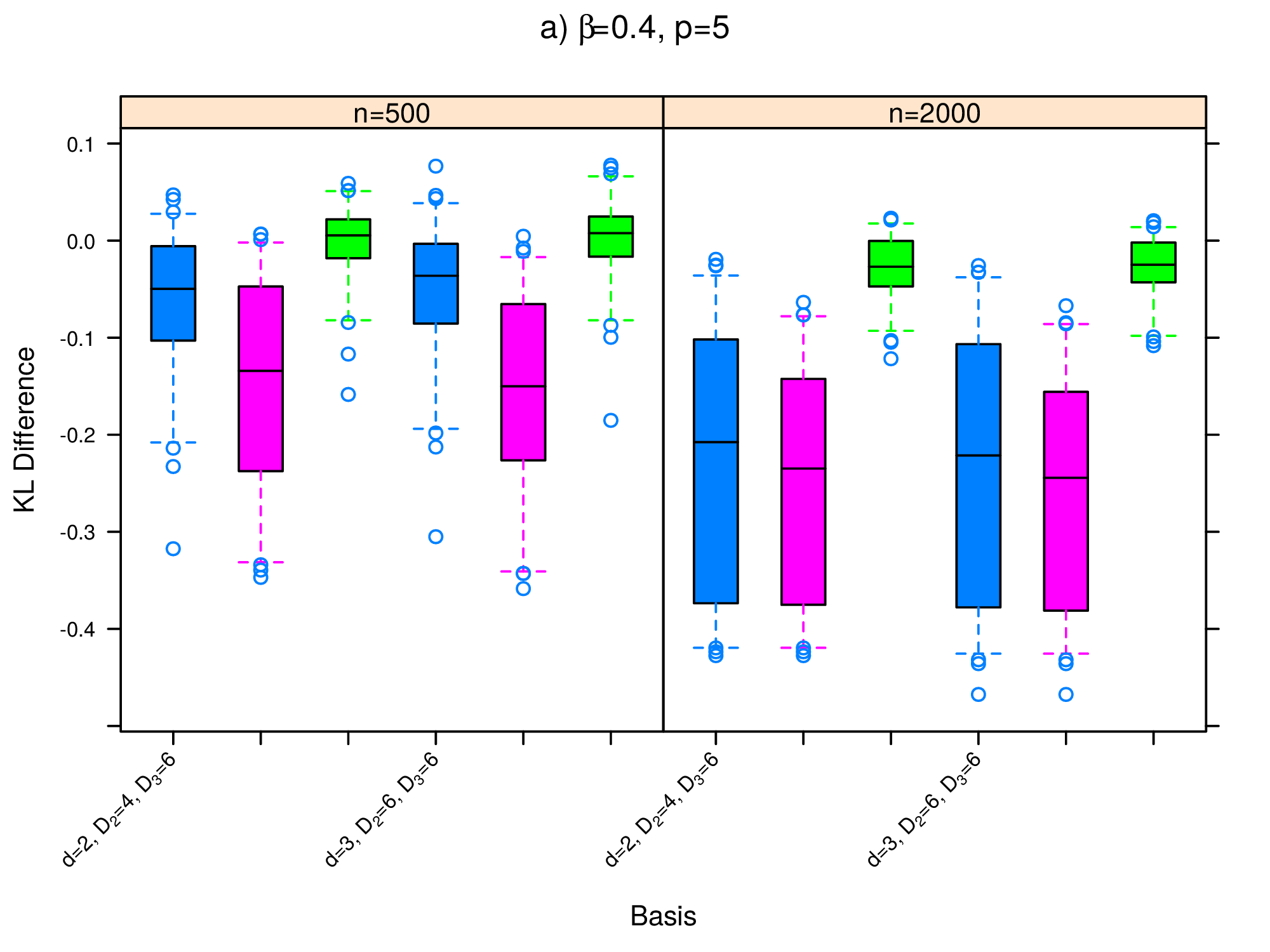}}
\subfloat{\includegraphics[width=8.25cm]{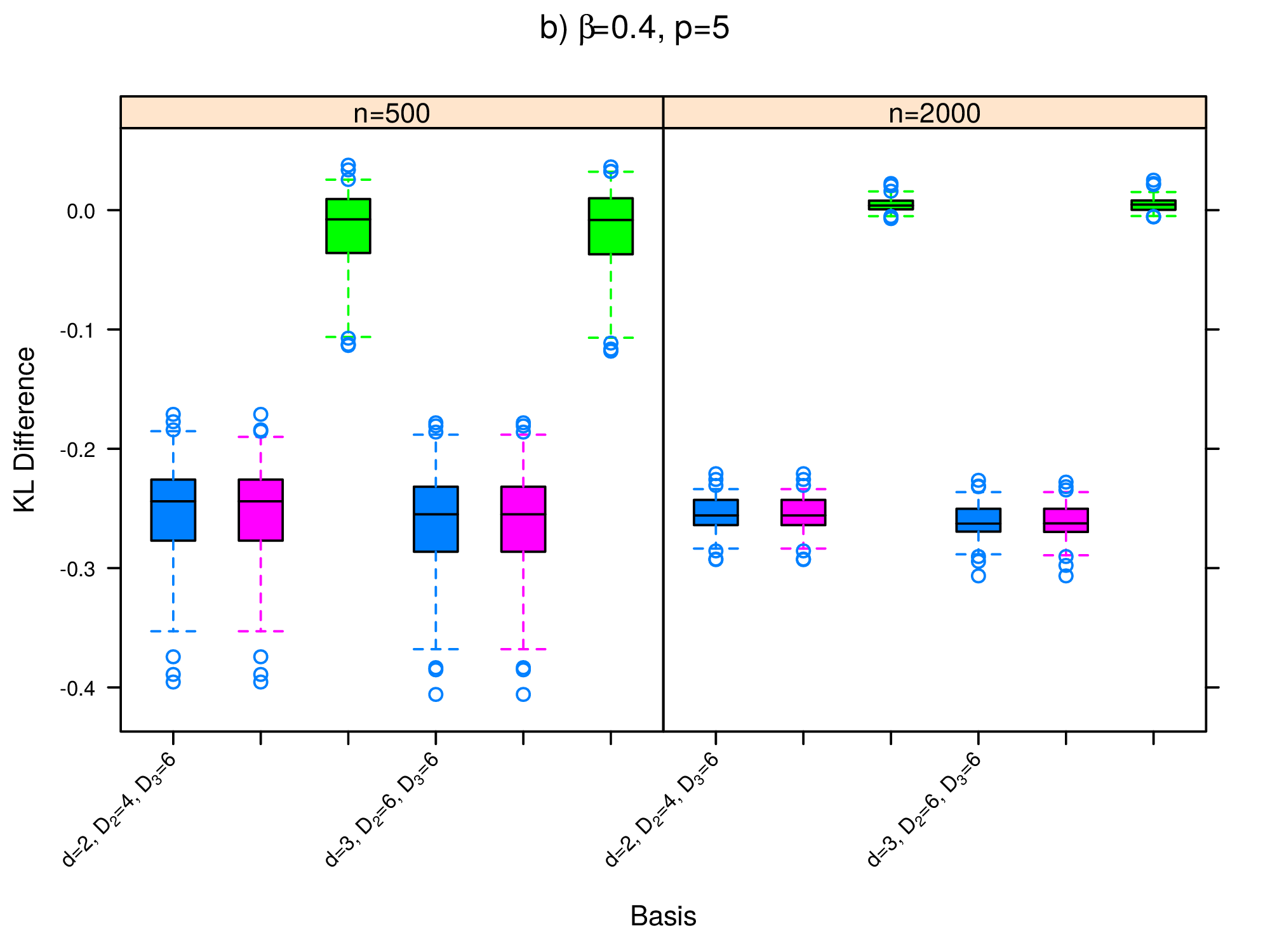}}\\
\subfloat{\includegraphics[width=8.25cm]{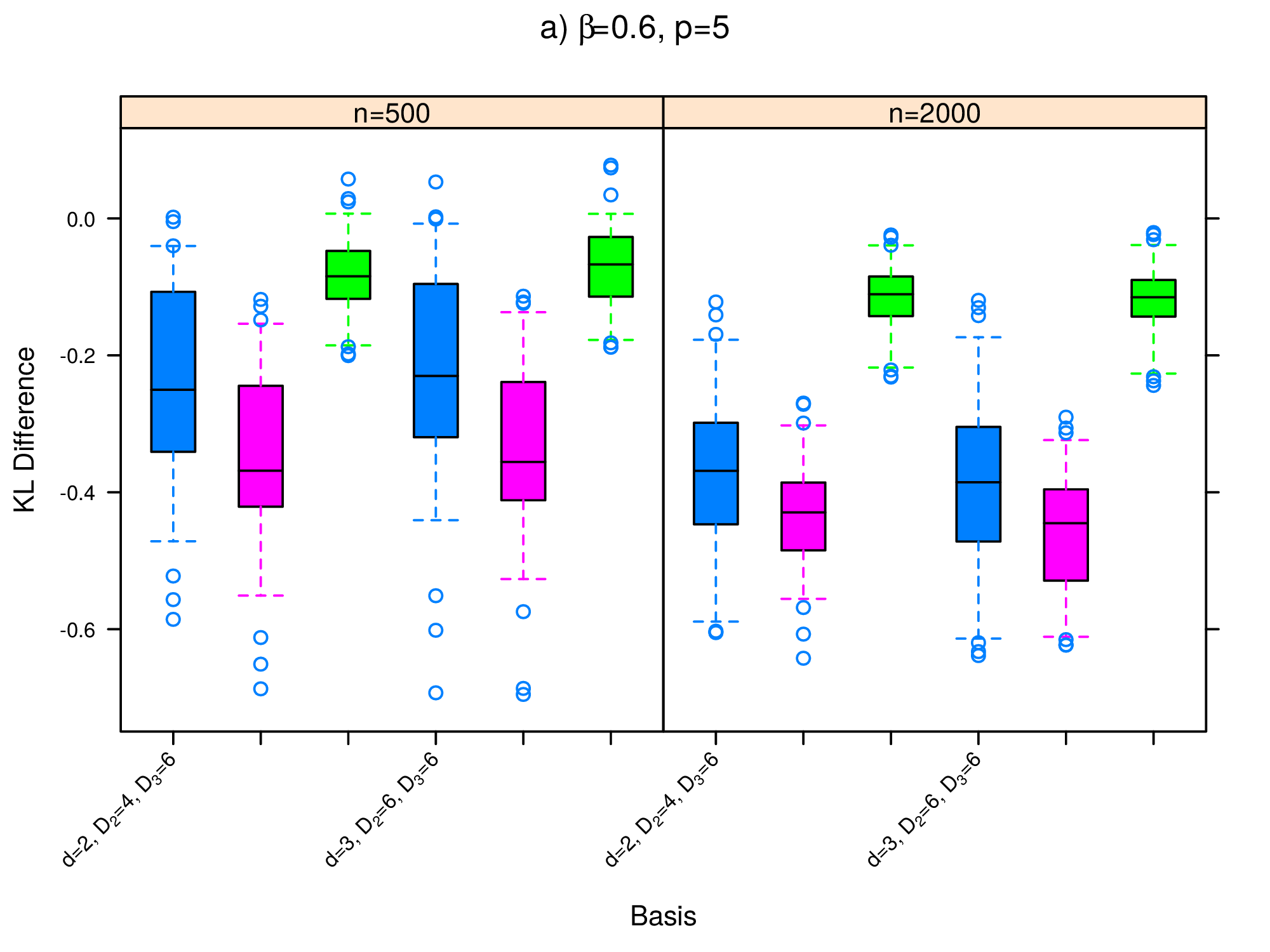}}
\subfloat{\includegraphics[width=8.25cm]{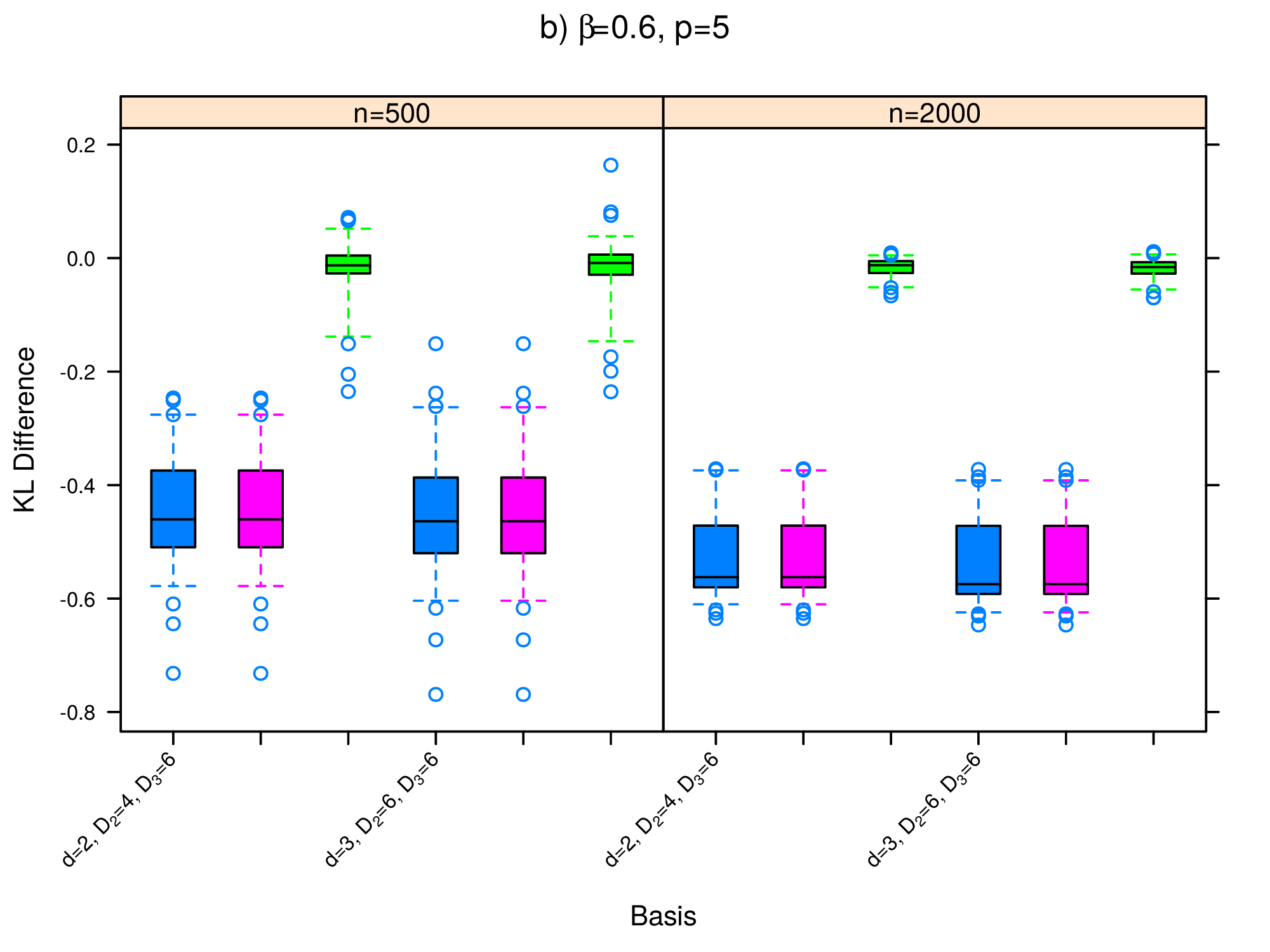}}\\
\caption{Box plots of the differences $KL_{non-par}-KL_{par}$ for five-dimensional non-simplified vine copulas. \fignote{} }\label{figp5}
\end{figure}

\begin{figure}[!ht]
\centering
\subfloat{\includegraphics[width=10.25cm]{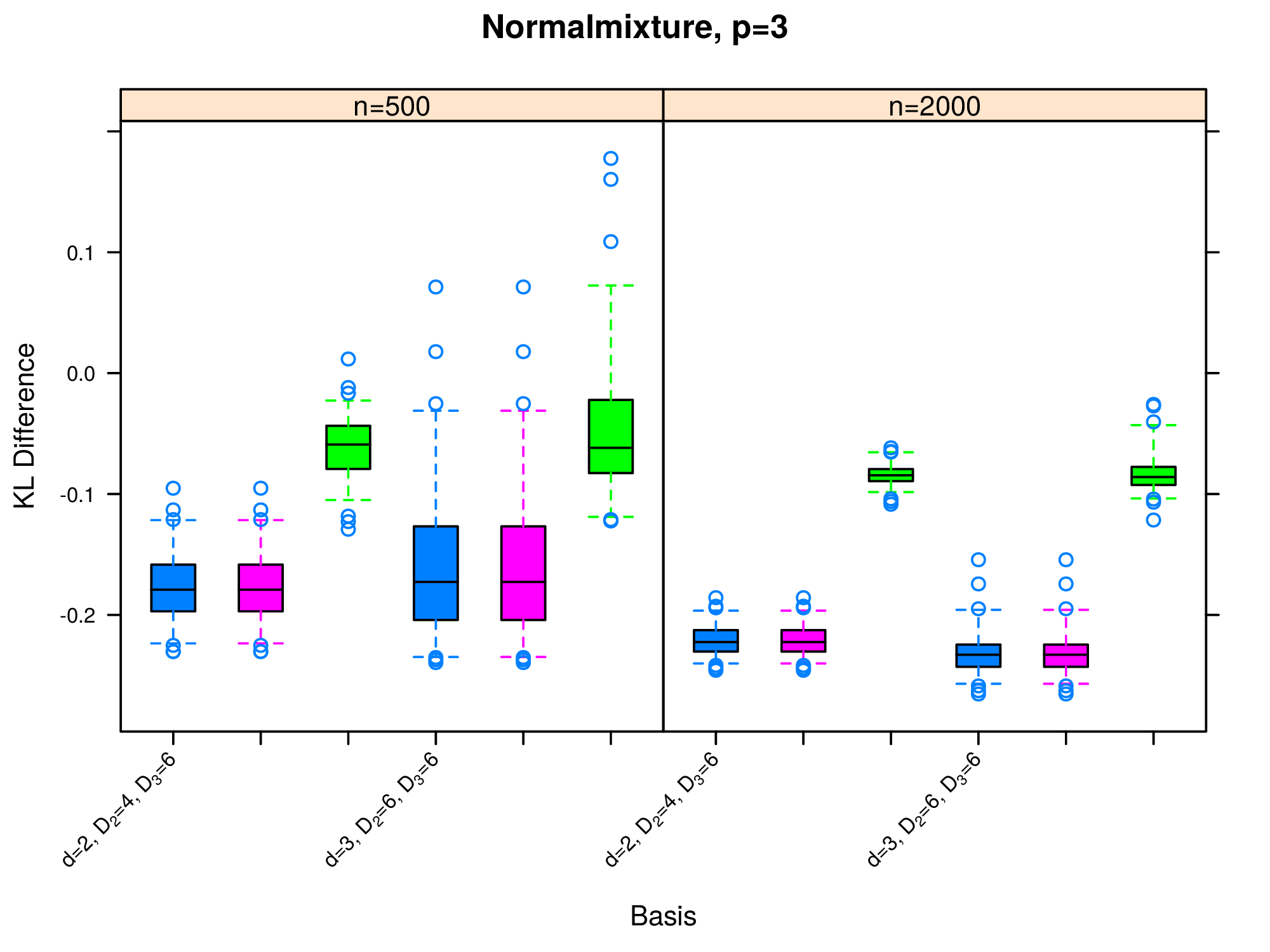}}\\
\subfloat{\includegraphics[width=10.25cm]{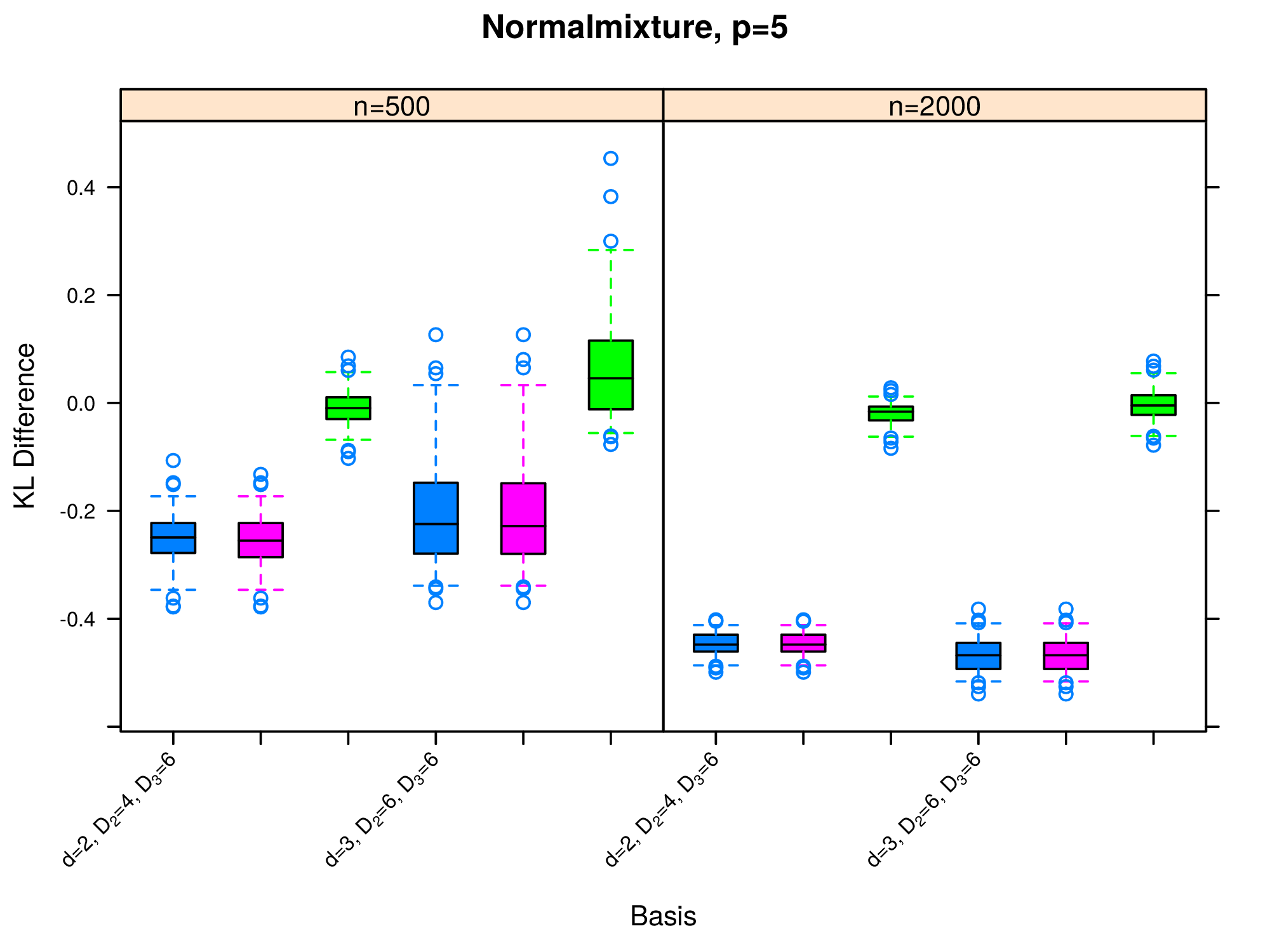}}
\caption{Box plots of the differences $KL_{non-par}-KL_{par}$ for three- and five-dimensional mixtures of normal distributions. \fignote{}}\label{fignorm}
\end{figure}

\begin{figure}[!ht]
\centering
\subfloat{\includegraphics[angle=-90,width=7.5cm]{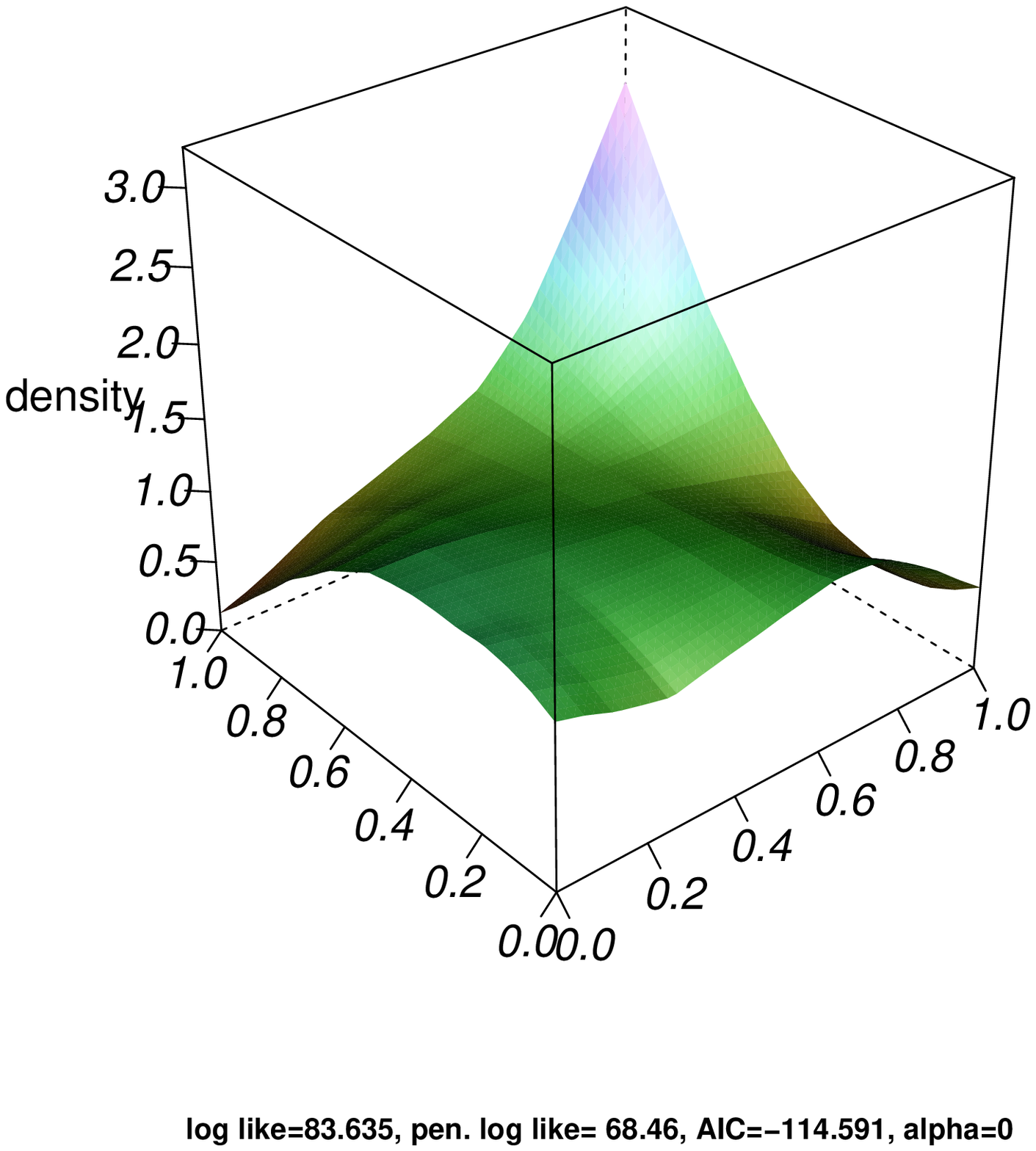}}
\subfloat{\includegraphics[angle=-90,width=7.5cm]{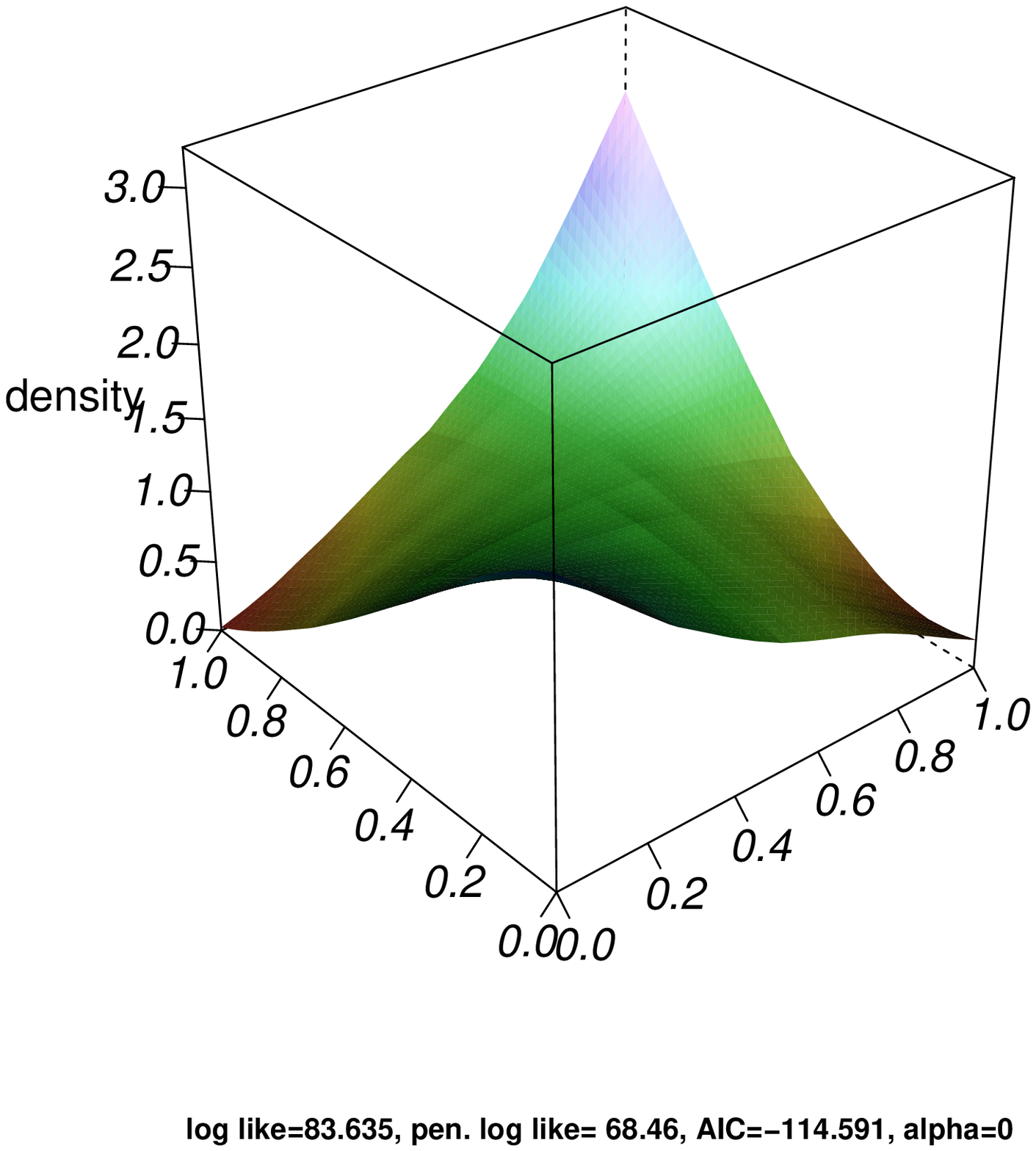}}\\
\subfloat{\includegraphics[angle=-90,width=7.5cm]{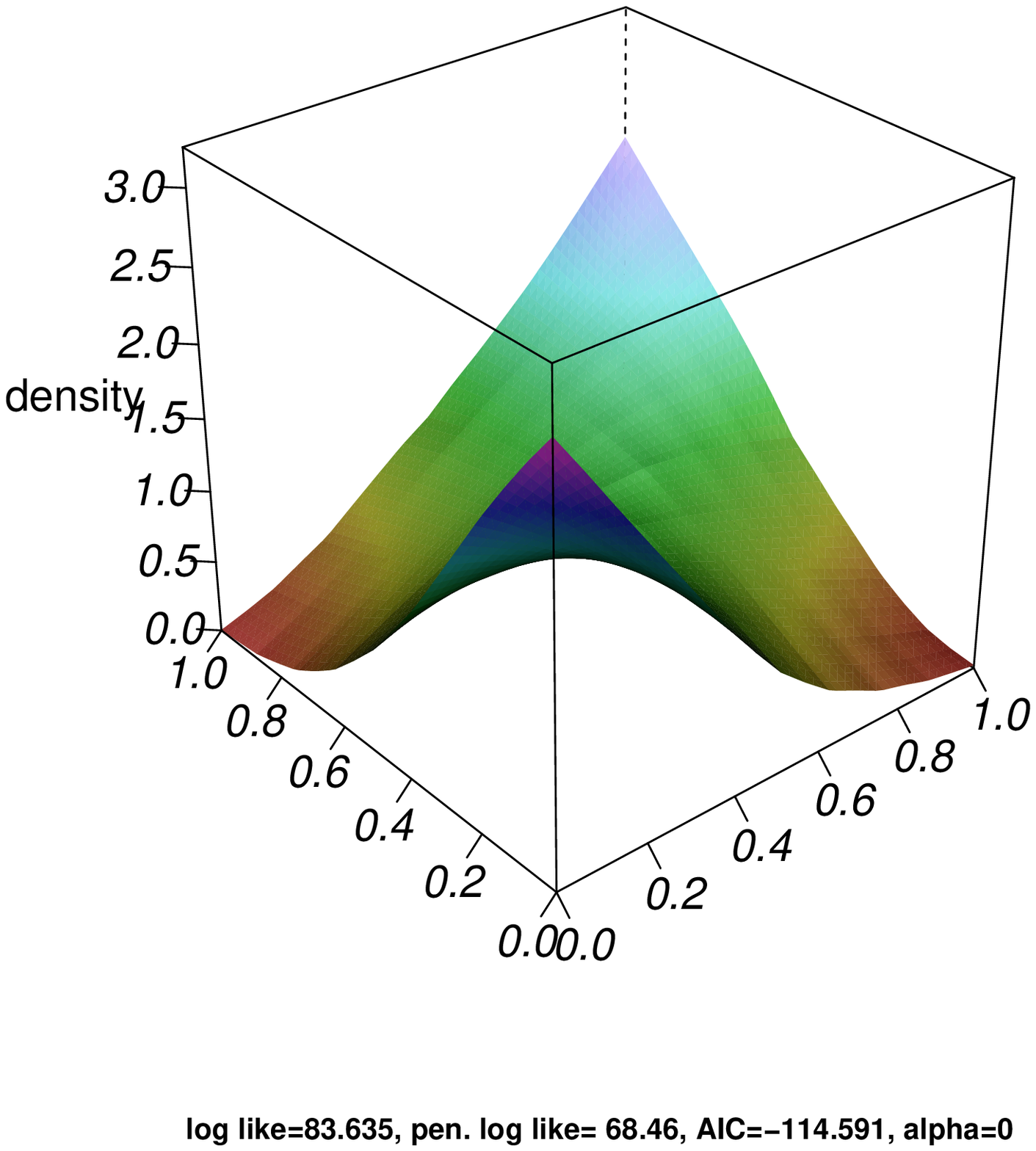}}
\subfloat{\includegraphics[angle=-90,width=7.5cm]{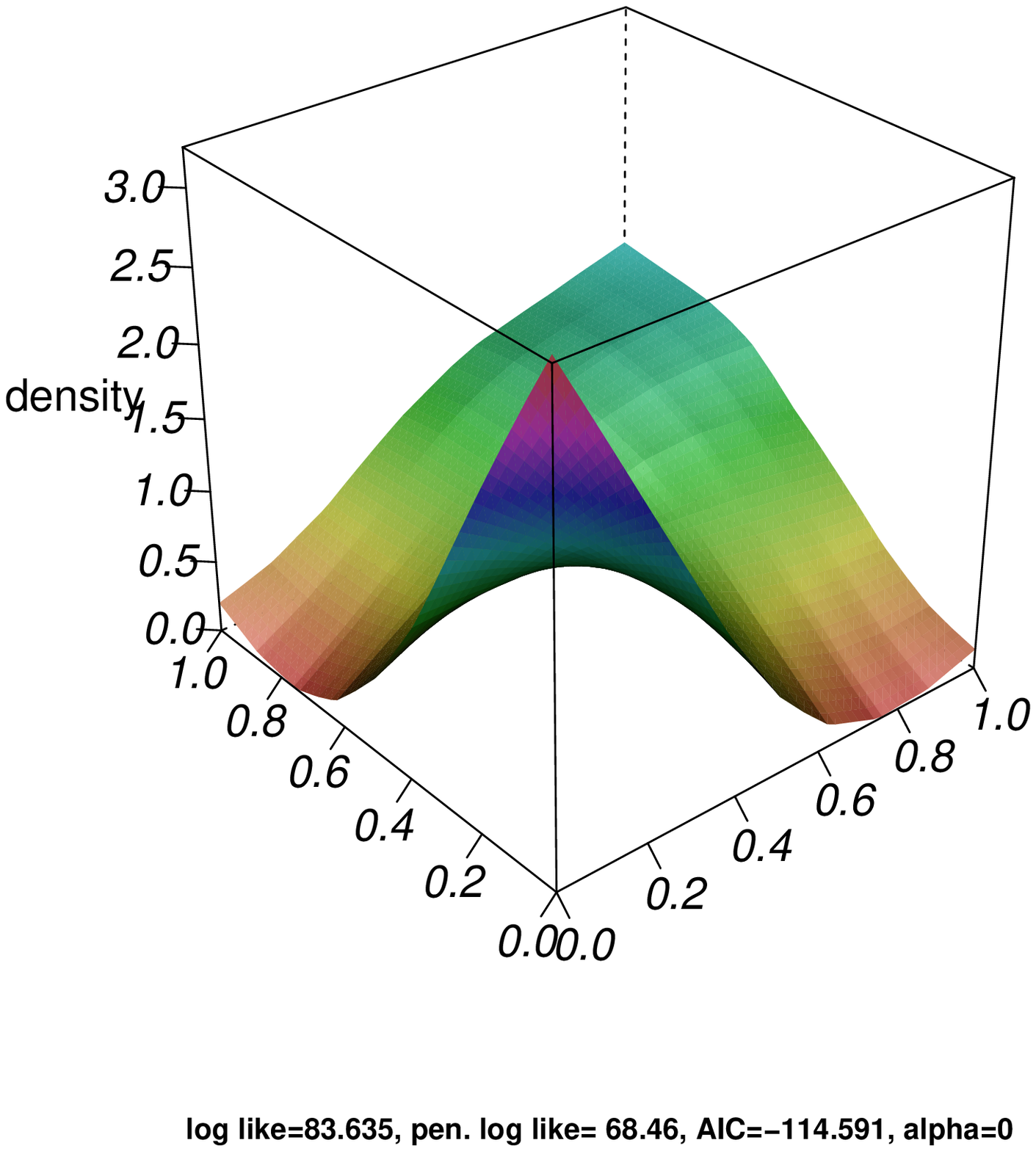}}\\
\subfloat{\includegraphics[angle=-90,width=7.5cm]{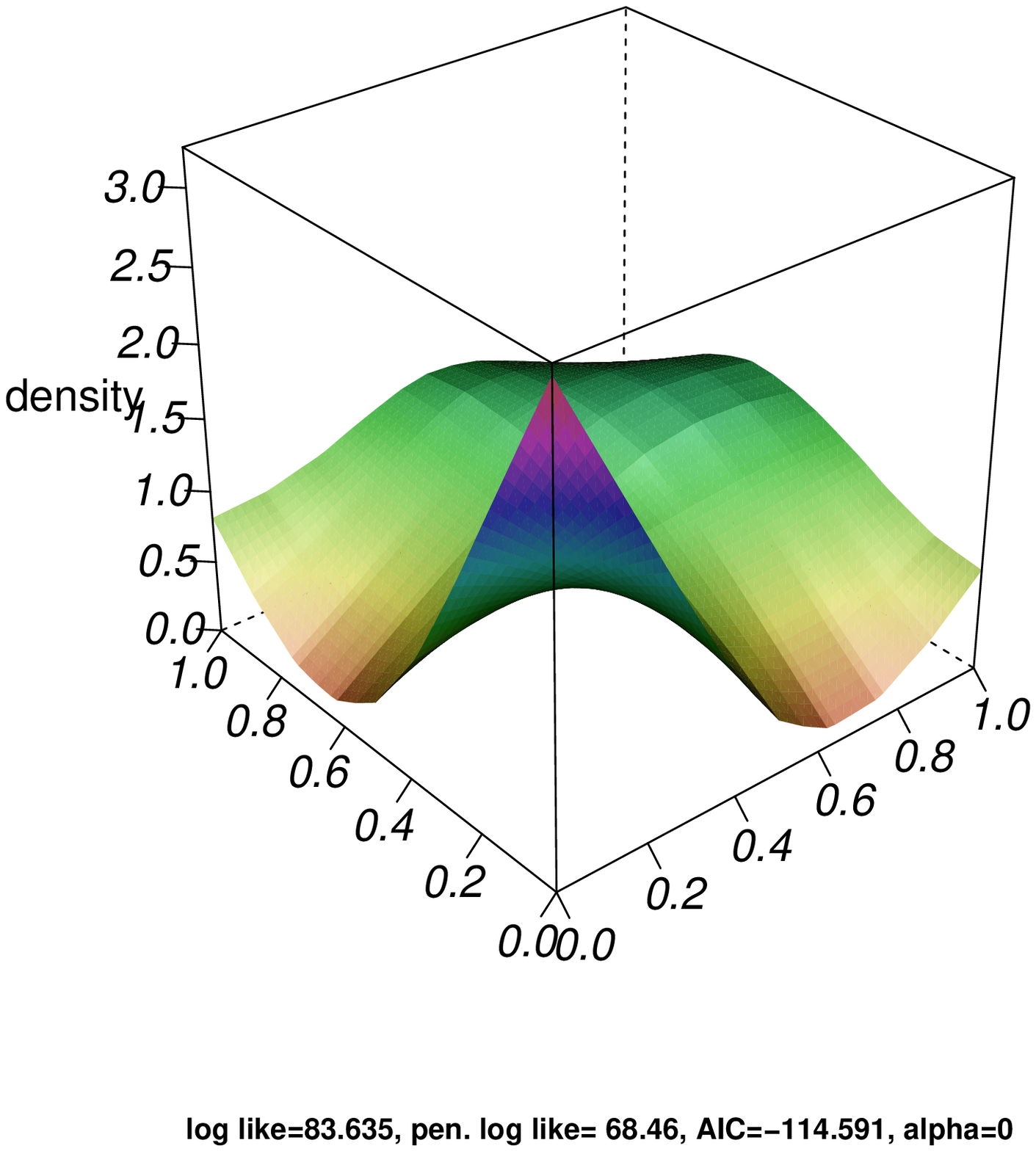}}
\subfloat{\includegraphics[angle=-90,width=7.5cm]{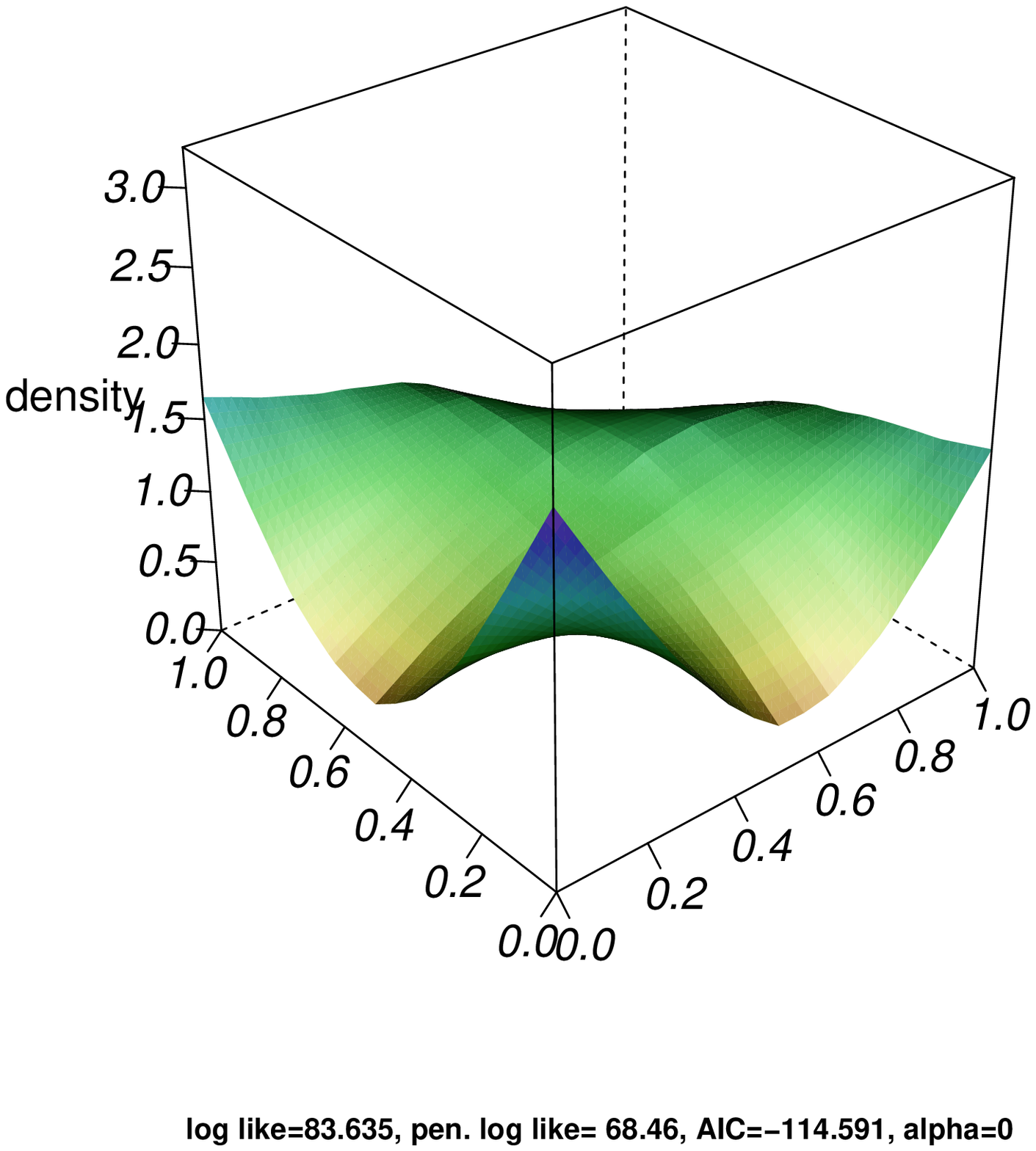}}\\
\subfloat{\includegraphics[angle=-90,width=7.5cm]{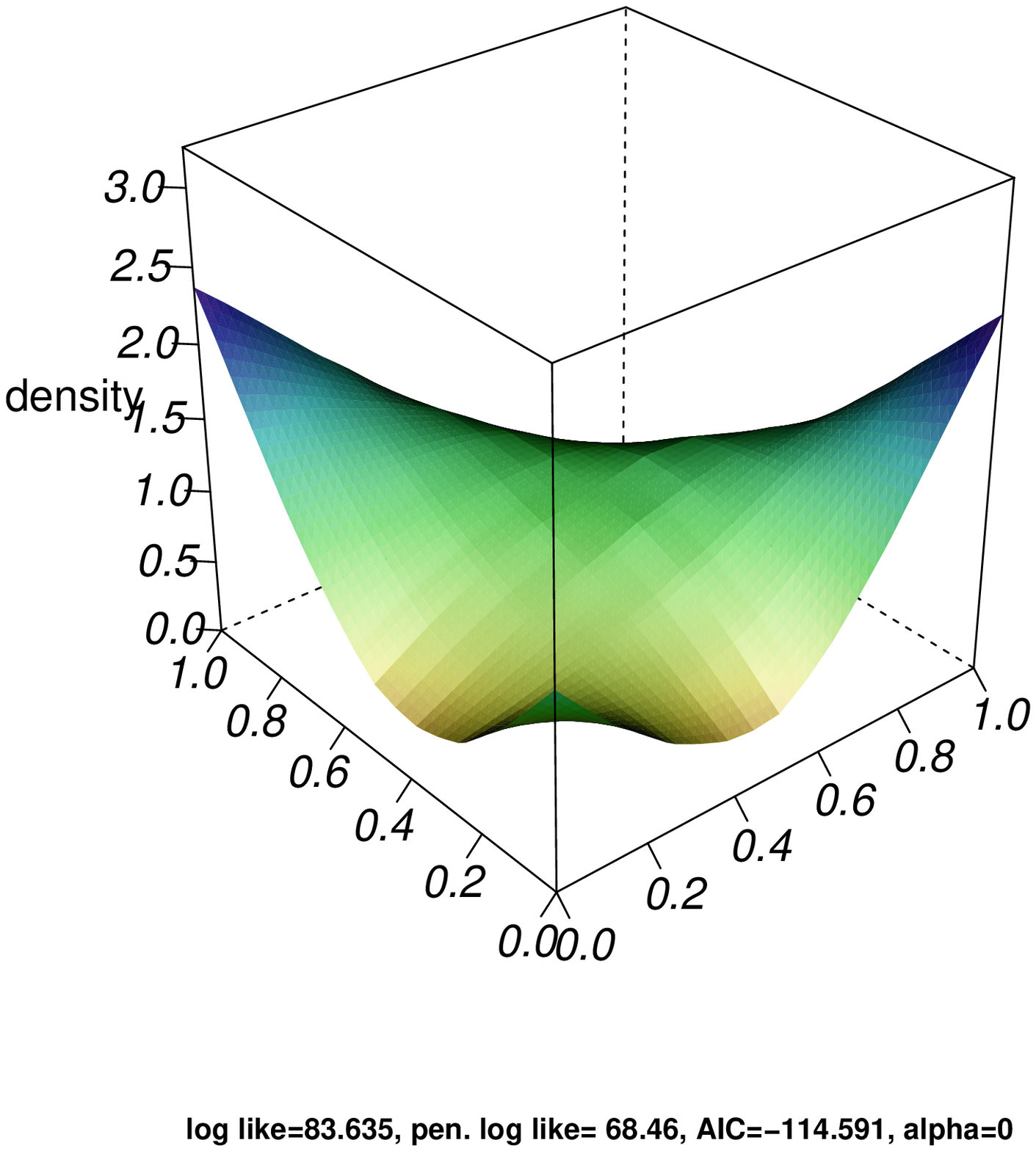}}
\subfloat{\includegraphics[angle=-90,width=7.5cm]{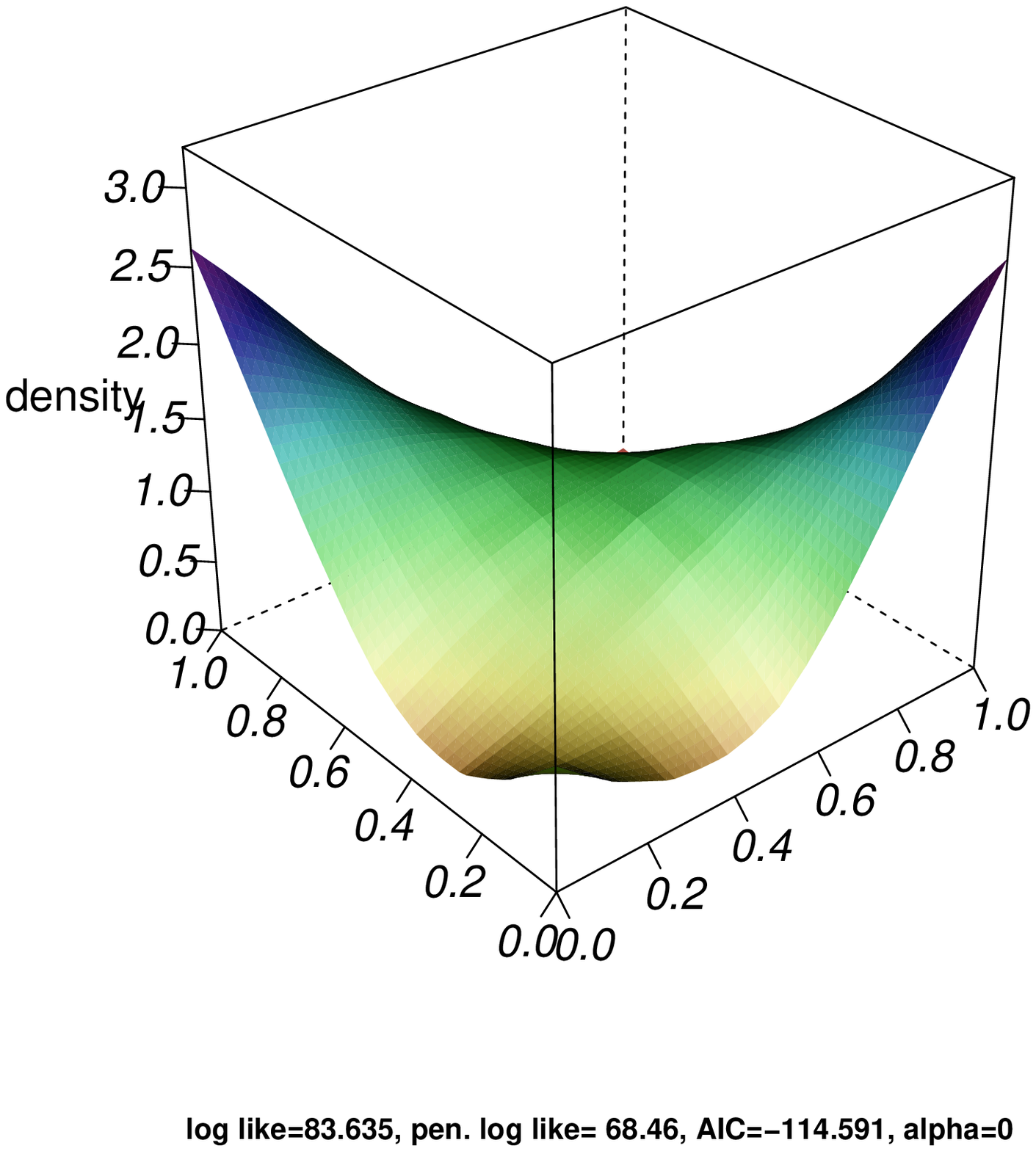}}
\caption{Conditional copula densities of the three-dimensional mixture of normal distributions for conditional arguments $0.01, 0.15, 0.29, 0.43, 0.57, 0.71, 0.85, 0.99$ (top left to bottom right).}\label{fignormp3-ex}
\end{figure}

\begin{figure}[!ht]
\centering
\includegraphics[width=16cm]{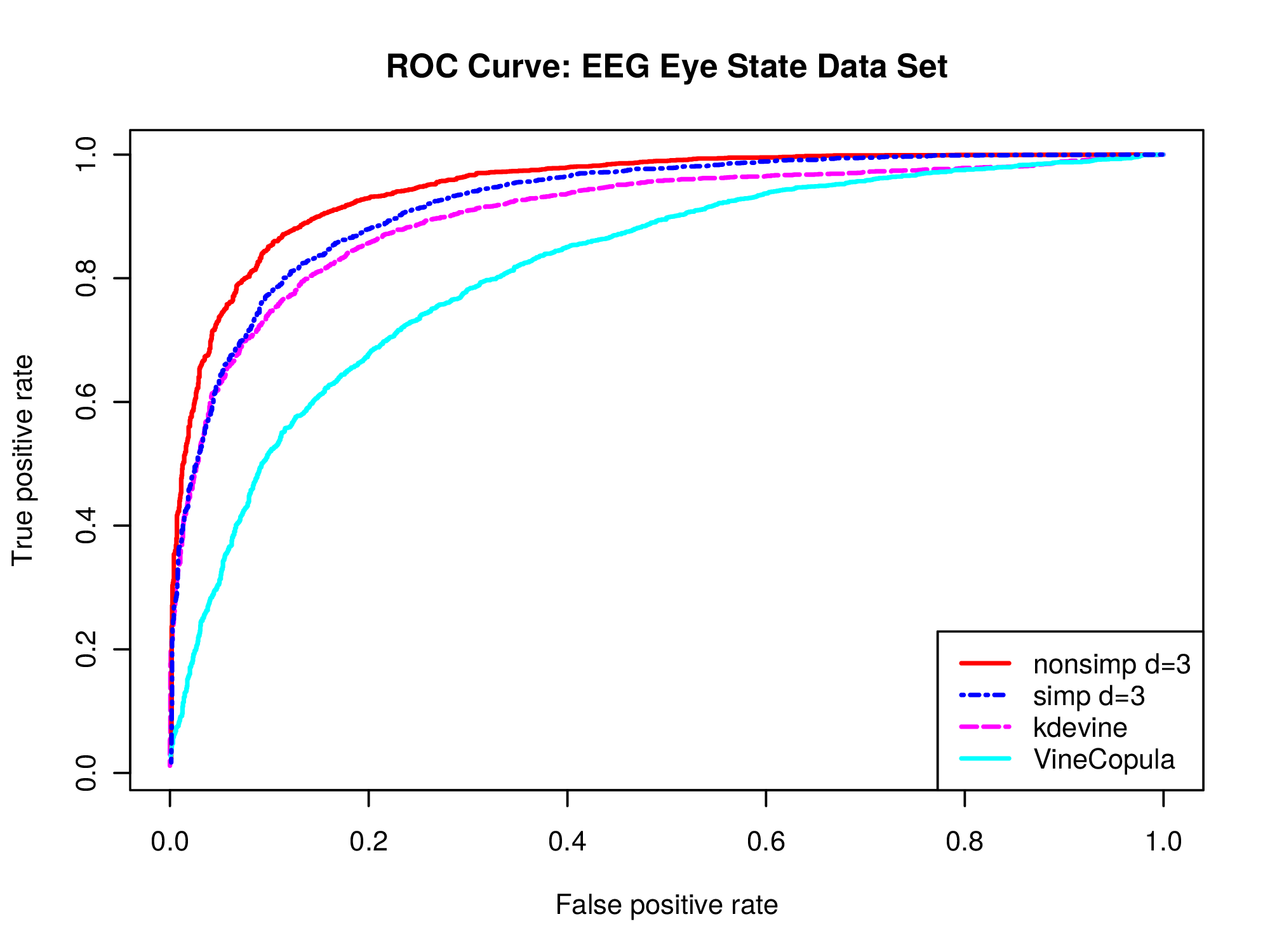}
\caption{ROC-Curve for classification of EEG Eye State Data Set.}\label{figclass}
\end{figure}

\onecolumn

\begin{table}[!ht]
\centering
\small
\caption{
Dimension of full tensor product basis and sparse hierarchical basis of linear B-splines for a bivariate (un)conditional bivariate copula density. $K=2^d+1$ is the number of marginal equidistant knots, $d$ is the degree of the univariate hierarchical B-spline basis and $D_2$ and $D_3$ denote the maximum cumulated hierarchy level, respectively. The full tensor product is abbreviated as full.}
\label{Table1}
\begin{tabular}{lll|cc|cc}
& & &\multicolumn{2}{c}{Unconditional copula}
& \multicolumn{2}{|c}{Cond. copula (one conditioning argument)
}
\\ \hline
$d$ & $D_2$& $D_3$ &basis & number of coefficients 
&basis & number of coefficients
 \\
\hline
2, (K=5) & 4& 4 &  full ($D_2=2d$) & 25 & sparse ($D_3=2d$) & 105\\
2, (K=5) & 4& 6 &  full ($D_2=2d$) & 25 & full ($D_3=3d$) & 125\\
3, (K=9) & 6& 6 &  full ($D_2=3d$) & 81 & sparse ($D_3=2d$) & 473\\
3, (K=9) & 6& 9 &  full ($D_2=3d$) & 81 & full ($D_3=2d$) & 729
 \end{tabular}
\end{table}

\begin{table}[!ht]
 \centering
\small
\caption{Log-likelihood values of fitted regular vine copulas for Uranium data set using (a) the estimator 'Test', (b) the estimator 'SimpA'  and (c) the parametric estimator 'VineCopula'.}
\begin{tabular}{l|l|c|c}\label{tab.uranium}
Type & Basis & Log-likelihood & Number Cond. Copula\\
\hline
(a)&\coefba & 1062.87 & 6\\
   &\coefaa &  978.76 & 6\\  
   \hline
(b)&\coefbS & 1029.53 & -\\
   &\coefaS & 926.25& -\\      
\hline
(c)& - & 918.97 & -\\
\hline
\end{tabular}
\end{table}

\begin{table}
\centering
\tiny
\caption{Mean elapsed computing time in seconds (standard deviation) for i) bivariate unconditional copula densities, ii) bivariate conditional copula densities with one conditioning argument and iii) the estimation of the whole simplified and non-simplified vine copulas with 'Test' estimator. Measured for $N=10$ runs.}\label{comp.time}
\begin{tabular}{|c|c|c||c|c||c|c||c|c|}
\hline
& \multicolumn{2}{|c||}{i) $c_{12}$}& \multicolumn{2}{|c||}{ii) $c_{12|3}$}& \multicolumn{2}{|c||}{vine copula $c_{1:3}$}& \multicolumn{2}{|c|}{vine copula $c_{1:5}$}\\
basis & $n=500$ & $n=2000$& $n=500$ & $n=2000$& $n=500$ & $n=2000$& $n=500$ & $n=2000$\\
\hline
\coefaS & 1.3 (0.6) & 0.9 (0.3) & - & - &  4.2 (1.2) & 3.4 (0.5) & 16.5 (1.6)& 14.0 (1.8)\\
\coefbS & 2.9 (1.7) & 1.7 (0.8) & - & - &  8.0 (2.3) & 6.0 (0.8) & 34.7 (4.4)& 27.8 (4.9)\\ \hline
\coefaa & - & - & 1.4 (0.3) & 2.0 (0.2) & 46.1 (22.8)& 19.1 (7.3) & 394.1 (80.7) & 177.2 (40.0)\\
\coefba & - & - & 34.6 (23.7)& 33.9 (5.5) & 52.1 (22.2) & 20.5 (5.5)& 400.4 (63.3) & 189.5 (40.0)\\ \hline
\end{tabular}
\end{table}
\end{document}